\documentclass[showpacs,eqsecnum,aps]{revtex4}
\usepackage{amsmath}
\usepackage{amssymb}
\usepackage{amsfonts}
\def\beq{\begin{equation}}
\def\eeq{\end{equation}}
\def\beqa{\begin{eqnarray}}
\def\eeqa{\end{eqnarray}}
\def\bsig{\mbox{\boldmath$\sigma$}}

\def\bphi{\mbox{\boldmath$\phi$}}

\def\btau{\mbox{\boldmath$\tau$}}
\def\brho{\mbox{\boldmath$\rho$}}
\def\bkappa{\mbox{\boldmath$\kappa$}}
\def\bLam{\mbox{\boldmath$\Lambda$}}
\usepackage[dvips]{graphicx,color}

\begin{document}

\title{Exchange current contributions in null-plane 
models of elastic electron deuteron scattering}

\author{Y. Huang}
\affiliation{
Department of Physics and Astronomy, The University of Iowa, Iowa City, IA
52242}

\author{W. N. Polyzou}
\affiliation{
Department of Physics and Astronomy, The University of Iowa, Iowa City, IA
52242}

\vspace{10mm}
\date{\today}

\begin{abstract}
We investigate the effects of two-body currents on elastic
electron-deuteron scattering in an exactly Poincar\'e invariant
quantum mechanical model with a null-plane kinematic symmetry.  While
calculations using single-nucleon currents as input produce good
qualitative agreement with experiment, the two-body current that
we investigate produces a good quantitative agreement between theory
and experiment for all three elastic scattering observables.

\end{abstract}

\vspace{10mm}


\maketitle

%\newpage

%****************************************************************************

%\narrowtext

%****************************************************************************

\section{Introduction}

In the one-photon-exchange approximation elastic electron-deuteron
scattering observables are functions of the current matrix elements, $\langle
(1,d) {P}'_d ,\mu'_d \vert I^{\mu} (0) \vert (1,d) {P}_d , \mu
\rangle$, where $\vert (1,d) {P}_d , \mu \rangle$ are deuteron four-momentum
eigenstates and $I^\mu(x)$ is the deuteron current
density operator.

In a relativistic quantum theory the deuteron eigenstates transform
irreducibly with respect to a dynamical unitary representation of the
Poincar\'e group.  In this work they are eigenstates of a two-nucleon
mass operator constructed by adding ``realistic'' nucleon-nucleon
interactions, such as Argonne V18 or CD Bonn
\cite{Wiringa95}\cite{cdb}, to the square of the non-interacting
invariant mass operator, $M^2 = M_0^2 + 4mv_{nn} $.  With this mass
operator the S-matrix elements of the resulting two-nucleon model are
consistent with the nucleon-nucleon scattering data used to determine
the interaction.

The representation of the dynamics can additionally be chosen so the
interaction $v_{nn}$ is invariant with respect to a non-interacting,
``kinematic'' subgroup of the Poincar\'e group.  Different choices of
kinematic subgroup \cite{Dirac49} are related by unitary
transformations that do not change the two-body S-matrix
\cite{Ekstein60}, but the choice of representation of the dynamics
effects the representation of the current operator.  In this work the
interaction is taken to be invariant with respect to a subgroup of
the Poincar\'e group that leaves a plane tangent to the light cone
invariant. This is called the null-plane kinematic subgroup.  It is the
largest possible kinematic subgroup and it has well-known advantages
for treating electron-scattering problems.

The hadronic current density operator is conserved and covariant with
respect to the dynamical representation of the Poincar\'e group.  It
is necessarily the sum of one-body and two-body operators.  Cluster
properties imply that one-body current operators, which are determined
by empirical nucleon form factors, are conserved and transform
covariantly with respect to the non-interacting representation of the
Poincar\'e group.  In a dynamical model the one-body parts of the
current are only covariant with respect to the kinematic subgroup.

Two-body parts of the current operator are needed to restore full
covariance and may include additional contributions of a dynamical
origin.  Deuteron matrix elements of an exactly covariant hadronic
current are constrained by the Poincar\'e transformation properties of
the current operator and deuteron state vectors.  These linear
relations can be used to express all of the current matrix elements in
terms of three linearly-independent current matrix elements, which are
linearly related to invariant deuteron form factors.  This current is
also covariant with respect to the null-plane kinematic subgroup.  The
kinematic subgroup is relevant because the one and two-body parts of
the current operator are separately covariant with respect to the
kinematic subgroup.  This more limited symmetry allows all current
matrix elements to be expressed in terms of four linearly independent
current matrix elements, using only null-plane kinematic
transformations.  The full covariance implies an additional dynamical
constraint among these four kinematically independent matrix elements.
In a dynamical model with a null-plane kinematic symmetry the
additional dynamical constraint is rotational covariance
\cite{Leutwyler78}\cite{Grach84}\cite{Chung88} \cite{CHH97} which was
appropriately referred to as the ``angular condition'' by Leutwyler
and Stern.

Any kinematically covaraint current operator can be used to define
matrix elements of a fully covariant current by evaluating any three
linearly independent current matrix elements using the kinematically
covariant current operator.  The remaining current matrix elements are
generated by the constraints implied by covariance, current
conservation, and parity.  This construction ensures that the fully
covariant current and kinematically covariant current operators agree
on the three independent matrix elements and {\it all matrix elements
related to them by kinematic Poincar\'e transformations}.  The
resulting covariant current depends on the choice of the three
linearly independent matrix elements, which are chosen among the four
kinematically independent matrix elements related by the angular
condition.

Null-plane ``impulse calculations'' compute the three independent
current matrix elements using the single-nucleon current operators.
When the null plane is oriented so the ``+'' component of the
momentum transfer, $Q^+=Q^0+Q^3$, is zero the independent matrix
elements can be chosen to be matrix elements of the $+$ component,
$I^+(0)= I^0(0) + I^3(0)$ of the current.  When the independent matrix
elements are taken to be matrix elements of the $+$ component of the
current the one-body matrix elements are (1) invariant with
respect to null-plane boosts (2) the single-nucleon form factors
exactly factor out of the integral for the current matrix element and
(3) the momentum transfer to the nucleons is identical to the momentum
transfer to the deuteron.

When used with realistic nucleon-nucleon interactions and empirical
nucleon form factors, null-plane impulse calculations give a good
qualitative description of all of the elastic electron-deuteron
scattering observables.  The results are not sensitive to the choice
of realistic interaction, empirical nucleon form factors, or the
choice of independent matrix elements used to construct the fully
covariant current.  Quantitative differences between the results of
null-plane impulse calculations and experiment, particularly in the
observables $A$ and $T^{20}$ cannot be explained by experimental
uncertainties or uncertainties in theoretical input.

In order to account for the difference between theory and experiment
it is necessary to introduce an additional two-body current operator
that has a non-vanishing contribution to the set of independent
current matrix elements used to generate the fully covariant current.
In this paper we investigate the consequences of including an
additional two-body current that has an operator structure motivated
by ``pair-current'' contributions in covariant formulations of elastic
electron-deuteron scattering.  In this work the two-body current is
constructed to be covariant with respect to the null-plane kinematic
subgroup.

Null-plane quantum mechanical models use Poincar\'e covariant
two-component spinors rather than Lorentz covariant four-component
spinors.  Dirac spinors, which are spinor representations of Lorentz
boosts, transform between the two- and four-component spinor
representations.  In the four-component representation, interactions
that have a $\gamma_5$ vertex include contributions that couple to the
photon through a causal propagator that do not appear in the
null-plane impulse calculation.  In the two-component spinor
representation this contribution can be included as a two-body
operator in the current.  These considerations determine the structure
of our model two-body current.  The operator is constructed to satisfy
exact null-plane kinematic covariance.  It is extended to a fully
covariant current in the same way that the one-body current in the
``null-plane impulse'' calculations is extended to a fully covariant
current.

The resulting model current, when added to the current generated by
the empirical nucleon form factors, leads to a good quantitative
description of all of the elastic electron-deuteron scattering
observables.  Ambiguities related to the implementation of the angular
condition are investigated.

Elastic electron-deuteron scattering has been studied by many
authors \cite{Gross65a}\cite{Terent76}\cite{Terent77}
\cite{acg79}\cite{Grach84}   
\cite{Chung88}\cite{Schiavilla91} \cite{VanOrden95}\cite{Phillips98}
\cite{Karmanov99} \cite{Arenh00}\cite{Lev00}\cite{Allen01}
\cite{VanOrden01}\cite{Phillips05}\cite{kruto06} using a variety of
different methods and assumptions. 
In all cases the problem is
to construct matrix elements of a conserved covariant current operator
%\beq
%\langle d, {P}+{Q}, \mu' \vert I^{\nu}(0) \vert
%d, {P} , \mu \rangle
%\label{a.1}
%\eeq
for a range of space-like momentum transfers, $Q$.
% where $P$ and $P+Q$ are the initial and final deuteron four momenta.  
The calculations
discussed in this work are based on quantum mechanical models where
the Poincar\'e group is an exact symmetry of the underlying theory and
the interaction has a null-plane kinematic symmetry.  Exactly
Poincar\'e invariant quantum mechanical models using other choices of
kinematic subgroups have also been applied to electron-deuteron
scattering.  Different choices of kinematic subgroup have dynamical
consequences for the impulse calculations.  Another class of
calculations are based on quasipotential treatments.  These
calculations extract the current matrix elements from
covariant amplitudes without directly using the underlying quantum
theory.  This is done by generalizing Mandelstam's \cite{Mandelstam55}
method to extract current matrix elements from Green functions using
Bethe-Salpeter wave functions.  ``Relativistic impulse
approximations'' in this formalism depend on the specific
quasipotential reduction, but they generally include a pair-current
contribution that is related to the two-body current used in this
paper.  Detailed reviews can be found in \cite{VanOrden01}
\cite{Gilman02}.

The Poincar\'e invariant dynamical models in this paper are formulated
in a representation with a null-plane kinematic symmetry
\cite{Leutwyler78} \cite{Coester82}\cite{Keister91}.  Unlike models
based on null-plane quantum field theory
\cite{Chang73}\cite{Chang73b}\cite{CHH97}, where few-degree of freedom
truncations break rotational covariance, the class of quantum
mechanical models that we consider are {\it exactly} rotationally
invariant.  The null-plane treatment of electron scattering has the
following advantages:

\begin{itemize}

\item [1.] For electron scattering the momentum transfer, $Q$, is
  spacelike, so the orientation of the null plane can be chosen so
  the $+$-component, $Q^+:= Q^0+Q^3$, of the momentum transfer is
  zero.  When $Q^+=0$ all current matrix elements can be constructed
  from three independent matrix elements of the $+$
  component, $I^+ (0) :=I^0 (0)+ I^3 (0)$, of the current:

\beq
\langle (1,d), \tilde{\mathbf{P}}', \mu' \vert I^+(0) \vert 
(1,d), \tilde{\mathbf{P}} ,\mu \rangle
\qquad \tilde{\mathbf{P}} := (P^+,P_1,P_2).   
\label{a.2}
\eeq
\item[] where we have labeled the deuteron four momenta by their
null-plane components, $\tilde{\mathbf{P}}$, and the spins are 
null plane spins (defined in section 2).

\item [2.]  There is a three-parameter subgroup of Lorentz boosts that
  leaves the null plane invariant.  Because these boosts form a
  subgroup, there are no Wigner rotations associated with null-plane
  boosts.  It follows that matrix elements of $I^+(0)$ with $Q^+=0$
  and {\it delta-function normalized initial and final states} are
  independent of frames related by null-plane boosts:

\beq
\langle (1,d), \tilde{\mathbf{P}}''', \mu' \vert I^+(0) \vert 
(1,d), \tilde{\mathbf{P}}'' ,\mu \rangle  
= 
\langle (1,d), \tilde{\mathbf{P}}', \mu' \vert I^+(0) \vert (1,d), 
\tilde{\mathbf{P}} ,\mu \rangle 
\label{a.3}
\eeq
\item[] where $P'''= \Lambda P'$, $P''=\Lambda P$ and $\Lambda$ is any
  null-plane preserving boost.  This means that matrix elements of
  $I^+(0)$ are equal to null-plane Breit-frame matrix elements with
  the same null-plane spin magnetic quantum numbers.  The null-plane
  Breit-frame matrix elements are defined to have momentum transfer
  perpendicular to the unit vector, ($\hat{\mathbf{z}}$), that defines
  the orientation of the null plane.  This is consistent with the
  requirement that $Q^+=0$.

\item [3.]  A consequence of property two is that the {\it one-body}
  current matrix elements of $I^+(0)$ with the physical momentum
  transfer {\it exactly} factor out of the corresponding nuclear
  current matrix elements:
\[
\langle \Psi, \tilde{\mathbf{P}}+\tilde{\mathbf{Q}} 
\vert I_i^+(0) \vert \Psi, \tilde{\mathbf{P}} \rangle =
\]
\beq
\langle \tilde{\mathbf{p}}_{i0}+\tilde{\mathbf{Q}} \vert 
I_i^+ (0) \vert \tilde{\mathbf{p}}_{i0} \rangle 
\int  d\tilde{\mathbf{p}}_1 \cdots d\tilde{\mathbf{p}}_n
\langle \Psi, \tilde{\mathbf{P}}+\tilde{\mathbf{Q}} \vert
\tilde{\mathbf{p}}_1 \cdots \tilde{\mathbf{p}}_i
+\tilde{\mathbf{Q}}
\cdots
\tilde{\mathbf{p}}_n \rangle 
\langle 
\tilde{\mathbf{p}}_1 \cdots \tilde{\mathbf{p}}_i
\cdots
\tilde{\mathbf{p}}_n \vert
\Psi, \tilde{\mathbf{P}} 
\rangle .
\label{a.4}
\eeq

\item[] This follows directly from (\ref{a.3}) because covariance with
  respect to null-plane boosts can be used to remove the dependence of
  the constituent current matrix elements on the internal momenta
  without changing the spin sums or normalization.  A proof can be
  found in \cite{Keister91}.  This means that the matrix elements of
  the one-body contributions to $I^+(0)$ can be expressed as sums of
  products of null-plane Breit-frame nucleon matrix elements
  multiplied by functions of $Q^2$ that only depend on the initial and
  final wave functions.

\end{itemize} 

With this formalism it is possible to satisfy exact rotational
covariance and the impulse contribution to the independent current
matrix elements involves only experimentally observable on-shell
single-nucleon matrix elements.

This paper is organized as follows.  In section two we introduce our
notation and define the null-plane kinematic subgroup.  In section
three we construct a dynamical representation of the Poincar\'e group
with a null-plane kinematic symmetry that provides a realistic
description of the two-nucleon system.  In section four we review the
experimental observables for elastic electron-deuteron scattering and
their relation to current matrix elements.  In section five we discuss
the nucleon currents that are used to construct the one-body part of
our current operator.  In section six we use the Wigner-Eckart theorem
for the Poincar\'e group to realize the dynamical constraints on the
current, and we discuss some of the ambiguities in the implementation
of the angular condition.  In section seven we define our dynamical
two-body current.  In section eight we summarize our results.
Appendix A summarizes how we relate our two- body current to the
pion-exchange part of the model interaction, and appendix B contains
additional details related to how this is done with the AV18
interaction.

\section{Null-plane kinematics}

In this section we discuss our notation and the null-plane
kinematic subgroup introduced 
by Dirac \cite{Dirac49}.  The null plane is the three-dimensional
hyperplane of points tangent to the light cone satisfying the
condition
\beq
\{ x \vert x^+ := t + \mathbf{x} \cdot \hat{\mathbf{e}}_3 = 0 \} .
\label{b.1}
\eeq
The null-plane components,
$\tilde{\mathbf{x}}= (x^+ , \mathbf{x}_{\perp})$  of the four 
vector $x^{\mu}$ are 
\beq
x^{\pm} :=  t \pm \mathbf{x} \cdot \hat{\mathbf{e}}_3
\qquad 
\mathbf{x}_{\perp} = (\mathbf{x} \cdot \hat{\mathbf{e}}_1,
\mathbf{x} \cdot \hat{\mathbf{e}}_2).
\label{b.2}
\eeq
Four vectors $x^{\mu}$ can be represented by $2\times 2$ 
Hermitian matrices in the null-plane components of $x^{\mu}$:
\beq
X = \left (
\begin{array}{cc} 
x^+ & x_{\perp}^* \\  
x_{\perp} & x^-
\end{array}
\right ) = x^{\mu} \sigma_{\mu} \qquad
x_{\perp}=x_1 + i x_2 \qquad x^{\mu} = {1 \over 2} \mbox{Tr}(\sigma_{\mu} X)
\label{b.3}
\eeq
where $\sigma_{\mu}= (I, \bsig )$ and $\bsig$ are the two $\times$ two
Pauli matrices.  Since the determinant of $X$ is minus the square of the
proper time, $-x^2$, if $\Lambda$ is a complex $2\times 2$ matrix with
unit determinant then the transformation
\beq
X\to X' = \Lambda X \Lambda^{\dagger} 
\label{b.4}
\eeq 
defines a real Lorentz transformation, 
\beq
\Lambda^{\mu}{}_{\nu} = 
{1 \over 2}\mbox{Tr}(\sigma_{\mu}\Lambda \sigma_{\nu} \Lambda^{\dagger} ).
\label{b.5}
\eeq
We use the notation $\Lambda$ to represent both the $2\times 2$ and
$4\times 4$ representation of Lorentz transformations; the implied
representation is easily determined.  Points on the null-plane can be
represented by triangular matrices of the form
\beq
X = \left (
\begin{array}{cc} 
0 & x_{\perp}^*  \\  
x_{\perp} & x^-
\end{array}
\right ) . 
\label{b.6}
\eeq
Poincar\'e transformations
\beq
X \to X' = \Lambda X \Lambda^{\dagger} +A  
\label{b.7}
\eeq
that leave the null plane invariant (i.e. preserve the form  of
(\ref{b.6})) have the form 
\beq
\Lambda  = \left (
\begin{array}{cc} 
\alpha  & 0  \\  
\beta & 1/\alpha
\end{array}
\right ) 
\qquad 
A  = \left (
\begin{array}{cc} 
0  & a_{\perp}^*  \\  
a_{\perp} & a^-
\end{array}
\right ) = a^{\mu}\sigma_{\mu}, 
\label{b.8}
\eeq
where $\alpha \not= 0$, $\beta$, and $a_{\perp}$ are complex and $a^-$
is real.  These transformations form a seven-parameter subgroup of the
Poincar\'e group that leaves the null plane invariant.  This subgroup
is called the {\it kinematic subgroup} of the null plane.  This
subgroup includes the three-parameter subgroup of translations on the
null plane, a three-parameter subgroup of null-plane-preserving
boosts as well as rotations about the $3$-axis.

The null plane-preserving-boost that transforms a rest 4-momentum 
$(m,\mathbf{0})$ to a value $p$ is the matrix-valued function of 
$\tilde{\mathbf{v}}=\tilde{\mathbf{p}}/m:= (p^+/m, p^1/m, p^2/m)$ given by
\beq
\Lambda_f (\tilde{\mathbf{p}}/m) := \Lambda_f (\tilde{\mathbf{v}}) :=
\left (
\begin{array}{cc} 
\sqrt{v^+}  & 0  \\  
v_{\perp}/\sqrt{v^+}  & 1/\sqrt{v^+}
\end{array}
\right ) .
\label{b.9}
%\label{b.10}
\eeq
Since the reality of $\alpha$ in (\ref{b.8}) is preserved under matrix
multiplication, the null-plane boosts, (\ref{b.9}), form 
a {\it subgroup} of the Lorentz group.

The action of a null-plane boost on an arbitrary four momentum vector is
determined by using (\ref{b.9}) in (\ref{b.4}).  
The resulting transformation property of the $+$ and $\perp$
components of the four momentum is
\beq
p^+ \to p^{+ \prime} = v^+ p^+ 
\qquad 
\mathbf{p}_{\perp} \to \mathbf{p}_{\perp}'=
\mathbf{p}_{\perp} + \mathbf{v}_{\perp} p^+ .
\label{b.10}
%\label{b.12}
\eeq
Since $p^-$ does not appear in (\ref{b.10}) the three components
$\tilde{\mathbf{p}}:= (p^+, \mathbf{p}_{\perp})$ are called a
``null-plane vector''.  The $-$ component can be calculated using the
mass-shell condition
\beq
p^- = {m^2 + \mathbf{p}_{\perp}^2 \over p^+} .
\label{b.11}
%\label{b.14}
\eeq
The null-plane spin of a particle of mass $m$ is defined
\cite{Leutwyler78}
\cite{Polyzou89}\cite{Keister91} so that (1) it agrees with ordinary
canonical spin in the particle's rest frame and (2) is invariant with
respect to null-plane boosts, (\ref{b.9}).

The null-plane boost (\ref{b.9}) differs from the more standard 
rotationless boost, which has the $2\times 2$ matrix form  
\beq
\Lambda_c ({\mathbf{k}}/m) = e^{{1 \over 2} \brho \cdot \bsig},
\label{b.19}
\eeq
where $\brho$ is the rapidity 
\beq
\brho = \hat{\mathbf{k}} \vert \brho \vert \qquad  
\cosh (\vert \brho \vert ) = \sqrt{\mathbf{k}^2+ m^2}/m 
\qquad \sinh (\vert \brho \vert ) = \mathbf{k}/m .
\label{b.20}
\eeq

The null-plane representation of the single-nucleon Hilbert space,
${\cal H}_1$, is the space of square integrable functions of the
null-plane vector components of the particle's four momentum and the
3-component of its null-plane spin:
\beq
\psi (\tilde{\mathbf{p}}, \mu) =  
\langle (j,m), \tilde{\mathbf{p}}, \mu \vert \psi \rangle 
\qquad
\int_0^\infty dp^+ \int_{\mathbb{R}^2} 
d\mathbf{p}_{\perp} \sum_{\mu=-j}^j  
\vert \psi (\tilde{\mathbf{p}}, \mu) \vert^2 < \infty .
\label{b.12}
%\label{b.15}
\eeq
The Poincar\'e group acts irreducibly on single-nucleon states 
\[
\langle (j,m) \tilde{\mathbf{p}}, \mu \vert U_1(\Lambda ,A) \vert \psi \rangle =
\]
\beq
\int_0^\infty dp^{+\prime} \int_{\mathbb{R}^2} 
d\mathbf{p}'_{\perp} \sum_{\mu'=-j}^j  
{\cal D}^{m,j}_{\tilde{\mathbf{p}},\mu; \tilde{\mathbf{p}}',\mu'}[\Lambda ,A]
\langle (j,m),
\tilde{\mathbf{p}}', \mu' \vert \psi \rangle ,
\label{b.13}
%\label{b.16}
\eeq
where the Poincar\'e group Wigner-${\cal D}$ function in the null-plane
irreducible basis \cite{Keister91} is
\[
{\cal D}^{m,j}_{\tilde{\mathbf{p}},\mu; \tilde{\mathbf{p}}',\mu'}
[\Lambda,A]:=  
\langle (j,m),\tilde{\mathbf{p}}, \mu \vert U(\Lambda ,A) 
\vert (j,m),\tilde{\mathbf{p}}', \mu' \rangle 
=
\]
\beq
\delta (\tilde{\mathbf{p}}  - \tilde{\bLam}(p') ) 
\sqrt{{p^+ \over p^{+\prime} }}
D^{j}_{\mu \mu'} [R_{fw} (\Lambda , \tilde{\mathbf{p}}') ]
e^{i p \cdot a}, 
\label{b.14}
%\label{b.17}
\eeq
$D^{j}_{\mu \mu'}[R]$ is the ordinary $SU(2)$ Wigner $D$-function
and 
\beq
R_{fw} (\Lambda , \tilde{\mathbf{p}}') := 
\Lambda_f^{-1} (\tilde{\mathbf{p}}/m) \Lambda 
\Lambda_f(\tilde{\mathbf{p}}'/m)
\eeq
is a null-plane Wigner rotation.  It is the identity matrix when
$\Lambda$ is a null-plane boost.

Our model Hilbert space for the two-nucleon system is the tensor
product of two single-nucleon Hilbert spaces.  The kinematic
representation of the Poincar\'e group on the two-nucleon Hilbert
space is the tensor product of two single-nucleon representations of
the Poincar\'e group:
\beq
{\cal H}={\cal H}_1 \otimes {\cal H}_1
\qquad 
U_0 (\Lambda,A) :=   U_1 (\Lambda,A) \otimes U_1 (\Lambda,A).
\label{b.15}
\eeq
The tensor product of two one-body irreducible representations of the
Poincar\'e group is reducible.  It can be decomposed into an
orthogonal linear superposition of irreducible representations using
the Poincar\'e group Clebsch-Gordan coefficients in the null-plane
basis,
\[
\Psi ((j_1,m_1), 
\tilde{\mathbf{p}}_1, \mu_1 , (j_2,m_2),\tilde{\mathbf{p}}_2, \mu_2 ) 
= 
\]
\beq
\sum \int 
\langle 
(j_1,m_1), \tilde{\mathbf{p}}_1, \mu_1 ;(j_2,m_2), \tilde{\mathbf{p}}_2, 
\mu_2   
\vert (j,k),l,s, \tilde{\mathbf{P}},\mu \rangle
d\tilde{\mathbf{P}} k^2 dk 
\Psi (
(j,m), l,s, \tilde{\mathbf{P}}, \mu   ) ,
\label{b.16}
%\label{b.19}
\eeq
where the Poincar\'e group Clebsch-Gordan coefficient \cite{Keister91}
in the null-plane basis is 
\[
\langle 
(j_1,m_1),\tilde{\mathbf{p}}_1, \mu_1 ;(j_2,m_2), \tilde{\mathbf{p}}_2, \mu_2   
\vert (j,k),l,s, \tilde{\mathbf{P}},\mu \rangle =
\]
\[
\delta ( \tilde{\mathbf{P}} - \tilde{\mathbf{p}}_1- \tilde{\mathbf{p}}_2)
{\delta (k - k(\tilde{\mathbf{p}}_1, \tilde{\mathbf{p}}_2))
\over k^2} 
\left \vert 
{\partial (\tilde{\mathbf{P}},\mathbf{k}) \over 
\partial (\tilde{\mathbf{p}}_1, \tilde{\mathbf{p}}_2 ) } 
\right \vert^{1/2}
\times 
\]
\[
\sum Y_{lm}(\hat{\mathbf{k}}(\tilde{\mathbf{p}}_1, \tilde{\mathbf{p}}_2)) 
D^{1/2}_{\mu_1, \mu_1'} [R_{fc} (\mathbf{k}/m_1)]
D^{1/2}_{\mu_2, \mu_2'} [R_{fc} (-\mathbf{k}/m_2)] \times
\]
\beq
\langle {1 \over 2}, \mu_1', {1 \over 2}, \mu_2' \vert 
s, \mu_s \rangle 
\langle l, m, s, \mu_s \vert 
j, \mu \rangle  
\label{b.17}
\eeq
%\label{b.20}
with 
\beq
\left \vert 
{\partial (\tilde{\mathbf{P}},\mathbf{k}) \over 
\partial (\tilde{\mathbf{p}}_1, \tilde{\mathbf{p}}_2 ) } 
\right \vert^{1/2} =
\sqrt{{(p_1^+ +p_2^+) \omega_1(\mathbf{k}) \omega_2(\mathbf{k})  \over
(\omega_1(\mathbf{k})+ \omega_2(\mathbf{k}))p_1^+ p_2^+ }}, 
\eeq
\beq
R_{fc} (\mathbf{k}/m_1) :=
\Lambda_f^{-1} (\mathbf{k}/m_1)\Lambda_c (\mathbf{k}/m_1)
\qquad
R_{fc} (-\mathbf{k}/m_2) :=
\Lambda_f^{-1} (-\mathbf{k}/m_2)
\Lambda_c (-\mathbf{k}/m_2)
\label{b.22}
\eeq
and $(\omega(\mathbf{k}), \mathbf{k}) = k(p_1,p_2) = \Lambda_f^{-1}
(\tilde{\mathbf{P}}/M_0) p_1$.  The rotation $R_{fc} (\mathbf{k}/m)$,
called a Melosh rotation, \cite{Melosh74}\cite{Keister91} transforms
the null-plane spins so they rotate under a single representation of
$SU(2)$ which allows them to be coupled using ordinary $SU(2)$
Clebsch-Gordan coefficients.  The absence of Wigner rotations in
(\ref{b.17}) is a consequence of the null-plane boosts forming a
subgroup.  The quantum numbers $l$ and $s$ are degeneracy parameters
that label different irreducible representations with the same mass
and spin that appear in the tensor product.  In (\ref{b.17}) the
two-body invariant mass $M_0$, which has continuous spectrum, is
replaced by $\mathbf{k}^2$ which is related to $M_0$ by
\beq
M_0=2 \sqrt{\mathbf{k}^2+ m^2}.
\label{b.18}
\eeq

In this basis two-nucleon wave functions have the form
\beq
\psi ( (j, k), l,s \tilde{\mathbf{P}},\mu ) =  
\langle  (j,k) ,l,s  \tilde{\mathbf{P}},\mu \vert \psi \rangle 
\qquad
\int_0^\infty dP^+ \int_{\mathbb{R}^2} 
d\mathbf{P}_{\perp} \int_0^\infty k^2 dk \sum_{\mu=-j}^j \sum_{s=0}^1 
\sum_{l= \vert j-s\vert}^{\vert j+s \vert}      
\vert \psi ( (j,k), l,s, \tilde{\mathbf{P}},\mu ) \vert^2 < \infty
\label{b.25}
%\label{b.5}
\eeq
and $U_0(\Lambda,A)$ acts irreducibly on these states: 
\[
\langle  (j,k),l,s, \tilde{\mathbf{P}},\mu  
\vert U_0(\Lambda ,A) \vert \psi \rangle =
\]
\[
\int_0^\infty dP^{+\prime} \int_{\mathbb{R}^2} 
d\mathbf{P}'_{\perp} \sum_{\mu'=-j}^j  
{\cal D} ^{M_0(k),j}_{\tilde{\mathbf{P}},\mu; \tilde{\mathbf{P}}',\mu'}[\Lambda ,A]
\langle 
(j,k),l,s, \tilde{\mathbf{P}}',\mu'  \vert \psi \rangle =
\]
\beq
\sum_{\mu'=-j}^j  
e^{i P \cdot a}
\sqrt{{P^{+\prime} \over P^+ }}
D^{j}_{\mu \mu'} [R_{fw} (\Lambda , \tilde{\mathbf{P}}'/M_0)]
\langle 
(j,k),l,s, \tilde{\mathbf{P}}',\mu'  \vert \psi \rangle 
\label{b.26}
\eeq
where $P= \Lambda P'$.  This basis is used in the formulation of our
dynamical model in the next section.

In the rest of this paper we use the following notation for 
covariant, canonical, and null-plane  
basis vectors for mass-$m$ spin-$j$ irreducible representation
spaces which are related by
\beq
\vert (j,m), p, \mu \rangle = 
\vert (j,m), \mathbf{p}, \mu \rangle \sqrt{{\omega_m (\mathbf{p})}\over m}  =
\sum_{\nu=-j}^j \vert (j,m), \tilde{\mathbf{p}}, \nu \rangle \sqrt{{p^+
\over m}} D^{j}_{\nu\mu} [R_{fc} (\mathbf{k}/m)].
\label{b.27}
\eeq

\section{dynamics}

Dynamical models of the two-nucleon system in Poincar\'e invariant
quantum mechanics are defined by a dynamical unitary representation of
the Poincar\'e group acting on the two-nucleon Hilbert space.  The mass
Casimir operator for this representation can be defined by adding a
realistic nucleon-nucleon interaction $v_{nn}$
\cite{Coester75b}\cite{Keister91} to the square of the non-interacting
two-nucleon mass operator as follows:
\beq
M^2 = M_0^2 + 4m v_{nn},
\label{c.1}
\eeq
where $m$ is the nucleon mass and for a dynamical model 
with a null-plane kinematic symmetry and $v_{nn}$ has the form
\beq
\langle 
(j',k'),l',s', \tilde{\mathbf{P}}',\mu' \vert v_{nn}
\vert (j,k),l,s ,\tilde{\mathbf{P}},\mu \rangle =
\delta (\tilde{\mathbf{P}}'- \tilde{\mathbf{P}})
\delta_{j'j} \delta_{\mu'\mu} 
\langle k',l',s' \Vert v^j \Vert k,l,s \rangle
\label{c.2}
\eeq
in the non-interacting irreducible basis (\ref{b.17}).
The interaction $v_{nn}$ is restricted so $M^2$ is a positive operator
and designed so $M^2$ commutes with the non-interacting operators
$j^2$ ,$\tilde{\mathbf{P}}$ and $j_z$ and is independent of the 
eigenvalues of $\tilde{\mathbf{P}}$ and $j_z$.

Simultaneous eigenstates of $M^2$, $j^2$ ,$\tilde{\mathbf{P}}$,
$j_z$  in the non-interacting irreducible basis 
have the form
\beq
\langle 
(j',k'),l',s', \tilde{\mathbf{P}}',\mu' \vert 
(j,\lambda),\tilde{\mathbf{P}},\mu  \rangle = 
\delta (\tilde{\mathbf{P}}'- \tilde{\mathbf{P}})
\delta_{j'j} \delta_{\mu'\mu} \phi_{\lambda,j}  (k,l,s)
\label{c.3}
\eeq
where the wave function $\phi_{\lambda,j} (k,l,s)$ is a solution of
the eigenvalue equation
\beq
(\lambda^2 - 4k^2 -  4m^2 ) \phi_{\lambda,j}  (k,l,s) =
\sum_{s'=0}^1 \sum_{l'=\vert j-s' \vert }^{\vert j+s' \vert }  
\int_0^{\infty} 
4m \langle k,l,s \Vert v_{nn}^j \Vert k',l',s' \rangle
k^{\prime 2}dk' 
\phi_{\lambda,j}  (k',l',s') .
\label{c.4}
\eeq
The eigenfunctions, $\phi_{\lambda,j} (k,s)$, of this mass
operator are also solutions to the non-relativistic Schr\"odinger
equation with interaction $v_{nn}$, which can be seen by dividing both
sides of (\ref{c.4}) by $4m$.  The deuteron is an even-parity bound
state with $j=s=1$ and parity limits the $l'$ sum in (\ref{c.4}) to
$l'\in \{ 0,2\}$.

If $\{M^2,j,\tilde{\mathbf{P}},\mu\}$ have the same interpretation as
$\{M_0^2,j,\tilde{\mathbf{P}},\mu\}$, then it follows from 
(\ref{c.2}) that the
eigenstates (\ref{c.3}) transform irreducibly with respect to a {\it
dynamical} representation of the Poincar\'e group defined by
replacing the eigenvalues of $M_0$ in  
${\cal D}^{M_0,j}_{\tilde{\mathbf{P}},\mu; \tilde{\mathbf{P}}',\mu'}[\Lambda ,A]$
by the eigenvalues $\lambda$ of
$M$:
\[
\langle 
(j',k'),l',s' \tilde{\mathbf{P}}',\mu' \vert U(\Lambda ,A) \vert  
(j,\lambda),\tilde{\mathbf{P}},\mu  \rangle =
\]
\[
\int \sum_{\mu''=-j}^j 
\langle 
(j',k'),l',s' \tilde{\mathbf{P}}',\mu' \vert 
(j,\lambda),\tilde{\mathbf{P}}'',\mu'' \rangle d\tilde{\mathbf{P}}''
\langle (j,\lambda) ,\tilde{\mathbf{P}}'',\mu''  \vert
U(\Lambda ,A) \vert  
(j,\lambda) ,\tilde{\mathbf{P}},\mu  \rangle =
\]
\beq
\phi_{j,\lambda} (k',l',s')  
{\cal D}^{\lambda,j'}_{\tilde{\mathbf{P}}',\mu'; \tilde{\mathbf{P}},\mu}
[\Lambda,A] .
\label{c.5}
\eeq
Since the eigenstates (\ref{c.3}) are complete, $U(\Lambda ,A)$,
defined by (\ref{c.5}), 
is a dynamical unitary representation of the Poincar\'e group on the
two-nucleon Hilbert space.  {\it When} $(\Lambda ,A)$ is an element of
the kinematic subgroup of the null plane, the Poincar\'e Wigner
${\cal D}$-function in the null-plane basis, ${\cal
  D}^{\lambda,j}_{\tilde{\mathbf{P}}',\mu';
  \tilde{\mathbf{P}},\mu}[\Lambda,A]$, is independent of the mass
eigenvalue, $\lambda$, and is thus identical to the Poincar\'e Wigner
${\cal D}$-function for the non-interacting irreducible
representation.  This shows that the null-plane kinematic subgroup
defined by (\ref{b.8}) is the kinematic subgroup of the dynamical
representation (\ref{c.5}).

The rotational covariance in (\ref{c.5}) is {\it exact}, but because ${\cal
  D}^{\lambda,j}_{\tilde{\mathbf{P}}',\mu';
  \tilde{\mathbf{P}},\mu}[\Lambda,A]$ depends explicitly on the mass
eigenvalue $\lambda$ when $\Lambda =R$ is a rotation, the rotations
are dynamical transformations.

The eigenfunctions of the interacting mass operator, $M$, are
identical functions of $k,l,s$ to the eigenfunctions of the
non-relativistic Schr\"odinger equation with interaction $v_{nn}$.
The identity
\beq
\Omega_{\pm} (H_{nr},H_{0nr} ) = \Omega_{\pm} (H_r,H_{0r} ),
\label{c.6}
\eeq 
where $\Omega_{\pm } (H,H_0 )$ are the scattering wave operators
for the non-relativistic and relativistic system respectively,   
follows from the elementary calculation:
\[
\Omega_{\pm} (H_r,H_{0r} ) = 
\lim_{t \to \pm \infty} e^{i (P_0^+ + P^-) t/2 }
e^{-i (P_0^+ + P_0^-) t/2 } = 
\lim_{t' \to \pm \infty} e^{i  P^- t' }
e^{-i P_0^- t' } =
\]
\[
\lim_{t' \to \pm \infty} e^{i  (M^2 + \mathbf{P}_{0\perp}^2)  t'/P_0^+ }
e^{-i  (M_0^2 + \mathbf{P}_{0\perp}^2)  t'/P_0^+ } =
\lim_{t'' \to \pm \infty} e^{i M^2  t'' }
e^{-i  M_0^2  t'' } =
\]
\[
\lim_{t'' \to \pm \infty} e^{i (\mathbf{k}^2/m +v_{nn}) 4mt'' + i4m^2 t'' }
e^{-i (\mathbf{k}^2/m ) 4mt'' - i4m^2 t'' }=
\]
\[
\lim_{t''' \to \pm \infty}
e^{i (\mathbf{P}^2 /4m + \mathbf{k}^2/m +v_{nn}) t'''  }
e^{-i (\mathbf{P}^2 /4m + \mathbf{k}^2/m ) t'''  } =
\]
\beq
\Omega_{\pm} (H_{nr},H_{0nr} )
\label{c.7}
\eeq
where the limits are strong or absolute abelian \cite{baum} limits and 
the times in (\ref{c.7}) were reparameterized as follows
\beq
t''' = 4 m t'' = 4m t'/P_{0}^+ = 2m t/P_{0}^+ .
\label{c.8}
\eeq    
This proof uses kinematic symmetries that follow from (\ref{c.2}) 
and the spectral condition, $P_0^+>0$.  It follows from eq. (\ref{c.7}) 
that relativistic and non-relativistic scattering matrices are 
{\it identical} functions of $\mathbf{k}$:
\beq
S(H_{nr},H_{0nr} ) = 
\Omega_{+}^{\dagger} (H_{nr},H_{0nr} )  \Omega_{-} (H_{nr},H_{0nr} ) 
= \Omega_{+}^{\dagger} (H_r,H_{0r} ) \Omega_{-} (H_r,H_{0r} )
= S (H_r,H_{0r} ) .
\label{c.8a}
\eeq
Since realistic $nn$ interactions \cite{cdb}\cite{Wiringa95} are
constructed by correctly transforming experimental differential 
cross-section data to the center-of-momentum frame and then fitting the
solution of the scattering problem to the transformed data, these
interactions can be used in (\ref{c.1}) {\it without modification}.
There is a small binding energy correction of about 1 part in 2000,
which we ignore.  The difference in the relativistic and
non-relativistic wave function is due to the Poincar\'e-group 
Clebsch-Gordan coefficients which are used to transform the wave function to
the tensor product representation which is used in the computation of
the current matrix elements.

We use this method to construct the deuteron eigenstates and the
representation of the Poincar\'e group that we use to evaluate the
deuteron current matrix elements.

\section{observables}

The differential cross section for elastic electron-deuteron scattering 
in the one-photon-exchange approximation,
\[
d\sigma = {(2\pi)^4 e
\over
\sqrt{(p_e \cdot p_d)^2 - m_e^2 m_d^2}} 
\vert \langle (j_e,m_e), 
{p}'_e, \mu'_e \vert I_{e \mu}(0)
(j_e,m_e), \vert {p}_e, \mu_e \rangle  
{(2 \pi)^3 \over q^2} \times
\]
\beq  
\langle (1,d), p'_d, \mu'_d \vert
I_d^{\mu} (0) \vert (1,d) , {p}_d, \mu_d \rangle 
\vert ^2 \delta^4 ( p_d' -p_d+p_e'-p_e) {d\mathbf{p}_e' \over \omega_e
 (\mathbf{p}'_e) } {d \mathbf{p}_d' \over \omega_d (\mathbf{p}'_d) } ,
\label{f.1}
\eeq 
is a quadratic function of the deuteron current matrix elements,
$\langle (1,d) {p}'_d, \mu'_d \vert
I_d^{\mu} (0) \vert (1,d), {p}_d, \mu_d \rangle$.
Covariance, current conservation, and discrete symmetries 
imply that only three of the deuteron current matrix elements are
linearly independent \cite{Yennie57}, which means that any deuteron 
elastic-scattering observable can be expressed in terms of three independent 
quantities for a given momentum transfer.  

Standard observables are the structure functions $A(Q^2)$ and $B(Q^2)$
and the tensor polarization $T_{20}(Q^2,\theta)$ at $\theta_{lab} =
70^o$.  The quantities $A(Q^2)$, $B(Q^2)$ are determined from the
unpolarized laboratory frame differential cross section using a
Rosenbluth \cite{Rosen50} separation:
\beq
{d\sigma \over d \Omega}({Q}^2,\theta) = {\alpha^2 \cos^2 (\theta/2)
\over 4E_i^2 \sin^4 (\theta/2)}{E_f \over E_i}
[A({Q}^2) + B({Q}^2) \tan^2 (\theta /2)]  
\label{f.2}
\eeq
while $T_{20}({Q}^2,\theta)$ is extracted from the difference in the 
cross sections for target deuterons having canonical spin polarizations 
$\mu_d=1$ and $\mu_d=0$ 
at a fixed laboratory scattering angle \cite{Haftel80}: 
\beq
T_{20}({Q}^2,\theta) = \sqrt{2} {{d\sigma \over d \Omega}_1({Q}^2,\theta)
-{d\sigma \over d \Omega}_0({Q}^2,\theta) \over
{d\sigma \over d \Omega}({Q}^2,\theta)  } .
\label{f.3}
\eeq

These experimental observables are  
related to the deuteron form factors \cite{Gross65a}
\cite{gourdin} 
\cite{Elias69}  \cite{ostebee75}
$G_0(Q^2)$, $G_1(Q^2)$, and $G_2(Q^2)$ by:
\beq
A({Q}^2)= G_0^2({Q}^2) + {2 \over 3} \eta 
G_1^2({Q}^2) + G_2^2({Q}^2)
\label{d.22a}
\eeq
\beq
B({Q}^2)= {4 \over 3}\eta (1+\eta) G_1^2({Q}^2) 
\label{d.23a}
\eeq
\beq
T_{20}({Q}^2,\theta)=- {G_2^2({Q}^2) +
\sqrt{8} G_0({Q}^2) G_2({Q}^2) + 
{1 \over 3} \eta G_1^2({Q}^2) [1+2(1+\eta) 
\tan^2 (\theta/2)] \over \sqrt{2} [A(Q^2)+ B(Q^2) 
\tan^2 (\theta /2)]} .
\label{d.24a}
\eeq
where $\eta={Q^2\over 4M_d^2}$ and $M_d$ is the deuteron mass.

The form factors are Poincar\'e invariant quantities that are linearly
related to current matrix elements.  They are traditionally expressed
in terms of canonical spin current matrix elements in the standard
Breit frame, with the momentum transfer chosen parallel to the axis of
spin quantization:
\beq 
G_0(Q^2) =  
{1\over 3}(\langle (1,d),P_b',0 \vert I^0(0) \vert (1,d),P_b ,0 \rangle + 
2\langle (1,d),P_b',1 \vert
I^0(0) \vert(1,d), - P_b,1 \rangle ) 
\label{d.22b}
\eeq 
\beq 
G_1(Q^2) = - 
\sqrt{{2 \over \eta}}\langle (1,d),P_b',1 \vert I^1(0) \vert
(1,d), P_b,0 \rangle 
\label{d.23b}
\eeq 
\beq 
G_2(Q^2) =
{\sqrt{2} \over 3}\langle (1,d),P_b',0 \vert I^0(0) \vert -
(1,d),P_b,0 \rangle - \langle (1,d),P_b',1 \vert
I^0(0) \vert (1,d),P_b,1 \rangle ) 
\label{d.24b}
\eeq 
where $P'_b= (M_d\sqrt{1+ \eta}, 0,0,{\mathbf{Q} \over 2})$ and
$P_b= (M_d\sqrt{1+ \eta}, 0,0,-{\mathbf{Q} \over 2})$
and the normalization of $G_2(Q^2)$ follows the conventions used 
in \cite{ostebee75}. 

The $0$-momentum
transfer limit of these form factors are related to the charge,
magnetic moment, and quadrupole moments \cite{Gross65a} 
\beq 
\lim_{Q^2 \to 0} G_0(Q^2) =1 \qquad \lim_{Q^2 \to 0} G_1(Q^2) 
= {M_d \over m_n} \mu_d \qquad
\lim_{Q^2 \to 0} {G_2(Q^2)\over Q^2} = {1 \over 2\sqrt{3}}Q_d .
\eeq 

Current covariance, current conservation, and discrete symmetries can 
be used to express the matrix elements in (\ref{d.22b}-\ref{d.24b}) 
in terms of matrix elements of the plus component of the current in 
the null-plane Breit frame, where the momentum transfer is in
$x$ direction and the spins are null-plane spins. 

The canonical spin basis vectors in (\ref{d.22b}-\ref{d.24b}) are
related to the null-plane basis vectors defined in (\ref{b.12}) by
(\ref{b.27}) which involves a change in normalization and a Melosh
rotation (\ref{b.22}) of the spins.
%\beq
%\vert \mathbf{p}, \mu \rangle_c = 
%\vert \tilde{\mathbf{p}}, \mu' \rangle_f 
%\sqrt{{p^+ \over \omega (\mathbf{p})}} 
%D^1_{\mu' \mu} [ \Lambda_c^{-1}(\mathbf{p}/m_d)  
%\Lambda_f (\mathbf{p}/m_d) ]. 
%\eeq

\section{Nucleon currents}

The impulse current is the sum of single-nucleon current operators for the 
proton and neutron.  For spin $1/2$ systems that are eigenstates of 
parity there are two linearly independent current 
matrix elements\cite{Yennie57}.  They are related
to the Dirac nucleon form factors $F_1(Q^2)$ and $F_2(Q^2)$ by
\beq
\langle (j_n,m_n),p' , \nu' \vert I^{\mu} (0) \vert
(j_n,m_n) p , \nu \rangle
=
\bar{u}(p') \Gamma^{\mu} u (p) 
\label{e.1}
\eeq
where
\beq
\Gamma^{\mu}  =  
%{1 \over 2} [\gamma^{\mu} , {1 \over 2m} \gamma\cdot Q] F_2 (Q^2) 
\gamma^{\mu} F_1 (Q^2) - 
{i  \over 2m } \sigma^{\mu \nu} Q_{\nu} F_2 (Q^2) 
\label{e.2}
\eeq
and $u(p)$ is a canonical spin Dirac spinor.  

For spacelike momentum transfers with $Q^+=0$ these form factors can be 
expressed in terms of the independent null-plane matrix elements    
\beq
\langle {1 \over 2} \vert I^+(0) \vert 
{1 \over 2}\rangle :=
\langle 
(j_n,m_n),
\tilde{\mathbf{p}}', {1 \over 2} \vert I^+(0) \vert 
(j_n,m_n),
\tilde{\mathbf{p}}, {1 \over 2}\rangle = F_1 (\mathbf{Q}^2) 
\label{e.3}
\eeq
\beq
\langle {1 \over 2} \vert I^+(0) \vert 
-{1 \over 2}\rangle :=
\langle 
(j_n,m_n),\tilde{\mathbf{p}}', {1 \over 2} \vert I^+(0) \vert 
(j_n,m_n),
\tilde{\mathbf{p}}, -{1 \over 2}\rangle =
- \sqrt{\tau}
F_2 (\mathbf{Q}^2) 
\label{e.4}
\eeq
where $\tau := {\mathbf{Q}^2 \over 4m^2}$.

The Dirac nucleon form factors $F_1(Q^2)$ and $F_2(Q^2)$  are related to the 
Sachs form factors by \cite{Sachs62}
\beq 
F_1(\mathbf{Q}^2)={G_e(\mathbf{Q}^2)+ \tau G_m(\mathbf{Q}^2) \over 1+\tau},
\label{e.5}
\eeq
\beq  
F_2(\mathbf{Q}^2)={G_m(\mathbf{Q}^2)- G_e(\mathbf{Q}^2) \over 1+\tau}.
\label{e.6}
\eeq
We consider recent parameterizations due to Bijker and Iachello
\cite{nnformfactor-BI}; Bradford, Bodek, Budd, and Arrington
  \cite{nnff-BBBA}; Budd, Bodek, and Arrington \cite{nnff-BBA}; Kelly
  \cite{nff-kelly} and Lomon
  \cite{nnformfactor-Lomon}\cite{nnformfactor-Lomon1}
  \cite{nnff-Lomon1}.
These parameterizations all determine the proton-electric form factors
using the polarization experiments \cite{Punjabi05}.  The input to our
calculations is the isoscalar form factors, which are the sum of the
proton and neutron form factors.

\section{Deuteron Currents: Dynamical Constraints} 

The hadronic current density operator, $I^{\mu}(x)$, transforms as a
four-vector density under the dynamical representation (\ref{c.5}) of
the Poincar\'e group,
\beq 
U(\Lambda ,A) I^{\mu}(x)U^{\dagger} (\Lambda ,A) =
(\Lambda^{-1})^{\mu}{}_{\nu} I^{\nu}(\Lambda x+ a) .
\label{d.2}
\eeq
The current operator must also satisfy current
conservation 
\beq
g_{\mu \nu} [P^{\mu},I^{\nu} (0)]_-=0,
\label{d.3}
\eeq
in addition to symmetries with respect to space reflections and time
reversal.  Here $g_{\mu \nu}$ is the Minkowski metric with signature
$(-+++)$.  Current covariance and current conservation are dynamical
constraints on the current operator.

Using the identity
\[
\langle (1,d), \tilde{\mathbf{P}}',\nu' \vert
I^{\mu} (x)  
\vert (1,d),\tilde{\mathbf{P}},\nu  \rangle =
\]
\beq
\langle (1,d),\tilde{\mathbf{P}}',\nu' \vert
U^{\dagger} (\Lambda ,A) U (\Lambda ,A)
I^{\mu} (x)  
U^{\dagger} (\Lambda ,A) U (\Lambda ,A)
\vert (1,d),\tilde{\mathbf{P}},\nu  \rangle 
\eeq
with current covariance (\ref{d.2}) and the transformation
properties (\ref{c.5}) of the deuteron eigenstates gives 
linear equations for each value of $\Lambda$ and $A$ that 
relate the different matrix elements:
%
%\[
%(\Lambda^{-1})^{\mu}{}_{\mu'} 
%\int \sum 
%e^{i (P''-P''')\cdot a} 
%\sqrt{{P^{\prime \prime \prime +}  \over P^{\prime +} }}
%D^1_{\nu' \nu'''}[R^{-1}_{wf} (\Lambda, P')]
%\times 
%\]
%\beq
%\langle (1,d), \tilde{\mathbf{P}}''',\nu''' \vert
%I^{\mu'} (\Lambda x +a )  
%\vert (1,d),\tilde{\mathbf{P}}'',\nu  \rangle
%D^1_{\nu'' \nu}[R_{wf} (\Lambda, P)]
%\sqrt{{P^{\prime \prime +}  \over P^{+} }}
%\label{d.3a}
%\eeq
%where
%\beq
%R_{wf} (\Lambda, P')=
%\Lambda_f^{-1}  (\tilde{\mathbf{P}}'''/m_d) 
%\Lambda \Lambda_f  (\tilde{\mathbf{P}}'/m_d)
%\eeq
%and 
%\beq
%R_{wf} (\Lambda, P)+
%\Lambda_f^{-1}  (\tilde{\mathbf{P}}''/m_d) 
%\Lambda \Lambda_f  (\tilde{\mathbf{P}}/m_d)
%\eeq
%are null-plane Wigner rotations.
Current conservation leads to the 
additional linear constraints on the 
current matrix elements:
\beq
g_{\alpha \mu} (P^{'\alpha} - P^{\alpha}) 
\langle (1,d),\tilde{\mathbf{P}}',\nu' \vert
I^{\mu} (0)  
\vert (1,d), \tilde{\mathbf{P}},\nu  \rangle = 0 .
\label{d.4a}
\eeq
It is well-known that these constraints, when combined with
space-reflection and time-reversal symmetries, \cite{Yennie57} can be
used to express any deuteron current matrix element as a linear
combination of three independent current matrix elements.  This is the
Wigner-Eckart theorem for the Poincar\'e group applied to a conserved
current.  The invariant form factors or independent current matrix
elements play the role of invariant matrix elements in the
Wigner-Eckart theorem.

If the coordinate axes are chosen so the $+$ component of the momentum
transfer is zero, which can always be done for electron scattering,
then it is possible to choose the independent matrix elements as
matrix elements of the ``$+$'' component of the current density,
$I^+(0)$ \cite{Chung88}.  Matrix elements of $I^+(0)$ in an
irreducible null-plane basis with normalization (\ref{b.12}) 
are invariant with respect to the group of
null-plane boosts.

It follows from this invariance that the matrix elements of $I^+(0)$
only depend on the square of the momentum transfer and the individual
null-plane spin components.  There are $3\times 3=9$ combinations of
initial and final spins.  Covariance with respect to the null-plane
kinematic subgroup relates all of them to four kinematically
independent current matrix elements.  The one and two-body parts of
the current operator transform independently under the action of the
kinematic subgroup.  The kinematic subgroup does not include the full
rotation group.  Rotational covariance
\cite{Leutwyler78}\cite{Grach84}\cite{Chung88} gives one additional
constraint among the kinematically-covariant matrix elements, called
the ``angular condition'', reducing the number of independent matrix
to three.  It relates matrix elements of the one and two-body parts 
of the current operator.

Because the constraints (\ref{d.2}) and (\ref{d.3}) are dynamical, it
is a non-trivial problem to construct fully-covariant current {\it
  operators}.  It is much easier to construct a set of deuteron matrix
elements of a conserved covariant current operator.  This can be done
by first computing a set of three independent current matrix
elements. The remaining matrix elements are generated from these three
matrix elements using the constraints implied by (\ref{d.2}) and
(\ref{d.4a}).  It is possible to do better by computing the three
independent matrix elements using a model current that is covariant
with respect to the null-plane kinematic subgroup.  Operators that are
covariant with respect to the kinematic subgroup are easily
constructed.  In this case the kinematically-covariant current and
fully-covariant current will agree not only on the three independent
currrent matrix elements, but also on any matrix elements generated
from the independent matrix elements by the null-plane kinematic
subgroup.  The angular condition is needed to obtain all of the
current matrix elements.  It is a dynamical constraint that generates
the fourth kinematically independent matrix element from the three
independent matrix elements in a manner consistent with rotational
covariance.  The remaining current matrix elements can be generated
from this fourth matrix element using only kinematic covariance.  This
gives all deuteron matrix elements of a conserved covariant current
operator.

While it is straight forward to construct kinematically covariant
current operators, there are many fully covariant current operators
that agree with a given kinematically covariant current operator on
different subsets of matrix elements.  This leads to ambiguities when
a kinematically covariant current is used to generate matrix elements
of a fully covariant current.  Model assumption are needed to
eliminate these ambiguities.  Specifically, a fully covariant current
operator can be chosen to agree with the kinematically covaraint
current operator on any set of three independent current matrix
elements and all matrix elements generated from them using kinematic
Poincar\'e transformations.  Because kinematic covariance cannot
generate all of the current matrix elements, the resulting covariant
current will depend on the choice of independent current matrix
elements used to generate the full current, which is the model
assumption.

``Null-plane impulse approximations'' choose the
kinematically-covariant current to be the sum of the single-nucleon
currents.  For the deuteron the elastic scattering observables
computed in the null-plane impulse approximation are not very
sensitive to the choice of independent current matrix elements used to
generate the full current.

In this paper we use the same method to compute matrix elements of a
dynamical two-body current operator.  A kinematically covariant model
two-body current is used to compute independent current matrix
elements.  The remaining current matrix elements are generated using
(\ref{d.2}) and (\ref{d.4a}).  As in the case of the null-plane
impulse approximation, a complete specification of a fully-covariant
model current requires both a kinematically covariant current and a choice
of kinematically independent current matrix elements.

If the deuteron eigenstates $\vert (1,d), \tilde{\mathbf{P}},\mu
\rangle $ are given a delta-function normalization, then the matrix
elements
\beq
I^+_{\mu, \nu} := \langle (1,d),\tilde{\mathbf{P}}',\mu \vert I^+(0) \vert
(1,d), \tilde{\mathbf{P}},\nu \rangle   
\label{d.4}
\eeq
with $Q^+= P^{+\prime}-P^+ =0$ and null-plane spins are invariant with
respect to null-plane boosts.  The four kinematically independent
matrix elements of $I^+(0)$ can be taken as $I^+_{1,1}, I^+_{1,0},
I^+_{0,0}$ and $ I^+_{1,-1}$.  They are related by rotational
covariance.  The result is that only three linear combinations of
these matrix elements can be taken as independent.

The three independent linear combinations of the current matrix
elements used to generate the fully covariant current 
in ref. \cite{Chung88} were:
\beq
I^+_{1,1}+ I^+_{0,0},  I^+_{1,0}, I^+_{1,-1} .
\label{d.5}
\eeq
We refer to this choice of independent current matrix elements as
choice I.  These independent matrix elements can be distinguished by
the number of spin flips between the initial and final states.

Rotational covariance, or equivalently the angular condition, relates
the difference $ I^+_{1,1}- I^+_{0,0} $ to the matrix elements
(\ref{d.5}):
\beq
I^+_{1,1}- I^+_{0,0} = 
-{ 1 \over 1+ \eta} [ \eta (I^+_{1,1}+ I^+_{0,0}) - 2 \sqrt{2 \eta}I^+_{1,0}+
I^+_{1,-1} ]
\label{d.6}
%\label{d.5}
\eeq
where 
\beq
\eta:= {Q}^2 /4M_d^2 .
%\label{d.6}
\label{d.7}
\eeq
This relation is dynamical because $\eta$ involves the deuteron mass 
eigenvalue, $M_d$.

Direct computation of the difference
$I^+_{1,1}- I^+_{0,0}$ using the kinematically covariant current
operator and the difference determined by the constraint (\ref{d.6})
of current covariance
\beq 
\Delta_- := I^+_{1,1}- I^+_{0,0} +{ 1 \over 1+ \eta} [ \eta
  (I^+_{1,1}+ I^+_{0,0}) - 2 \sqrt{2 \eta}I^+_{1,0}+ I^+_{1,-1}]
\label{d.8}
%\label{d.7}
\eeq
gives a measure of the size of the dynamical
contribution to the current operator that is generated by the
rotational covariance constraint.  If this difference is small
compared to the size of independent matrix elements, then there will
not be too much sensitivity to the choice of independent matrix
elements; on the other hand, if this difference is large, there will
be an increased sensitivity to the choice of independent current
matrix elements.

Both impulse or impulse plus an exchange current can be used to
compute the independent current matrix elements.  Our calculations,
shown in fig. 1, indicate that $\Delta_-$ is larger when our model
exchange current contributions are added to the independent matrix
elements.  This means that the presence of the exchange current
enhances the sensitivity of the results to the choice of independent
matrix elements.  Because of this increased sensitivity, we
investigate the impact of using different choices of independent
linear combinations of current matrix elements to generate fully
covariant current matrix elements on the sensitivity of the elastic
electron-deuteron scattering observables.

To motivate the other choices that we use note that any matrix element
of a conserved covariant current in a set of deuteron eigenstates can
be expressed in terms of a rank 3 covariant current tensor
\cite{gourdin}.
The deuteron null-plane spin operator can be expressed in terms of the
Poincar\'e group generators as 
\beq
(0, \mathbf{j}_f) := - {1 \over 2M_d} 
\Lambda^{-1}_f({\tilde{\mathbf{P}}/M_d})^{\mu}{}_{\nu} 
\epsilon^{\mu}{}_{\alpha \beta \gamma}P^{\alpha}J^{\beta\gamma} 
\label{d.8b}
\eeq   
where $P^{\alpha}$ is the deuteron four momentum operator,
$J^{\beta\gamma}$ is the deuteron angular momentum tensor, and
$\Lambda^{-1}_f({\tilde{\mathbf{P}}/M_d)}^{\mu}{}_{\nu}$ is a $4\times
4$ Lorentz-transform-valued matrix of the operator $P^{\mu}$
\cite{Coester82}\cite{Keister91}.  The quantity $(0, \mathbf{j}_f)$ is
not a four vector because
$\Lambda^{-1}_f({\tilde{\mathbf{P}}/M_d})^{\mu}{}_{\nu}$ is a matrix
of operators.  If the operator
$\Lambda_f({\tilde{\mathbf{P}}/M_d})^{\mu}{}_{\nu}$ is applied to both
sides of (\ref{d.8b}) the result is a four vector, which up to the
factor $M_d$ in the denominator, is the Pauli-Lubanski vector
for the deuteron. 

It follows from
(\ref{d.8b}) that it is possible to replace the deuteron spins by four
vectors by multiplying the current matrix element by the inverse of
the boost used in (\ref{d.8b}); the
resulting vector is orthogonal to the four momentum.  If this boost is
applied to the initial and final deuteron states in a current matrix
element then the result is a rank-three tensor density, $T^{\mu}
_{\rho \sigma}(P',P)$ satisfying \cite{gourdin}\cite{acg79}
\beq
P^{\rho \prime } T^{\mu} _{\rho \sigma}(P',P) =   
T^{\mu} _{\rho \sigma} (P',P) P^{\sigma } = 
(P_{\mu}'- P_{\mu})T^{\mu} _{\rho \sigma}(P',P) =0.
\label{d.9}
\eeq
Formally this tensor density is related to the null-plane 
current matrix elements by
\[ 
T^{\mu}_{\rho\sigma}(P',P) :=
\Lambda_f(\tilde{\mathbf{P}}'/M_d)^i{}_\rho O^{*}_{\nu' i} \sqrt{P^{+\prime}}
\langle d, \tilde{\mathbf{P}}', \nu' \vert I^\mu (0) \vert
d, \tilde{\mathbf{P}}, \nu \rangle \sqrt{P^{+}} O_{\nu j}
\Lambda_f(\tilde{\mathbf{P}}/M_d)^j{}_{\sigma} 
\]
\beq 
\label{d.10}
\eeq 
where
$O_{\nu j}$ is the unitary
matrix
\beq
O_{\nu j} = 
\left (
\begin{array}{ccc}
-{1 \over \sqrt{2}} & -{i \over \sqrt{2}} & 0 \\
0 &0 & 1 \\ 
{1 \over \sqrt{2}} & -{i \over \sqrt{2}} & 0 \\
\end{array} 
\right )
\label{d.11} 
\eeq 
that converts Cartesian components of vectors to spherical components 
and the $i,j$ sums go from 1 to 3.   The factors  $\sqrt{P^{+}}$
give
$
\vert
\tilde{\mathbf{P}}, \nu \rangle \sqrt{P^{+}}
$
a covariant normalization.

The general form of this Lorentz covariant tensor density 
$T^{\mu}_{\rho\sigma}(P',P)$ can be parameterized by the available 
four vectors and three invariant form factors
$F_1(Q^2) ,F_2 (Q^2) $ and ,$F_3 (Q^2)$:
\[
T^{\mu}_{\rho\sigma}(P',P) = 
{1 \over 2} (P'+P)^{\mu} E(P')_{\rho\alpha}
[ g^{\alpha \beta} F_1 (Q^2)  
+ {Q^{\alpha}Q^{\beta}\over M_d^2}F_2 (Q^2)
]
E(P)_{\beta \sigma}
\]
\beq
-{1 \over 2} E(P')_{\rho\alpha}[Q^{\alpha}g^{\mu \beta} - 
g^{\mu \alpha}Q^{\beta}] F_3 (Q^2)    
E(P)_{\beta\sigma}
\label{d.12}
\eeq
where 
\beq
E(P)_{\alpha \beta} = g_{\alpha \beta} - {P_{\alpha}P_{\beta}\over P^2}  
\label{d.13} 
\eeq
is the covariant projector on the subspace orthogonal to $P$.  The
projectors are not needed when this tensor is contracted with vectors
orthogonal to the final or initial four-momentum. 

We use 
this covariant representation of the 
current matrix elements 
to identify preferred choices of independent current
matrix elements.
Independent matrix elements can be constructed by contracting this 
current tensor with different
sets of ``polarization 4-vectors''.  These are 4-momentum dependent 
vectors $v^{\nu} (P)$ that are orthogonal $P^{\mu}$.
The contraction has the form
\[
v'_{af} (P')^{\rho}  T^{\mu}_{\rho\sigma}(P',P)
v_{bi} (P)^{\sigma} = 
{1 \over 2} (P'+P)^{\mu} 
[ v_{af}'(P') \cdot v_{bi} (P) F_1 (Q^2)  
+ { (v'_{af} (P')\cdot Q )(Q\cdot v_{bi}(P))\over M_d^2}F_2 (Q^2)
]
\]
\beq
-{1 \over 2} [ 
( v'_{af} (P') \cdot Q) v_{bi}(P)^{\mu} -
 v_{af}(P')^{\mu} (Q \cdot v_{bi} (P))
 ] F_3 (Q^2) .    
\label{d.14}
\eeq
In this case the three invariant scalar products 
\beq
v_{af}'(P') \cdot v_{bi} (P)
\qquad
v'_{af} (P')\cdot Q 
\qquad 
Q\cdot v_{bi}(P)
\label{d.15}
\eeq
can be used in (\ref{d.15}) and (\ref{d.10}) to relate the form
factors to the current matrix elements.  Three pairs of polarization
vectors give three equations that can be solved to express
$F_1(Q^2)$, $F_2(Q^2)$ and $F_3(Q^2)$ in terms of the current matrix
elements.
Given the form factors the rest of the current matrix elements are
determined by the current tensor.

The form factors $F_1(Q^2)$, $F_2(Q^2), F_3(Q^2)$ defined by
(\ref{d.10}) and (\ref{d.12}) are related to the 
form factors $G_1(Q^2)$, $G_2(Q^2), G_3(Q^2)$ 
\cite{ostebee75} which are defined in terms of 
canonical-spin Breit-frame matrix 
elements with momentum transfer in the 3 direction:
\beq
%F_1 (Q^2) - 4 \eta^2 (1+ \eta) F_2(Q^2) =
((1+2 \eta) F_1 + 4 \eta (1-\eta) F_2 + 2 \eta F_3) =  
_c\langle (1,d), {\mathbf{Q} \over 2}, 0 \vert I^0(0) \vert 
(1,d), -{\mathbf{Q} \over 2}, 0 \rangle_c = G_0(Q^2) + \sqrt{2} G_2 (Q^2) 
\label{d.16a}
\eeq
\beq
F_1 (Q^2) = _c\langle (1,d), {\mathbf{Q} \over 2}, 1 \vert I^0(0) \vert 
(1,d), -{\mathbf{Q} \over 2}, 1 \rangle_c = G_0(Q^2) - {1 \over \sqrt{2}} G_2 (Q^2) 
\label{d.17a}
\eeq
\beq
F_3 (Q^2) = \sqrt{2 \over \eta} 
_c\langle (1,d), {\mathbf{Q} \over 2}, -1 \vert I^1(0) \vert 
(1,d), -{\mathbf{Q} \over 2}, 0 \rangle_c = 
G_1 (Q^2) .
\label{d.18a}
\eeq

The enhanced sensitivity to the choice of linearly independent
matrix elements when the exchange current is included, as shown in fig.
1,  suggests that in addition to the choice (\ref{d.5}) of linearly
independent matrix elements used in \cite{Chung88}, other choices
should be considered.

Frankrurt, Frederico, and Strikman \cite{Frankfrut93} defined
independent current matrix elements using polarization vectors
(\ref{d.14}) constructed from the last three columns
$v_{1c}(P_b),v_{2c}(P_b),v_{3c}(P_b)$, of the canonical boost in the
null-plane Breit frame
\beq
(v_{1c} (P_b) , v_{2c} (P_b)  v_{3c} (P_b))  =
\left (
\begin{array}{ccc}
-\sqrt{\eta}  & 0 & 0 \\
\sqrt{1+\eta}      & 0 & 0\\
0      & 1 & 0 \\
0 & 0 & 1 \\
\end{array}
\right ) .
\label{d.16}
\eeq
These are
automatically orthogonal to the momentum.  Three independent 
current matrix elements are extracted from the contractions
$
v_{1c} (P_b')^{\rho}  T^{+}_{\rho\sigma}(P_b',P_b)
v_{1c} (P_b)^{\sigma}
$, \phantom{a} 
$v_{2c} (P_b')^{\rho}  T^{+}_{\rho\sigma}(P_b',P_b)
v_{2c} (P_b)^{\sigma}
$ and 
$
v_{3c} (P_b')^{\rho}  T^{+}_{\rho\sigma}(P_b',P_b)
v_{1c} (P_b)^{\sigma}
$.  
These can be computed consistently using a kinematically covariant
model current in all frames related to the null-plane Breit frame by
null-plane kinematic transformations.  Their preference for this
choice is to avoid using the $3'3$ matrix element, which is maximally
suppressed in the infinite momentum frame.  In \cite{Leutwyler78}
operators are classified as good, bad or terrible according to their
scaling properties with respect to Lorentz boosts to the infinite
momentum frame.  The $3'3$ matrix element is terrible in this
classification.  We call this choice of independent current matrix
elements choice II.

We are indebted to Fritz Coester \cite{Fc08} for suggesting a third
choice of independent current matrix elements that emphasizes the
kinematic symmetries of the null plane.  This choice uses columns of
the null-plane boost as polarization vectors.
\beq
(v_{1f} (P) , v_{2f} (P),  v_{3f} (P))  =
\left (
\begin{array}{ccc}
P^x/M_d  & P^y/M_d & P^0/M_d-M_d/P^+ \\
1      & 0 &  P^x/M_d\\
0      & 1 &  P^y/M_d \\
-P^x/M_d & -P^y/M_d& P^z/M_d+M_d/P^+ \\
\end{array}
\right ) 
\label{d.17}
\eeq
The independent combinations are 
$
v_{1f} (P')^{\rho}  T^{+}_{\rho\sigma}(P',P)
v_{1f} (P)^{\sigma}
$,
$v_{2f} (P')^{\rho}  T^{+}_{\rho\sigma}(P',P)
v_{2f} (P)^{\sigma}
$, and 
$
v_{3c} (P')^{\rho}  T^{+}_{\rho\sigma}(P',P)
v_{1c} (P)^{\sigma}
$.
When the first two terms are evaluated in the null-plane Breit frame
the scalar products (\ref{d.15}) that relate the contractions to the
form factors are mass independent, which is preserved under kinematic
transformations.  This means that the boost parameters that
appear in the calculations of the one-body current matrix elements 
are independent of the constituent masses.  To construct
a third independent linear combination that exploits this mass
independence would require using matrix elements of $I^2 (0)$.  Rather
than using $I^2(0)$, $F_3$ is extracted using polarization vectors that are
mutually orthogonal, with one of them also orthogonal to $Q$, which gives
matrix elements that are independent of $F_1$ and $F_2$.  This results 
in the same expression for $F_3=G_1$ used in the other two schemes. 
We call this choice of independent current matrix
elements choice III.

These three choices lead to expressions for deuteron form factors
based on three different covariant current operators generated using
three different choices of independent linear combinations of matrix
elements of a kinematically covariant model current $I^+(0)$.  These
different current operators lead to distinct expressions for the
deuteron form factors in terms of the null-plane current matrix
elements of $I^+(0)$.  The results, expressed in terms of the standard
deuteron form factors and null-plane Breit frame matrix elements of
the $+$ component of the current are:

\noindent Choice I:
\beq
(1 + \eta) G_{0}(Q^2) = ({1 \over 2} - {\eta \over 3})( I_{11}+I_{00})
+ {5\sqrt{2 \eta} \over 3} I_{10}  + ({2 \eta \over 3} - 
{1 \over 6})I_{1-1}     
\label{d.18}
\eeq
\beq
(1 + \eta) G_{1}(Q^2) = I_{11}+I_{00} - I_{1-1}
-(1 -\eta) \sqrt{{2 \over \eta}} I_{10} 
\label{d.19}
\eeq
\beq
(1 + \eta) G_{2}(Q^2) = -{\sqrt{2}\eta \over 3} (I_{11}+I_{00})
+ {4\sqrt{\eta}\over 3} I_{10}  - {\sqrt{2} \over 3} (2+ \eta) I_{1-1} 
\label{d.20}
\eeq

\noindent Choice II ($G_{0}(Q^2)$ of choice I replaced by):

\beq
(1 + \eta) G_{0}({Q}^2)=  ({2 \eta \over 3} +1)
I^+_{1,1}  - {\eta \over 3} I^+_{00} 
+ {2 \sqrt{2 \eta} \over 3}  I^+_{1,0} + 
({2 \eta +1 \over 3} ) I^+_{1,-1}] 
\label{d.21}
\eeq

\noindent Choice III ($G_{0}(Q^2)$ and $G_{2}(Q^2)$ of choice I replaced by): 

\beq
G_{0}({Q}^2) = 
(1+ {2 \eta \over 3} )  I^+_{11} +{1 \over 3} I^+_{1,-1} 
-{2 \eta \over 3}G_{1}({Q}^2) 
\eeq
\beq
G_{2}({Q}^2) = 
{2 \sqrt{2} \over 3}(\eta I^+_{1,1}- I^+_{1,-1}- \eta G_{1}({Q}^2) ) 
\label{d.22}
\eeq
Since $B(Q^2)$ only depends on $G_1(Q^2)$ which uses the same linear
combination of current matrix elements for all three choices, the
computation of $B(Q^2)$, is unchanged.  These choices lead to a
$B(Q^2)$ that is in good agreement with experimental data.

The different linear combinations of null-plane current matrix
elements used in choices I , II and III have a non-trivial 
effect on the scattering observables $A(Q^2)$ and $T_{20}(Q^2,\theta)$.
Selecting one of these choices is a model assumption.

\section{Deuteron Currents: Dynamical Exchange Currents} 

The realistic interactions \cite{Wiringa95}\cite{cdb} used to
construct the mass operator (\ref{c.1}) include a one-pion-exchange 
contribution plus a short-range contribution that is 
designed so the two-body cross sections fit experiment.

``Required'' two-body currents, discussed in the previous section, are
needed to satisfy the constraints of current covariance and current
conservation.  These constraints are satisfied by directly
computing a set of linearly-independent matrix elements and then using
the constraints to determine the remaining current matrix elements.
Even if the independent current matrix elements are computed using the
one-body impulse current, the matrix elements generated by the
constraints will have two-body contributions.  The interaction
dependence arises because the covariance and current conservation
constraints involve the deuteron mass, which is the eigenvalue of the
dynamical two-body mass operator (\ref{c.4}).
 
In addition to the two-body currents directly generated by covariance
and current conservation, realistic interactions contain terms that
involve the exchange of charged mesons, leading to the exchange of the
nucleon charges, and the possibility of two-body currents associated
with these exchanges.  These currents give additional
contributions to the linearly independent matrix elements of $I^+(0)$
that are needed to compute electron-scattering observables in
null-plane quantum mechanics.  The form of the dynamical exchange
current contribution to $I^+(0)$ is model dependent, but we assume
that the most important contribution is motivated by one-pion exchange
physics.

To motivate the structure of our two-body current consider 
the pseudoscalar pion-nucleon vertex 
\beq
{\cal L}(x) = -i {g_{\pi}} :\bar{\Psi} (x)
\gamma_5 \Psi (x) \btau \cdot \bphi (x) :. 
\label{r.1}
\eeq
When this vertex is coupled to a photon-nucleon
vertex with a causal propagator, the propagator
connecting these two vertices can be
decomposed into a sum of two terms involving the 
covariant spinor projectors
$\Lambda_+(\mathbf{p}) = u(\mathbf{p})\bar{u}(\mathbf{p})$ and
$\Lambda_-(-\mathbf{p}) = - v(-\mathbf{p})\bar{v}(-\mathbf{p})$, where
${u}(\mathbf{p})$ and ${v}(\mathbf{p})$ are Dirac spinors.
This results in a sum of two terms with the following spin structures
\beq
\bar{u}(\mathbf{p}) \Gamma^{\mu}  u(\mathbf{p}')
\bar{u}(\mathbf{p}') \gamma^5 u(\mathbf{p}'')
\eeq
\beq
-\bar{u}(\mathbf{p}) \Gamma^{\mu}  v(-\mathbf{p}')
\bar{v}(-\mathbf{p}') \gamma^5 u(\mathbf{p}'') .
\eeq
In this form these operator structures are expressed as $2 \times 2$
matrices in the nucleon spins.  In Poincar\'e invariant quantum models
the first term is included in the one-body part of the current.  The
second term, which leads to the ``pair current'' in covariant
formulations, is not included in the one-body part of the current in
Poincar\'e invariant quantum models.  It can be included as a
contribution to the two-body part of the current.  To do this it is
useful to convert the $v$-spinors to $u$ spinors using
\beq
v (-\mathbf{p}) \bar{v}(-\mathbf{p})  =
\gamma_5 \beta u (\mathbf{p}) \bar{u}(\mathbf{p})\beta \gamma_5  
\label{r.12}
\eeq
which replaces the operator structure in the second term by
\beq
\bar{u}(\mathbf{p}) \Gamma^{\mu} \beta \gamma_5 u (\mathbf{p}') 
\bar{u}(\mathbf{p}')\beta  u(\mathbf{p}''). 
\label{r.12a}
\eeq
The factors $\beta$ appearing in this expression are spinor 
space-reflection operators.  Since space reflections do not leave the null
plane invariant, they become dynamical operators in null-plane 
quantum mechanics.  

The canonical Dirac spinors can be converted to null-plane spinors by
multiplying the spins by Melosh rotations, $u_f (\mathbf{p}) = u
(\mathbf{p})D^{1/2}[R_{cf}(p)]$.  These rotations cancel in the
combinations that appear in the projectors
$\Lambda_{\pm}(\mathbf{p})$, so the projectors can be
expressed directly in terms of null-plane Dirac spinors.
 
To construct a kinematically covaraint current we first make 
the model assumption that $\beta$ should be replaced by an invariant 
operator that agrees with $\beta$ in the rest frame of the 
initial (or final deuteron).  It requires 
$\beta \to -P_d \cdot \gamma/M_d$.

Because the null-plane boosts do not have Wigner rotations, the null-plane
Dirac spinors satisfy the following covariance condition with respect to
null-plane boosts $\Lambda$:
\[
{u}_f(\tilde{\mathbf{p}}) =
S (\Lambda^{-1}) {u}_f(\tilde{\bLam}{p}) \qquad 
\bar{u}_f(\tilde{\mathbf{p}}) =
\bar{u}_f(\tilde{\bLam} p)S(\Lambda ) .
\]
This leads to the identity
%
%which leads to the identity
%\beq
%\bar{u}_f(\mathbf{p}) \Gamma^{\mu} \beta \gamma_5 u_f (\mathbf{p}') 
%\bar{u}_f(\mathbf{p}')\beta  u_f(\mathbf{p}'')= 
%\bar{u}_f(\tilde{\bLam}{p}) \Gamma^{\nu}\Lambda_{\nu}{}^{\mu}
%\gamma^{\alpha}\Lambda_{\alpha}{}^{0}  
%\gamma_5 u_f (\tilde{\bLam}{p}') 
%\bar{u}_f(\tilde{\bLam}{p}')\gamma^{\beta}\Lambda_{\beta}{}^{0}  
%u_f(\tilde{\bLam}{p}''). 
%\label{g.12ab}
%\eeq
%Becasue the $+$ component of the current only gets rescaled under null
%plane boosts, this covariance simplifies when $\mu=+$
\[
{m \over \sqrt{p^+}} \bar{u}_f(\mathbf{p}) \Gamma^{+} (-P_d\cdot \gamma/M_d)  
\gamma_5
u_f (\mathbf{p}') \bar{u}_f(\mathbf{p}')(-P_d\cdot \gamma/M_d) u_f(\mathbf{p}'') {m
\over \sqrt{p^{+\prime}}}  =
\]
\beq 
{m \over \sqrt{(\Lambda p)^+}}
\bar{u}_f(\tilde{\bLam}{p}) \Gamma^{+}
(-\Lambda P_d\cdot \gamma/M_d) \gamma_5 u_f
(\tilde{\bLam}{p}')
\bar{u}_f(\tilde{\bLam}{p}')(-\Lambda P_d\cdot \gamma/M_d)
 u_f(\tilde{\bLam}{p}'') {m \over \sqrt{(\Lambda p'')^+}} .
\label{r.12abc}
\eeq
which defines a null-plane-boost-invariant current kernel
that agrees with (\ref{r.12a}) in the rest frame of the 
initial deuteron.  For the adjoint we use
the rest frame of the final deuteron.

The operator (\ref{r.12a}) factors into a product of $2 \times 2$ spin
matrices, $\bar{u}(\mathbf{p}) \Gamma^{\mu} (- P_d\cdot \gamma/M_d) 
\gamma_5 u
(\mathbf{p}')$ and $\bar{u}(\mathbf{p}')(- P_d\cdot \gamma/M_d) 
u(\mathbf{p}'')$. The
term $\bar{u}(\mathbf{p}')(- P_d\cdot \gamma/M_d) 
u(\mathbf{p}'')$ replaces
$\bar{u}(\mathbf{p}')\gamma_5 u(\mathbf{p}'')$ in one vertex of the
one-pion-exchange interaction, while the factor $\bar{u}(\mathbf{p})
\Gamma^{\mu} (- P_d\cdot \gamma/M_d) \gamma_5 u (\mathbf{p}')$ 
has the appearance of a modified one-body current.

To be consistent with our input two-body interaction we modify the
one-pion exchange contribution of the model interaction by replacing
the part of the phenomenological interaction that comes from
$\bar{u}(\mathbf{p}')\gamma_5 u(\mathbf{p}'')$ by a modified
interaction that comes from $\bar{u}(\mathbf{p}') (-\Lambda P_d\cdot
\gamma/M_d) u(\mathbf{p}'')=
\bar{u}(\mathbf{p}') \beta u(\mathbf{p}'')
$ in the deuteron rest frame.  We then
apply this modified interaction to the deuteron wave function and 
use kinematic null-plane boosts to transform the result to an arbitrary frame.
 
The resulting matrix elements of the kinematically covariant two-body
current has the following form
\[
\langle (1,d), \tilde{\mathbf{P}}', \nu'  \vert I_{ex}^\mu(0) \vert
(1,d), \tilde{\mathbf{P}}, \nu  \rangle := 
\]
\[
\int \langle (1,d),\tilde{\mathbf{P}}', \nu'  \vert 
(j_n,m_n),\tilde{\mathbf{p}}_1', \nu_1', 
(j_n,m_n),\tilde{\mathbf{p}}'_2 , \nu_2' 
\rangle 
(-{1 \over 2m}) \bar{u}_{nf}(p_1')\Gamma^{\mu} \gamma_5 
({P \cdot \gamma \over M_d}) {u}_{nf}(p_1'') 
d\tilde{\mathbf{p}}_1'd \tilde{\mathbf{p}}_1'' d\tilde{\mathbf{p}}_2'
\times 
\]
\beq
\langle (j_n,m_n),
\tilde{\mathbf{p}}_1'', \nu_1'', (j_n,m_n), \tilde{\mathbf{p}}'_2 , \nu_2'
\vert U(\Lambda_f (P/M_d)) \tilde{v}_{\pi} \vert 
(1,d), \tilde{\mathbf{P}}_0, \nu  \rangle + (1 \to 2 ) + h.c.
\label{j.1}
\eeq
where in this expression we replaced the Dirac spinors with a 
covariant normalization used in 
projection operator $\Lambda_{\pm} (\mathbf{p})$ by spinors with a 
non-covariant normalization:
\beq
u_{nf} (p) := \sqrt{m\over p^+} u_f(p),  
\eeq
to simplify the notation.  We also remark that the projectors can be
expressed in terms of canonical or null-plane spinors with covariant
normalization, $u_c(\mathbf{p})\bar{u}_c(\mathbf{p}) =
u_f(\mathbf{p})\bar{u}_f(\mathbf{p})$.
Details of the construction are discussed in appendix A and B.

The quantity 
\beq
\langle (1,d), \tilde{\mathbf{P}}, \nu  \vert 
(j_n,m_n), \tilde{\mathbf{p}}_1, \nu_1, (j_n,m_n),
\tilde{\mathbf{p}}_2 , \nu_2 
\rangle 
\eeq
is the deuteron eigenstate in the tensor product basis, 
$\tilde{\mathbf{P}}_0 = (M_d,0,0)$ is the rest frame 
value of the deuteron null-plane momentum,  
$U(\Lambda_f (P/M_d))$ represents a kinematic null-plane boost to the 
Breit frame.  The modified pion-exchange interaction $\tilde{v}_{\pi}$
has the form (see Appendix A): 
\[
\langle \tilde{\mathbf{P}}', {\mathbf{k}}'
 , \nu_1', \nu_2'
\vert \tilde{v}_{\pi} \vert
\tilde{\mathbf{P}}, {\mathbf{k}} , \nu_1, \nu_2\
\rangle  \to 
\]
\beq
\delta (\tilde{\mathbf{P}}' - \tilde{\mathbf{P}}) 
{ g^2_{\pi} \over (2 \pi)^3} 
{2m \over 2m} \btau_1 \cdot \btau_2   
v_{\pi} (\mathbf {k} - \mathbf{k}' ) 
{(\mathbf{k}' - \mathbf{k}) \cdot \bsig_2 \over 2m}  
\label{j.4}
\eeq
where $v_{\pi} (\mathbf {k} - \mathbf{k}' )$ is the coefficient 
function of the one-pion-exchange contribution to the operator 
\beq
\delta (\tilde{\mathbf{P}}' - \tilde{\mathbf{P}}) 
{ g^2_{\pi} \over (2 \pi)^3} 
{(\mathbf{k} - \mathbf{k}') \cdot \bsig_1 \over 2m}   \btau_1 \cdot \btau_2   
v_{\pi} (\mathbf {k} - \mathbf{k}' ) 
{(\mathbf{k}' - \mathbf{k}) \cdot \bsig_2 \over 2m}  
\label{j.5}
\eeq
in the phenomenological interaction.  For the Argonne V18 interaction
$v_{\pi} (\mathbf {k} - \mathbf{k}' )$ is extracted from the
one-pion-exchange contribution to the tensor $\times \, (\btau_1 \cdot
\btau_2)$ and spin-spin $\times \, (\btau_1 \cdot \btau_2)$ parts of
the interaction using the method discussed in \cite{Riska85} and
\cite{Schi90}.  The extraction is discussed in Appendix B.  The
resulting interaction, $v_{\pi} (\mathbf {k} - \mathbf{k}' )$, differs
from $1/ ( m_\pi^2 + (\mathbf {k} - \mathbf{k}' )^2)$ by the effects
of the short-distance cutoff in the AV18 interaction.

This exchange current is constructed to transform covariantly with
respect to the null-plane kinematic subgroup.  We add this current to
the impulse current when computing the three independent matrix
elements.  These matrix elements are then used to generate the remaining
current matrix elements using the constraints of current covariance and
current conservation as discussed in the previous section.

Each term in the expression (\ref{j.1}) for the 
exchange current matrix element can be represented as 
the product of $(-{1 \over 2m})$ with the modified current 
$
\bar{u}_{nf}(p_1')\Gamma^{\mu} \gamma_5 
({P \cdot \gamma \over M_d}) {u}_{nf}(p_1'') 
$
evaluated between a deuteron eigenstate and a pseudo-state defined by
applying the rotationally invariant modified interaction $\tilde{v}_{\pi}$
in (\ref{j.4}) to the rest deuteron state and kinematically boosting
the result to the Breit frame.  This defines a kinematically covariant
exchange current that we use to compute the independent 
matrix elements of $I^+(0)$.
 
\section{results} 

The input to our calculation of the elastic electron-deuteron
scattering observables is (1) the choice of nucleon form factors
\cite{nnformfactor-BI} \cite{nnformfactor-Lomon1}\cite{nnff-Lomon1}
\cite{nnff-BBA}\cite{nnff-BBBA}\cite{nff-kelly}, (2) the choice of
nucleon-nucleon interaction \cite{Wiringa95}\cite{cdb}, and (3) the choice
of independent linear combinations of current matrix elements used to generate
the full current,
\cite{Chung88}\cite{Frankfrut93} \cite{Terent77}\cite{Karmanov92}.

The extraction of proton-electric form factors based on polarization
measurements compared with measurements based on the Rosenbluth
separation were found to be inconsistent\cite{Punjabi05} ; these
inconsistencies have been explained \cite{tgm} by including
two-photon-exchange corrections in the Rosenbluth separation.  This
has led to a modification of the phenomenological parameterizations of
the proton electric form factor.  All of the nucleon form factors that
we use are consistent with the extractions based on the polarization
measurements.

Neutron electric form factor data is only available for a limited
range of momentum transfers.  The high-momentum transfer behavior of
the different parameterizations is a consequence of different
theoretical assumptions.  This leads to some variation of the
parameterizations for momentum transfers above $Q^2 \sim 1 (GeV)^2$.
Different parameterizations of the neutron electric form factor are
compared to a dipole form factor in fig 2.  The form factors that we
compare are recent parameterizations given by Lomon
\cite{nnformfactor-Lomon1}, Budd, Bodeck and Arrington (BBA)
\cite{nnff-BBA}, Bradford Budd, Bodeck and Arrington (BBBA)
\cite{nnff-BBBA}, Kelly \cite{nff-kelly}, and Bijker and Iachello (BI)
\cite{nnformfactor-BI}.  All of these parameterizations agree for
$Q^2<1$.  The curves in figure 2 are given as ratios to dipole form
factors, which emphasize differences in the parameterizations.  The
input to our calculations are the isoscalar linear combinations of the
nucleon form factors, $F_{1N}(Q^2)$ and $F_{2N}(Q^2)$.  These form
factors are plotted in figures 3 and 4 for the same parameterizations
that are compared in figure 2.  These plots show no significant
variation among the various nucleon form factors.  Our calculations of
elastic-scattering observables are not very sensitive to these
differences.

As shown in section VII, our model exchange current breaks up into
a product of an effective ``one-body'' current and an interaction.
The interaction is a modification of the one-pion exchange
contribution to the tensor part of the phenomenological interaction.
Most of our calculations are based on the Argonne V18 interaction.  We
extract the one-pion exchange contribution to the V18 potential by
first discarding the short-range parts of the interaction that
contribute to the spin-spin and tensor forces, then we use a method
developed by Riska \cite{Riska85} and Schiavilla, Pandharipande, and
Riska \cite{Schi90} to extract the one-pion exchange contribution to
the tensor force from the remaining parts of the interactions.  This
procedure is discussed in Appendix B.  The one-pion-exchange
potential that we extract differs from
${1 \over m_{\pi}^2 + (\mathbf{k}-\mathbf{k}')^2} $ by the effects of
the short distance configuration-space cutoff that appears in the
Argonne V18 interaction.  The Fourier transform of the extracted
interaction is compared to the pion-exchange potential without the
cutoff in figure 5.  The dotted curve includes the cutoff parameters
that are used in the Argonne V18 interaction.  The most important
differences are for momenta above 1-2 $fm^{-1}$.

Figures 6,7,8 show the three deuteron form factors, $G_0$, $G_1$ and
$G_2$ with and without the exchange current included for the
two-nucleon form factors (BI and BBBA) that have the largest 
high-momentum transfer difference in the neutron electric form factor.  The
independent matrix elements (\ref{d.5}) are calculated using the
one-body parts of the current and with the exchange current
added.  The remaining current matrix elements are determined by the
constraints of current conservation and current covariance.  The
figures show that the addition of the exchange current
contributions lead to an enhancement of $G_0$ above the minimum at
$Q^2=1GeV^2$. The minimum of $G_1$ shifts to the right, and $G_2$ is
enhanced.  Except for momentum transfers $Q^2$ between 6 and 7 these
form factors are not very sensitive to the different assumptions made
about the neutron electric form factor.

Data for the observables $A$ are labeled
Stanford Mark III~\cite{buchanan65}, CEA ~\cite{Elias69},
Orsay~\cite{Benak66}, SLAC E101~\cite{Arnold75}, Saclay
ALS~\cite{Platch90}, DESY ~\cite{Galster71},
Bonn~\cite{cramer85}, Mainz~\cite{simon81}, JLab Hall C
~\cite{JlabHallc99}, JLab Hall A ~\cite{JlabHalla99} and
Monterey~\cite{Berard74}.  Data for $B$ are 
labeled: SLAC NPSA NE4~\cite{Bosted90},
Martin~\cite{martin77}, Bonn~\cite{cramer85}, Saclay
ALS~\cite{auffret85}, Mainz~\cite{simon81}, Stanford Mark
III~\cite{buchanan65}.  Data for 
$T_{20}$  are labeled: Novosibirsk-85 ~\cite{Dmitr85}
~\cite{Novos86}, Novosibirsk-90~\cite{Gilman90},
Bates-84~\cite{Schul84}, Bates-91~\cite{The91} and JLab Hall
C~\cite{Jlabt20}.

Calculations of $A$, $B$, and $T_{20}$ using the independent matrix
elements (\ref{d.5}) (choice I) and the five parameterizations of the
nucleon form factors used in figures 2-4 are shown in figures 9-11.
While the null-plane impulse calculations give a qualitative
understanding of the data, it is clear from these calculations that
the null-plane impulse approximation is inadequate.

Figures 12,13,14 show the effects of including the phenomenological
pion-exchange current defined in section VII.  We see that $A$ and $B$
provide acceptable fits to the data when the pion exchange current is
included.  The results are insensitive to the assumptions used in
parameterizing the high-momentum transfer behavior of the neutron
electric form factor.  $T_{20}$ is closer to the data, but it is still
below the most recent Jlab hall C data \cite{Jlabt20} between $Q^2$ of
$.5$ and $2 GeV^2$.

The calculation displayed in figures 12,13,14 are based on the choice I of
independent current matrix elements given by (\ref{d.5}).  The
presence of the exchange current (\ref{j.1}) increases the sensitivity
to the choice of the independent current matrix elements.  This is
shown in figure 1.  What is plotted is the difference
$I^+_{11}(0)-I^+_{00}(0)$ with and without the exchange current using
a direct calculation of the difference or by generating the difference
using current conservation and current covariance.  The difference
between the dashed curve and solid curve shows that the required
two-body contributions to the current in this difference is small when
the independent current matrix elements are computed in the impulse
approximation.  Comparing the dotted and dash-dot curve indicates that
much larger required two-body contributions to the current are needed
when the exchange current contributions are included in all matrix
elements.  This suggests that there will be a non-trivial sensitivity
to the choice of independent current matrix elements used to 
generate the fully covariant exchange current.

To test this we examined the two other choices, II and III, of
independent matrix elements discussed in section VI.  These methods
relate form factors to independent current matrix elements by
contracting different sets of polarization vectors into the current
tensor.  In both approaches there are preferred polarization vectors;
in one case the vectors are chosen to minimize the dependence on
matrix elements that are maximally suppressed in the infinite momentum
frame ($P^+ \to 0$), while the other choice minimizes the mass
dependence in the contractions used to define the independent current
matrix elements.  Both choices were discussed in section VI.

Figures 15,16,17, and 18 show the deuteron elastic scattering
observables $A(Q^2)$ and $T_{20}(Q^2)$ for both choices, II and III,
of polarization vectors.  In both cases $G_1$ is computed using the
same linear combination of current matrix elements used for choice I.
Since $B$ only depends on $G_1$, $B$ is identical for all three
choices.  All three choices of independent matrix elements give
different predictions for $A$ and $T_{20}$.  For choices II and III
there is a mild enhancement of $A$ at higher momentum transfers
compared to choice of independent matrix elements given by
(\ref{d.5}).  There is also a larger effect on the tensor polarization
that brings the curve to within the experimental error bars.

The result is that both choice II and III of independent current
matrix elements give consistent results for the elastic scattering
observables and they both provide a good description of the existing
data over a wide range of momentum transfers.  It is clear that there
is a non-trivial sensitivity to the choice of independent current
matrix elements when these results are compared to the corresponding
results based on choice I.

Another potential source of sensitivity to the input is the
choice of nucleon-nucleon interaction.  Any phase equivalent change in
the nucleon-nucleon interaction is automatically accompanied by 
a corresponding change in the current operator.  For interactions 
with a long-range meson exchange tail one might expect that 
the same data could be understood by simply adjusting the 
cutoff parameter.   For typical soft interactions that are useful 
in low-energy problems, one expects that a more significant 
modification of the current would be necessary.   

In figs. 19,20 and 21 we compare calculations of $A(Q^2)$, $B(Q^2)$,
and $T_{20}(Q^2)$ using the CD Bonn wave functions with and without
the exchange current.  In these calculations the exchange current is
still based on the Argonne V18 cutoffs.  The calculations show the
generalized impulse calculations and calculations where the exchange
current is added to the impulse current,  without adjusting the cutoff
parameters. The calculations clearly show that there is more
sensitivity to the choice of nucleon-nucleon interactions than to the
choice of nucleon form factor.

Good consistency with all experimental observables is obtained using
the V18 interaction with nucleon form factors \cite{nnff-BBBA} and the
choice of independent current matrix elements suggested by Frankfrut,
Frederico and Strickman or Coester, discussed in section VI.  These
calculations are shown in figs. 22,23 and 24.  For these choices the
model exchange current explains the difference between the generalized
impulse approximation and the experimental data.  Figure 25 shows the
ratio of the experimental values of A(Q$^2$) to the calculated values
using exchange current II.  The data is presented on a linear scale to
better illustrate the comparison between theory and experiment.  The
solid line represents the calculation with exchange current II while
the diamonds show the ratio of exchange current III to exchange
current II.  There are no significant differences between the two
exchange currents for momentum transfers below 4 (GeV)$^2$.  While our
exchange current, which used the one-pion exchange part of the Argonne
V18 interaction, required the numerical computation of Fourier
transforms, the resulting interaction contribution to the exchange
current differed very little from a simple static momentum-space pion
exchange interaction, fig 5, which indicates the simplicity of our
exchange current.

Our results indicate that this simple exchange current is
sufficient to provide a good quantitative understanding of elastic
electron-deuteron  scattering for a wide range of momentum
transfers.  The sensitivities to both the choice of interaction and
the choice of independent matrix elements are the largest
uncertainties in the calculations, and these uncertainties are all
larger than the experimental uncertainties.  For momentum transfers
where data is available there is very little sensitivity to the
uncertainties in the neutron electric form factors.

Finally we compare the magnetic and quadrupole moments of the deuteron
with and without the exchange current and using the three different
choices of independent current matrix elements.  The results are shown
in table 1. Our calculations do not exhibit any sensitivity to the
choice of nucleon form factor, which are sufficiently well constrained
by experiment at low momentum transfers.  The exchange current
contributions affect the results for all of the moments. In both cases
they are closer to experiment than the moments computed using the
generalized impulse approximation.  The magnetic moment is in good
agreement (to within computational accuracy) with experiment, while
the quadrupole moment differs from the experimental result by a few
percent.

The conclusion of our research is that a simple exchange current
motivated by one-pion exchange and the freedom to define a conserved
covariant current operator by choosing a preferred set of independent
current matrix elements is sufficient to provide a good fit to all
three elastic scattering observables using Poincar\'e invariant
quantum theory with a null-plane kinematic symmetry.  The model
exchange current has the spin structure of a ``pair'' current, designed
with a null-plane kinematic symmetry.  The quality of the nucleon form
factors has progressed to the point that our results are insensitive
to the choice of nucleon form factor.

While is it straightforward to include low-order pion-exchange physics
in a more general class of models, the strategy for making the best
choice of independent current matrix elements in a general class of
electron-nucleus reactions requires more investigation.  The
principles used to derive choices I, II and III of independent current
matrix elements all can be generalized to treat initial and final
states with different spins.  Whether there is one consistent set of
principles that works universally for all reactions is not yet known.

A second observation is that our model, with one of the two preferred
choices of independent current matrix elements, provides a better
description of all three observables than methods based on truncations
of null-plane field theory or instant form relativistic quantum
mechanics.  Our model has features of both - unlike the instant-form
model our model has the full-null plane kinematic symmetry with all of
the advantages discussed at the beginning of this paper. Unlike
truncations of a null-plane field theory, which emphasize cluster
properties at the expense of exact Poincar\'e invariance, our model is
exactly Poincar\'e invariant.  We expect that the Poincar\'e
invariance constraint to be more important for momenta near or
slightly above the deuteron mass scale, since the deuteron 
mass scale is involved in the implementation of the symmetry.

This research provides a useful first step in trying to devise a more
systematic treatment of model exchange currents in Poincar\'e
invariant quantum mechanics with a null-plane kinematic symmetry.  It
leads to a simple current that provides a significant improvement in
all three elastic scattering observables when compared with the
corresponding impulse calculations, however additional research is
still needed in order to determine if these methods can be
successfully applied to larger class of reactions.

This work was performed under the auspices of the U.~S.  Department of
Energy, Office of Nuclear Physics, under contract
No. DE-FG02-86ER40286.  The authors would like to express our
gratitude to Fritz Coester who made many material suggestions that
significantly improved the quality of this work.

\appendix
\section{Current construction}

The steps motivating the structure of the model exchange current 
(\ref{j.1}) are
summarized in this appendix.  
We consider a pseudoscalar pion-nucleon vertex:
\beq
{\cal L}(x) = -i {g_{\pi}} :\bar{\Psi} (x)
\gamma_5 \Psi (x) \btau \cdot \bphi (x) : 
\label{g.1}
\eeq
where $g_{\pi} = 2m {f_{\pi} \over m_\pi}$ is the pseudoscalar
pion-nucleon coupling constant and $m$ is the nucleon mass.
In what follows we use canonical Dirac spinors with a non-covariant 
normalization
\beq
u_{nc} (\mathbf{p}) := \sqrt{m\over \omega(\mathbf{p}) } u_c(\mathbf{p})  
\eeq
and use the notation 
\beq
v_{\pi}(\mathbf{k}' , \mathbf{k} )  := {g_{\pi}^2  \over (2 \pi)^{3}}
{\btau_1 \cdot \btau_2
\over m_{\pi}^2 + (k'-k)^2 -i0^+} .
\eeq
In this notation the tree-level one-pion-exchange transition amplitude 
is given by the rotationally invariant kernel:
\[
\langle \mathbf{k}',\mu_1', \mu_2'  \vert v_{\pi} \vert \mathbf{k},
\mu_1,\mu_2 
\rangle :=
\]
\beq
\bar{u}_{nc}(\mathbf{k}' )\gamma_5 {u}_{nc}(\mathbf{k} ) v_{\pi} 
(\mathbf{k}' , \mathbf{k} )
\bar{u}_{nc}(-\mathbf{k}' )\gamma_5 {u}_{nc}(-\mathbf{k} ) 
\label{g.3}
\eeq
where 
%$(k'-k)^2 := (\mathbf{k}- \mathbf{k}')^2 - (\omega_m
%(\mathbf{k})- \omega_m(\mathbf{k}'))^2$ and 
we have assumed a
plane wave normalization $\langle \mathbf{k}' \vert \mathbf{k} \rangle
= \delta (\mathbf{k}' - \mathbf{k})$.

The tree-level one-pion-exchange transition amplitude in the presence of an 
external electromagnetic field,  
using the vertex (\ref{g.1}) with 
\beq
A^{\mu}(q) := {1 \over (2 \pi)^4} \int
e^{- i q \cdot y} A_{\mu}(y) d^4y, 
\label{g.5}
\eeq
includes four terms, one of which is 
\[
\langle \mathbf{p}_1', \mu_1', \mathbf{p}_2', \mu_2' \vert I^{\mu}(0)^{\mu}
\vert \mathbf{p}_1, \mu_1, \mathbf{p}_2, \mu_2 \rangle = 
\]
\[  
\bar{u}_{nc}(\mathbf{p}_1') \Gamma^{\mu} u_{nc}(\mathbf{r}) 
\bar{u}_{nc}(\mathbf{r}) \gamma_5  u_{nc}(\mathbf{p}_1) 
{1 \over E_{12} - \omega(\mathbf{r}) - \omega(\mathbf{p}_2') +i0^+} \times
\]
\[
v_{\pi}(\mathbf{p}_2' , \mathbf{p}_2 )
\bar{u}_{nc}(\mathbf{p}_2') \gamma_5  u_{nc} (\mathbf{p}_2) 
+
\]
\[
\bar{u}_{nc}(\mathbf{p}_1') \Gamma^{\mu} v_{nc}(-\mathbf{r}) 
\bar{v}_{nc}(-\mathbf{r}) \gamma_5  u_{nc}(\mathbf{p}_1) 
{1 \over E_{12} - \omega(\mathbf{p}_2') + \omega(\mathbf{r}) -i0^+}
\times
\]
\beq
v_{\pi}(\mathbf{p}_2' , \mathbf{p}_2 )
\bar{u}_{nc}(\mathbf{p}_2') \gamma_5  u_{nc} (\mathbf{p}_2) 
+ \cdots 
\label{g.6}
\eeq
where 
\beq
E_{12} =  \omega(\mathbf{p}_1) + \omega(\mathbf{p}_2) 
\label{g.7}
\eeq
is the initial energy,
\beq
\mathbf{r}= \mathbf{p}_1 + \mathbf{p}_2 - \mathbf{p}_2' .
\label{g.8}
\eeq
and 
\beq
\Gamma^{\mu} =  \gamma^{\mu} F_1 (Q^2) - {i \over 2 m} \sigma^{\mu \nu},
Q_{\nu} F_2 (Q^2)
\label{g.8a}
\eeq
is the nucleon impulse current.   The $+\cdots$ represents the contribution 
from the other three terms related by either Hermitian conjugation 
and/or exchanging the proton and neutron.

To motivate the structure our model exchange current we first evaluate this
covariant expression in the rest frame of the initial two-body system
so the pion-exchange interaction appears as a rotationally
invariant function of the relative momenta.  This allows us to relate
the interaction part of this kernel to the interaction kernel in
(\ref{g.3}).  Next we treat the initial energy of the two-body
system, $E_{12}$, as a parameter that can be expressed in terms of the
mass of the initial state (deuteron) and the kinematically conserved
momenta.  We use the assumed null-plane kinematic symmetry of the 
interaction to express the energy denominators in terms of masses 
and null-plane kinematic variables;  
$E_{12} \to {1 \over 2} (P^+ + {M^2 + \mathbf{P}_{\perp} \over
P^+})$, $E_{120} \to {1 \over 2} (P^+ + {M_0^2 + \mathbf{P}_{\perp} \over
P^+})$.  These are all model assumptions.  
In the rest frame of the initial deuteron $P^-=M_d$
which gives 
\beq E_{12} -\omega(\mathbf{r}) -\omega(\mathbf{p}_2') \to
E_{12} - E_{120}  \to {1 \over 2} {M_d^2
-M_0^2(\mathbf{k}') \over P^+_{rest}} \to {M_d^2 -M_0^2(\mathbf{k}')
\over 2 M_d } \approx {M_d^2 -M_0^2(\mathbf{k}') \over 4 m }
\label{g.9}
\eeq
and 
\beq
 E_{12} - \omega(\mathbf{p}_2') + \omega(\mathbf{r})  \to E_{12} \to 
{1 \over 2} (P^+ + {M^2 + \mathbf{P}_{\perp}
\over P^+}) \to
{1 \over 2} (M_d + {M_d^2 
\over M_d}) = M_d \approx 2m
\label{g.10}
\eeq
Using (\ref{g.9}) and (\ref{g.10}) in (\ref{g.6}) gives
\[
\langle \mathbf{p}_1', \mu_1', -\mathbf{k}', \mu_2' \vert I^{\mu}(0)^{\mu}
\vert \mathbf{k}, \mu_1, -\mathbf{k}, \mu_2 \rangle = 
\]
\[  
\bar{u}_{nc}(\mathbf{p}_1') \Gamma^{\mu} u_{nc}(\mathbf{k}') 
{4m \over M_d^2 - M_0^2  -i0^+}
\bar{u}_{nc}(\mathbf{k}') \gamma_5  u_{nc}(\mathbf{k}) 
\times
\]
\[
v_{\pi}(\mathbf{k}' , \mathbf{k} )
\bar{u}_{nc}(-\mathbf{k}') \gamma_5  u_{nc} (-\mathbf{k}) 
-
\]
\[
\bar{u}_{nc}(\mathbf{p}_1') \Gamma^{\mu} v_{nc}(-\mathbf{k}') 
{1 \over 2m} 
\bar{v}_{nc}(-\mathbf{k}') \gamma_5  u_{nc}(\mathbf{k}) 
\times
\]
\beq
v_{\pi}(\mathbf{k}' , \mathbf{k} )
\bar{u}_{nc}(-\mathbf{k}') \gamma_5  u_{nc} (-\mathbf{k}) 
+ \cdots 
\label{g.11}
\eeq
The terms in this expression are easy to interpret.  The first two
lines represent the product of a one-body current matrix element, the
propagator, ${4m \over M_d^2 - M_0^2 -i0^+}$, associated with
equation (\ref{c.4}) and the rotationally invariant kernel of the
pion-exchange interaction (\ref{g.3}).  This term is already included
in the one-body contribution to the current matrix element.

The last two lines have a similar form except one of the
$u_{nc}(\mathbf{k})$ spinors in the current and the rotationally
invariant interaction kernel are replaced by $v_{nc}(-\mathbf{k})$
spinors.  In addition, the propagator term is replaced by the factor
$1/2m$.  We find it convenient for computational purposes to split the
$v$ spinor terms that give the Dirac projector
$\Lambda_-(-\mathbf{k}) =- v_c(-\mathbf{k}) \bar{v}_c(-\mathbf{k}) $
and to use $\gamma_5 \beta$ to convert the $v_{nc}(-\mathbf{k})$ to
$u_{nc}(\mathbf{k})$.  The
two factors of $\beta$ break kinematic covariance.
We restore manifest kinematic covaraince by using this
expression to define the current matrix elements in the rest frame of
the initial system, and we transform to the Breit frame by requiring 
kinematic covaraince.

Using   
\beq
v_{nc} (-\mathbf{p}) \bar{v}_{nc}(-\mathbf{p})  =
\gamma_5 \beta u_{nc} (\mathbf{p}) \bar{u}_{nc} (\mathbf{p})\beta \gamma_5  
\label{g.12}
\eeq
the last two lines of (\ref{g.11}) become 
\[
\bar{u}_{nc}(\mathbf{p}_1') \Gamma^{\mu}   \gamma_5 \beta u_{nc} (\mathbf{k}') 
{1 \over 2m} 
\times
\]
\beq
\bar{u}_{nc}(\mathbf{k}')\beta u(\mathbf{k}) 
v_{\pi}(\mathbf{k}' ), (\mathbf{k} ))
\bar{u}_{nc}(-\mathbf{k}') \gamma_5  u_{nc} (-\mathbf{k}) 
+ \cdots . 
\label{g.13}
\eeq
This separates into the product of an effective one-body current,
\beq
\bar{u}_{nc}(\mathbf{p}_1') \Gamma^{\mu} \gamma_5 
\beta u_{nc} (\mathbf{k}') 
\label{g.14}
\eeq
a factor $1/2m$, and a modified one-pion-exchange interaction
\beq
\bar{u}_{nc}(\mathbf{k}')\beta u_{nc}(\mathbf{k}) 
v_{\pi}(\mathbf{k}' ), (\mathbf{k} ))
\bar{u}_{nc}(-\mathbf{k}') \gamma_5  u_{nc} (-\mathbf{k}) .
\label{g.15}
\eeq
The interaction term is identical to the rotationally invariant kernel
of the one-pion-exchange interaction (\ref{g.3}) with the replacement:
$\bar{u}_{nc}(\mathbf{k}')\gamma_5 u_{nc}(\mathbf{k}) \to
\bar{u}_{nc}(\mathbf{k}')\beta u_{nc}(\mathbf{k})$.  This replacement 
preserves the rotational invariance of this kernel in the rest frame, 
but it is not covariant with respect to null-plane boosts.

This splitting has the advantage that the computation of the exchange
current matrix element has the same structure as the computation of an
impulse current matrix element with the current replaced by
(\ref{g.14}) and the deuteron wave function $\vert \psi \rangle$
replaced by ${1\over 2m} \tilde{v} 
\vert \psi \rangle$ where $\tilde{v}$ is the modified
interaction (\ref{g.15}).  The input to this calculation 
is defined in rest frame of the initial deuteron.  We get the 
null-plane Breit frame result by requiring 
kinematic covariance. 
 
The effective one-body current (\ref{g.14}) can be extended to 
a kinematically covariant operator that agrees with 
(\ref{g.14}) in the deuteron rest frame by replacing 
the $\beta$ in (\ref{g.14}) by 
\beq
\beta \to - P\cdot \gamma /M_d    
\label{g.16}
\eeq
This leads to a kinematically covariant modified current kernel 
\beq
\bar{u}_{nc}(\mathbf{p}_1') \Gamma^{\mu} \gamma_5 {(P\cdot \gamma) 
\over M_d}   
u_{nc} (\mathbf{k}'). 
\label{g.17}
\eeq
Kinematic covariance is all that this needed for a consistent 
computation of the current matrix elements.

The modified interaction (\ref{g.15}) is rotationally invariant so
when it is applied to the rest deuteron eigenstate the resulting
pesudo-state has the same spin as the deuteron.  This can be
consistently defined in any other kinematically related frame using
null-plane boosts.  This, along with the replacement (\ref{g.16})
restores the manifest kinematic covariance.

The last step is to replace the modified interaction (\ref{g.15}) by
the corresponding modified one-pion-exchange part of a realistic
model interaction.  In a typical realistic interaction the
pseudoscalar pion-exchange interaction is obtained by replacing the
spinor terms in the rotationally invariant kernel (\ref{g.3}) by
\beq
\bar{u}_{nc}(\mathbf{k}')\gamma_5 u_{nc}(\mathbf{k})
\to {\bsig \cdot (\mathbf{k} - \mathbf{k}') \over 2m}
\label{g.18} 
\eeq
This expression is obtained by retaining the leading term in a 
$\mathbf{k}/m$ expansion of the spinor term, which is normally 
justified because the one-pion exchange interaction also includes 
a high-momentum or short-distance cutoff.  The net effect is that the 
resulting interaction, when included in the full nucleon-nucleon 
interaction, provides a good description of the two-nucleon bound state 
and scattering observables. 

Expanding
$\bar{u}(\mathbf{k}')\beta u(\mathbf{k}) $ to the same order in
$\mathbf{k}/m$
gives 1.
This suggests that the modified interaction (\ref{g.15}) 
can be modeled by replacing the one-pion exchange 
contribution to the realistic interaction  
\beq
\bsig_1 \cdot (\mathbf{k} - \mathbf{k}') v_\pi (\mathbf{k}-\mathbf{k}')  
\bsig_2 \cdot (\mathbf{k}' - \mathbf{k}) \btau_1 \cdot \btau_2 
\label{g.17a}
\eeq
by
\beq 
2m v_\pi (\mathbf{k}-\mathbf{k}')
\bsig_2 \cdot (\mathbf{k}' - \mathbf{k}) \btau_1 \cdot \btau_2 .
\label{g.18a}
\eeq 
The one-pion exchange interaction (\ref{g.3}) only contributes to the
part of the tensor force in the nucleon-nucleon interaction that
multiplies the isospin exchange operator $\btau_1 \cdot \btau_2$.  The
tensor interaction also has contributions from the vector exchanges
which also contribute to the spin-spin interaction.  Riska
\cite{Riska85} and Schiavilla, Pandharipande, and Riska \cite{Schi90}
introduced a method for isolating the pion-exchange contribution to
the tensor force of a phenomenological interaction using linear
combinations of the tensor and spin-spin interactions. The resulting
interaction is
\beq
v_{ps}(\mathbf{k}-\mathbf{k}') = {1 \over 3} (2 v_t
(\mathbf{k}-\mathbf{k}') - v_{ss}(\mathbf{k}-\mathbf{k}')) 
\label{g.18b}
\eeq
where $v_t$ and $v_{ss}$ are tensor and spin-spin contributions to the
charge exchange part of the Argonne V18 interaction.
We extract this from the pion-exchange contribution to the Argonne V18
potential.  This is compared to ${1 \over m_{\pi}^2 +
(\mathbf{k}-\mathbf{k}')^2}$ in figure 5.  The difference between
these curves is due to the short-distance cutoff used in the AV18
interaction.

When this interaction is applied to the deuteron bound state vector
the result is a spin 1 ``pseudo wave function''. The resulting
``pseudo vector'' can be defined in the Breit frame by requiring that
it transforms covariantly with respect to the null-plane kinematic
subgroup.  This kinematic covariance ensures that the current kernel
is kinematically covariant provided the pseudo-current is modified
following (\ref{g.16}).  The kinematic covariance of the current is
needed for a consistent calculation matrix elements of $I^+(0)$.

We can now write the form of the Breit frame matrix elements of this
one-pion exchange contribution to the exchange current:
\[
\langle (1,d),\tilde{\mathbf{P}}',  \mu' \vert I_{ex}^+ (0) \vert 
(1,d), \tilde{\mathbf{P}}, \mu \rangle   =
\]
\[
\sum \int \langle (1,d),\tilde{\mathbf{P}}', \mu' \vert 
(j',k'), \tilde{\mathbf{P}}''' , l',s', \mu''' \rangle 
k^{\prime 2} dk' d\tilde{\mathbf{P}}'''
\times
\]
\[
\langle (j',k'), \tilde{\mathbf{P}}''' , l',s', \mu''' \vert
(j_p,m_p), \tilde{\mathbf{p}}_p', \mu_p' , (j_n,m_n),
\tilde{\mathbf{p}}_n', \mu_n' 
\rangle 
d\tilde{\mathbf{p}}_p' d\tilde{\mathbf{p}}_n'
\times
\]
\[
\langle  \tilde{\mathbf{p}}'_p, \mu'_p , \tilde{\mathbf{p}}'_n, \mu'_n
\vert I_{ex-eff}^+(0) \vert 
(j_p,m_p) \tilde{\mathbf{p}}_p, \mu_p , 
(j_n , f_n) \tilde{\mathbf{p}}_n, \mu_n
\rangle 
d\tilde{\mathbf{p}}_p d\tilde{\mathbf{p}}_n
\times 
\]
\beq
\langle (j_p,m_p), \tilde{\mathbf{p}}_p, \mu_p , 
(j_n,m_n), 
\tilde{\mathbf{p}}_n, \mu_n
\vert (j,k) \tilde{\mathbf{P}}'' ,l,s, \mu''
\rangle 
k^{2} dk d\tilde{\mathbf{P}}''
\langle  (j,k) \tilde{\mathbf{P}}'' , l,s, \mu'' 
\vert  \tilde{\mathbf{P}},  \mu ,\chi 
\rangle 
\label{g.19}
\eeq
where the terms in this expression are the deuteron wave 
function in the free-particle irreducible null-plane basis 
\beq
\langle (1,d) \tilde{\mathbf{P}}',  \mu',d \vert 
(1,k'),  l',s', \tilde{\mathbf{P}}''',  \mu''' \rangle
=
\delta (\tilde{\mathbf{P}}'- \tilde{\mathbf{P}}''')
\delta_{j'j'''} \delta_{\mu'\mu'''}
\phi^*_{j'} (k',l',s') \qquad (j'=s'=1),  
\label{g.20}
\eeq
the Poincar\'e group Clebsch-Gordan coefficients in the null-plane basis 
\[
\langle \tilde{\mathbf{P}}''' , k' ,j', l',s', \mu''' \vert
\tilde{\mathbf{p}}_p', \mu_p' , \tilde{\mathbf{p}}_n', \mu_n' 
\rangle = 
\]
\[
\delta ( \tilde{\mathbf{P}}'''- 
\tilde{\mathbf{p}}_p'- \tilde{\mathbf{p}}_n') 
{\delta (k' - k(\tilde{\mathbf{p}}_p',\tilde{\mathbf{p}}_n')) 
\over k^{\prime 2}}  
\sqrt{{\partial ( \tilde{\mathbf{P}} , \mathbf{k} ) \over
\partial (\tilde{\mathbf{p}}_p', \tilde{\mathbf{p}}_n') }}
\langle j, \mu \vert l, m_l, s, m_s \rangle
\times
\]
\[
\langle s, m_s \vert {1 \over 2}, \mu_p, {1 \over 2}, \mu_n \rangle 
Y_{l m_l}^* (\hat{\mathbf{k}}( \tilde{\mathbf{p}}_p' \tilde{\mathbf{p}}_n'))
\times
\]
\beq  
D^{1/2}_{\mu_p \mu_p'} [\Lambda_c^{-1} (\mathbf{k}/m) \Lambda_f (\mathbf{k}/m)]
D^{1/2}_{\mu_n \mu_n'} [\Lambda_c^{-1} (-\mathbf{k}/m) \Lambda_f (-\mathbf{k}/m)], 
\label{g.21}
\eeq
the proton effective current
\beq
\langle  \tilde{\mathbf{p}}'_p, \mu'_p , \tilde{\mathbf{p}}'_n, \mu'_n
\vert I_{ex-eff}^+(0) \vert 
\tilde{\mathbf{p}}_p, \mu_p , \tilde{\mathbf{p}}_n, \mu_n
\rangle =
\label{g.22}
\eeq
\beq
\delta (\tilde{\mathbf{p}}_n - \tilde{\mathbf{p}}_n)
\bar u_{nf}(\tilde{\mathbf{p}}'_p,\mu'_p) 
\Gamma_p^\mu (q) \gamma^5 (\eta^0 \gamma^0 - \eta^1 \gamma^1) 
u_{nf}(\tilde{\mathbf{p}}_p,\mu_p),
\label{g.23}
\eeq
where the null plane Dirac spinors are related to the 
canonical Dirac spinors by a Melosh rotation
\beq
u_f(\tilde{\mathbf{p}}_p,\mu_p)_{\mu}   =
u_c(\tilde{\mathbf{p}}_p,\mu_p)_{\mu'}  
D^{1/2}_{\mu' \mu} [\Lambda_c^{-1} (\mathbf{p}_p/m) \Lambda_f (-\mathbf{p}_p/m)],
\label{g.24}
\eeq
another Poincar\'e Clebsch-Gordan coefficient
\beq
\langle \tilde{\mathbf{p}}_p, \mu_p , \tilde{\mathbf{p}}_n, \mu_n
\vert \tilde{\mathbf{P}}'' , k ,j, l,s, \mu''
\rangle = 
\label{g.25}
\eeq
\beq
\delta ( \tilde{\mathbf{P}}'''- 
\tilde{\mathbf{p}}_p'- \tilde{\mathbf{p}}_n') 
{\delta (k' - k(\tilde{\mathbf{p}}_p', \tilde{\mathbf{p}}_n')) 
\over k^{\prime 2}}  
\sqrt{{\partial ( \tilde{\mathbf{P}} , \mathbf{k} ) \over
\partial (\tilde{\mathbf{p}}_p', \tilde{\mathbf{p}}_n') }}
\times
\label{g.26}
\eeq
\[
D^{1/2}_{\mu_p \mu_p'} [\Lambda_f^{-1} (\mathbf{k}/m) \Lambda_c (\mathbf{k}/m)]
D^{1/2}_{\mu_n \mu_n'} [\Lambda_f^{-1} (-\mathbf{k}/m) \Lambda_c (-\mathbf{k}/m)]
\times
\]
\beq 
\langle s, m_s \vert {1 \over 2}, \mu_p', {1 \over 2}, \mu_n' \rangle 
\langle j, \mu \vert l, m_l, s, m_s \rangle 
Y_{l m_l} (\hat{\mathbf{k}}( \tilde{\mathbf{p}}_p', \tilde{\mathbf{p}}_n')),  
\label{g.27}
\eeq
and the pseudo wave function
\[
\langle  (k,j),l,s,\tilde{\mathbf{P}}'' , \mu'' 
\vert  \tilde{\mathbf{P}}, \mu ,\chi 
\rangle =  
\]
\[
\delta (\tilde{\mathbf{P}}'' - \tilde{\mathbf{P}})
\int \sum
\langle j, \mu'' \vert l, m_l, s ,m_s \rangle 
\langle s, m_s  \vert {1 \over 2},  \mu_p ,{1 \over 2}, \mu_n \rangle  
{1 \over (2 \pi)^{3/2}} 
Y^*_{l m_l}( \hat{\mathbf{k}})  d\hat{\mathbf{k}}
\times 
\]
\beq
(\mathbf{k} - \mathbf{k}')\cdot \bsig_{\mu_n \mu'_n} 
v_{ps}(\mathbf{k}' - \mathbf{k}) \btau_1 \cdot \btau_2
d \mathbf{k}' 
Y_{l' m_l'}( \hat{\mathbf{k}}')
\langle s', m'_s, \vert  {1 \over 2},  \mu_p' ,{1 \over 2}, \mu_n' \rangle 
\langle j, \mu, \vert, s', m'_s, l', m'_l \rangle 
\phi_{lsj} (k')  
\label{g.28}
\eeq
where $v_{ps}$ is given by (\ref{g.18}).
The full exchange current is the sum of the above quantity and
the three other terms related by taking Hermitian conjugates or 
exchanging the particle that couples to the photon.
The computation of the pseudo wave function is discussed in 
Appendix B.

\section{Calculation of pseudo wave function}

The calculation of the pseudo wave function$~(\ref{g.28})$ requires
the one-pion exchange part of the tensor interaction.
Riska~\cite{Riska85} and Schiavilla, Pandharipande, and Riska
\cite{Schi90}outlined a method to extract the pseudoscalar
contribution to the tensor force as a linear combination of the radial
coefficient functions in the Argonne V18 interaction.  The method is
based on the observation that both pseudoscalar and vector meson
exchange contribute to both the tensor and spin-spin interaction in
the static limit\cite{Riska85}.

The Fourier transform of the interaction has the structure 
\beq
v_{nn}(\bkappa) = \Omega_t v_t (\bkappa^2) + 
\Omega_{ss} v_{ss} (\bkappa^2)
+ \cdots 
\label{i.1}
\eeq
where $\bkappa:= \mathbf{k}'-\mathbf{k}$,
\beq 
\Omega_{t}:= (\bsig_{1} \cdot \bsig_2 \bkappa^2-3 \bsig_1 \cdot \bkappa \bsig_2 \cdot  \bkappa)(   \btau_1 \cdot \btau_2)
\label{i.2}
\eeq
\beq
\Omega_{s}:=\bkappa^2 (\bsig_1 \cdot \bsig_2)(  \btau_1 \cdot \btau_2)
\label{i.3}
\eeq
and we have explicitly exhibited the tensor and spin-spin contributions
to the interaction.

The coefficient functions in (\ref{i.1}) are Fourier transforms 
of the coefficient functions the corresponding 
terms in the Argonne V18 interaction, 
\beq 
v_{ss}(\bkappa^2)=\frac{4\pi}{\bkappa^2}
\int^{\infty}_{0} dr r^2 v^{\sigma \tau}(r)(j_0(\kappa r)-1),
\label{i.4}
\eeq
\beq 
v_{t}(\bkappa^2)=\frac{4 \pi}{\bkappa^2}\int_{0}^{\infty} 
dr r^2 v^{t\tau}(r)j_2(\kappa r),
\label{i.5}
\eeq
where the factor $1/ \bkappa^2$ is due to the difference in
the conventions used to define tensor operator in  
momentum and configuration space.

These coefficient functions have contributions
from both pseudoscalar and vector meson exchange.  To separate
them, following \cite{Riska85}, define the tensor and spinor operators
and note that 
\beq
(\bsig_1 \cdot \bkappa \bsig_2 \cdot \bkappa) ( (  \btau_1 \cdot \btau_2)) = 
- \frac{1}{3}
({\Omega}_{t} -  \Omega_{ss})
\label{i.6}
\eeq
while vector meson exchange gives the combination \cite{Riska85}
\beq
((\bsig_1 \times \bkappa) \cdot (\bsig_2 \times \bkappa)
)(  \btau_1 \cdot \btau_2) =
\frac {1}{3} [2\Omega_{ss} + \Omega_{t} ] .
\label{i.7}
\eeq
This implies that 
\beq
v_t \Omega_t + v_{ss} \Omega_{ss} 
= v_{ps} \frac{1}{3}
({\Omega}_{ss} -  \Omega_{t}) + 
v_{v} \frac {1}{3} (2\Omega_{s} + \Omega_{t} ) =
\label{i.8}
\eeq
\beq
\frac{1}{3} (v_{v}-v_{ps}) \Omega_t + {1\over 3} 
(v_{ps} + 2 v_{v}) \Omega_s .
\label{i.9}
\eeq
Using these expressions we can isolate the pseudoscalar and vector
contribution to the interaction using the following linear combinations of the 
tensor and spin-spin interactions:
\beq
v_{ps} = v_{ss} - 2 v_t
\label{i.10}
\eeq
\beq
v_{v} = v_{ss} + v_t .
\label{i.11}
\eeq
For the Argonne V18 interaction, which is used in our calculations,
there remains a small difference between the pseudoscalar interaction
calculated using (\ref{i.10}) and the contribution that comes
directly from terms that can be directly identified with the one-pion
exchange contribution.  We only retain the one-pion exchange
contribution to these terms, this still
includes the short-distance cutoff used
in the Argonne V18 r-space potential.

The interactions 
$v^{\sigma \tau}(r)$ and $v^{t\tau}(r)$ have the form 
\beq
v^{\sigma \tau}(r) \equiv {f^2\over 9}\left\{\left({m_0\over m_\pm}\right)^2m_0 Y(\mu_0,r)
+2m_\pm Y(\mu_\pm,r)\right\}+v^c(r)
\label{i.12}
\eeq
\beq
v^{t\tau}(r) \equiv {f^2\over 9}\left\{\left({m_0\over m_\pm}\right)^2m_0 T(\mu_0,r)
+2m_\pm T(\mu_\pm,r)\right\}+v^t(r) .
\label{i.13}
\eeq
Here $Y(\mu ,r)$ and $T(\mu ,r)$ are the Yukawa and tensor
functions with the exponential cutoff of the Urbana and Argonne
$V_{14}$ models:
\beq
Y(\mu ,r) = {e^{-\mu r} \over \mu r} (1 - e^{-cr^2}),
\label{i.14}
\eeq
\beq
T(\mu ,r) = (1 + {3 \over \mu r} + {3 \over (\mu r)^2 } )Y(\mu r)
(1 - e^{-cr^2}),
\label{i.15}
\eeq
The terms $v^c(r)$ and $v^t(r)$ are short-range phenomenological 
interactions that we discard for the computation of the
exchange current.
 
In order to carry out the angular part of the integral in (\ref{g.28}), we 
expand the pseudoscalar potential $v_{ps}(\mathbf{k} - \mathbf{k}')$ in 
partial waves:    
\beq
v_{ps} (| \mathbf{k}-\mathbf{k}'|) =
\sum_{lm} v_l(\mathbf{k}^2,\mathbf{k}^{\prime 2})Y^*_{lm}(\hat{\mathbf{k}})
Y_{lm}(\hat{\mathbf{k}}'),  
\label{i.16}
\eeq
where 
\beq
v_l (\mathbf{k}^2,\mathbf{k}^{ \prime 2} ) = 2 \pi \int_{-1}^1
P_l (u) v_{ps}(\sqrt{\mathbf{k}^2 + \mathbf{k}^{\prime 2} - 2 kk'' u} )du.  
\label{i.17}
\eeq
We expand 
the vector $\mathbf{k}$ and $\mathbf{k}'$  in spherical 
harmonics
\beq
\mathbf{k} = k(\sqrt{{2 \pi \over 3}} (Y_{1-1}(\hat{\mathbf{k}})
- Y_{11}(\hat{\mathbf{k}})), i\sqrt{{2 \pi \over 3}} (Y_{1-1}(\hat{\mathbf{k}})
+ Y_{11}(\hat{\mathbf{k}})), \sqrt{{4 \pi \over 3}} Y_{10}(\hat{\mathbf{k}}))
\label{i.18}
\eeq
to obtain the following expression for the 
pseudo wave function:
\beqa 
&\quad & \chi_{\tilde P_d,\mu_d}^{(c)}(\vec {k}'',\mu''_p,\mu'_n) \\
&\equiv & \frac{1}{(2\pi)^3}   
\int d\hat{\mathbf{k}} k^2 dk  \sum_{l=0,2} \sum_{\mu_l=-l}^l \sum_{l'=0}^\infty 
\sum_{\mu_l'=-l'}^{l'} v_{l'} (\mathbf{k}^2,\mathbf{k}^{''2}  )
Y_{l'\mu_l'}(\hat{\mathbf{k}}'') Y^*_{l' \mu_l'} (\hat{\mathbf{k}})
Y_{l\mu_l}(\hat{\mathbf{k}}) \nonumber \\
&\times &
[k''(\sqrt{{2 \pi \over 3}} (Y_{1-1}(\hat{ \mathbf{k}''})
- Y_{11}(\hat{\mathbf{k}''})), i\sqrt{{2 \pi \over 3}} (Y_{1-1}(\hat{\mathbf{k}''})
+ Y_{11}(\hat{\mathbf{k}})), \sqrt{{4 \pi \over 3}} Y_{10}(\hat{\mathbf{k}})) \nonumber \\
&-&
k(\sqrt{{2 \pi \over 3}} (Y_{1-1}(\hat{ \mathbf{k}})
- Y_{11}(\hat{\mathbf{k}})), i\sqrt{{2 \pi \over 3}} (Y_{1-1}(\hat{\mathbf{k}})
+ Y_{11}(\hat{\mathbf{k}})), \sqrt{{4 \pi \over 3}} 
Y_{10}(\hat{\mathbf{k}}))] \nonumber \\
&\times & {\langle \mu'_n  \vert \vec \sigma \vert \mu_n \rangle}
\langle {s_p},\mu''_p,{s_n},\mu_n \vert s,\mu_s \rangle 
\langle  s, \mu_s, l,\mu_l\vert j,\mu \rangle .
Y_{\mu_l}^l(\hat{\mathbf{k}}) u_l(k), \nonumber
\label{i.19}
\eeqa
The angular integrals are evaluated using:
\beq
\int d\hat{\mathbf{k}}
Y^{*l'}_{\mu_l'} (\hat{\mathbf{k}})
Y^l_{\mu_l} (\hat{\mathbf{k}}) = \delta_{l'l} \delta_{\mu_l'\mu_l},
\label{i.20}
\eeq
\beqa
&\quad & \int d\hat{\mathbf{k}}
Y^{*l'}_{\mu_l'} (\hat{\mathbf{k}})
Y^l_{\mu_l} (\hat{\mathbf{k}})
Y^1_{\mu_l''} (\hat{\mathbf{k}}) \\
& = & 
\sqrt{{(2\cdot 1 +1) (2 \cdot l +1) \over 4 \pi (2\cdot l'+1)}}
\langle 1, \mu_l'', l, \mu_l \vert l', \mu_l' \rangle
\langle 1, 0, l, 0 \vert l', 0 \rangle. \nonumber
\label{i.21}
\eeqa

After carrying out the integral on $\hat{\mathbf{k}}$:
the expression for the pseudo wave function becomes 
\beqa
&\quad & \chi_{\tilde P_d,\mu_d}^{(c)}(\vec {k}'',\mu''_p,\mu'_n) \\
&\equiv & \frac{1}{(2\pi)^3}   
\int k^2 dk  \sum_{l=0,2} \sum_{\mu_l=-l}^l \sum_{l'}  \sum_{\mu_l'=-l'}^{l'}
v_{l'} (\mathbf{k}^2,\mathbf{k}^{\prime \prime 2}, )
Y_{l'\mu_{l'}}(\hat{\mathbf{k}}'') u_l (k)
\nonumber \\ &\times & 
\langle {s_p},\mu''_p,{s_n},\mu_n \vert s,\mu_s \rangle 
\langle  s, \mu_s, l,\mu_l\vert j,\mu \rangle 
\nonumber \\ &\times & 
[
\sqrt{{2 \pi \over 3}} \langle \mu_n' \vert \sigma_{x} \vert \mu_n \rangle 
\quad [\quad k''(Y_{1-1}(\hat{\mathbf{k}''}) - Y_{11}(\hat{\mathbf{k}''}) )\delta_{l'l}\delta_{\mu_{l'}\mu_l}
\nonumber \\ &+& 
k\sqrt{{(2\cdot 1 +1) (2 \cdot l +1) \over 4 \pi (2\cdot l'+1)}}
\langle 1, 0, l, 0 \vert l', 0 \rangle
(-\langle 1, -1, l, \mu_l \vert l', \mu_{l'} \rangle
+\langle 1, 1, l, \mu_l \vert l', \mu_{l'} \rangle)
\quad ]
\nonumber \\ &+& 
i\sqrt{{2 \pi \over 3}} 
\langle \mu_n' \vert \sigma_{y} \vert \mu_n \rangle
\quad [ \quad
k''( Y_{1-1}(\hat{\mathbf{k}''}) + Y_{11}(\hat{\mathbf{k}''}) )\delta_{l'l}\delta_{\mu_{l'}\mu_l}
\nonumber \\ &+& 
k \sqrt{{(2\cdot 1 +1) (2 \cdot l +1) \over 4 \pi (2\cdot l'+1)}}
\langle 1, 0, l, 0 \vert l', 0 \rangle
(-\langle 1, -1, l, \mu_l \vert l', \mu_{l'} \rangle
-\langle 1, 1, l, \mu_l \vert l', \mu_{l'} \rangle)
\quad ]
\nonumber \\ &+&
\sqrt{{4 \pi \over 3}} 
\langle \mu_n' \vert \sigma_{z} \vert \mu_n \rangle
\quad [ \quad 
k'' Y_{10}(\hat{\mathbf{k}''})\delta_{l'l} \delta_{\mu_{l'}\mu_l} 
\nonumber \\ &-&
k \sqrt{{(2\cdot 1 +1) (2 \cdot l +1) \over 4 \pi (2\cdot l'+1)}}
\langle 1, 0, l, 0 \vert l', 0 \rangle 
\langle 1, 0, l, \mu_l \vert l', \mu_{l'} \rangle
\quad ]\quad 
]\nonumber
\label{i.22}
\eeqa
The sum over $l'$ includes only a finite number of values.  It can
only be $0$ or $1$ for $l=0$ and $1$, $2$ or $3$ when $l=2$.  
This limits the sum on $l'$ to the first four partial waves,
$l'=0,1,2,3$.  The final expression of the pseudo wave function can be
written in the form:
\beq
\chi_{\tilde P_d,\mu_d}^{(c)}(\vec {k}'',\mu''_p,\mu'_n) 
= \sum _{ll'}I_{ll'}(k'') f_{ll'}(\hat k'',\mu''_p,\mu'_n) ,
\label{i.23}
\eeq
where $f_{ll'}(\hat k'',\mu''_p,\mu'_n)$ are angle-dependent coefficients 
and $I_{ll'}(k'')$ are the relevant scalar quantities which have the following form: 
\beq
I_{ll'}(k'') \equiv \int k^2 dk v_l (k,k'') u_{l'} (k) 
\label{i.24}
\eeq
or
\beq
I_{ll'}(k'') \int k'^3 dk v_l (k,k'') u_{l'}(k).
\label{i.25}
\eeq
The allowed combinations of $l$ and $l'$ pairs requires the 
following integrals $I_{ll'}(k'')$:
\beq
I_{00}(k'')\equiv \int v_0(k,k'') u_0 (k) k^2 dk 
\label{i.26}
\eeq
\beq
I_{10}(k'')\equiv \int v_1(k,k'') u_0 (k) k^3 dk 
\label{i.27}
\eeq
\beq
I_{12}(k'')\equiv \int v_1(k,k'') u_2 (k) k^3 dk 
\label{i.28}
\eeq
\beq
I_{22}(k'')\equiv \int v_2(k,k'') u_2 (k) k^2 dk 
\label{i.29}
\eeq
\beq
I_{32}(k'')\equiv \int v_3(k,k'') u_2 (k) k^3 dk 
\label{i.30}
\eeq

The total exchange current contribution is
\begin{eqnarray}
& \quad & \langle  \tilde{\mathbf{P}}_d', \mu_d',d 
\vert I^+_{ex}(0) \vert  \tilde{\mathbf{P}}_d , \mu_d ,d \rangle 
=  \sum \int d{\mathbf{k}'} J_i  
\\ & \times &
[  
\Psi^*_{\tilde{\mathbf{P}}'_d, \mu'_d} (\mathbf{k}', \mu'_p,  \mu_n)
[ \bar u_f(\tilde{\mathbf{p}}'_p,\mu'_p) 
\Gamma_p^\mu (q) \gamma^5 
%(\eta^0 \gamma^0 - \eta^1 \gamma^1) 
(- P \cdot \gamma/M_d)
u_f(\tilde{\mathbf{p}}_p,\mu_p) ]
\chi_{\tilde{\mathbf{P}}_d,\mu_d}^{(f)}(\mathbf{k}, \mu_p,  \mu_n) \nonumber \\
&+& 
\chi^{(f)*}_{\tilde{\mathbf{P}}'_d, \mu'_d}(\mathbf{k},  \mu'_p,  \mu_n) 
[ \bar u_f(\tilde{\mathbf{p}}'_p,\mu'_p)
%(\eta^0 \gamma^0 - \eta^1 \gamma^1)
(- P \cdot \gamma/M_d)
\gamma^5 \Gamma_p^\mu (q)u_f(\tilde p_p,\mu_p)] %  
\Psi_{\tilde{\mathbf{P}}_d,\mu_d}(\mathbf{k},  \mu_p,  \mu_n)
] \nonumber  \\
&+&
[ p \leftrightarrow n ] \nonumber
\label{i.31}
\end{eqnarray}
where $J_i$ is the jacobian 
\beq 
J_i=\vert \frac {\partial (\tilde{\mathbf{p}}''_p, \tilde{\mathbf{p}}''_n)} 
{\partial (\tilde{\mathbf{P}}'',\mathbf{k}'')}\vert^{\frac{1}{2}}  
\vert \frac {\partial (\tilde{\mathbf{P}}',\mathbf{k}')}{\partial 
(\tilde{\mathbf{p}}'_p, \tilde{\mathbf{p}}'_n)} \vert^{\frac{1}{2}}
\label{i.32}
\eeq

%%%%%%%%%%%%%%%%%%%%%%%%%%%%%%%%%%%%%%%%%%%%%%%%%%%%%%%%%%%%%%%%%%%%%%%%%%%%%%

\bibliography{deut_paper_6x}

\clearpage
%%%%%%%%%%%%%%%%%%%%%%%%%%%%%%%%%%%%%%%%%%%%%%%%%%%%%%

%\setlength{\captionmargin}{23pt}
\begin{table}%[t]
\begin{center}
\begin {tabular}{rrrrr}
\hline
\hline
$\quad$  & IM & IM+Exchange  & IM(WSS) & IM+MEC(WSS) \\
\hline
$Q_d$  & 0.2698 & 0.2752   &  0.270 &  0.275 \\
\hline
$\mu_d $  & 0.8535 & 0.8596  & 0.847 & 0.871 \\
\hline
\hline
\end{tabular}%}
\caption
%[deuteron magnetic and quadrupole moments evaluated in the impulse approximation and including the ``pair'' current contribution constructed with(WS) and without(NS) the residual short-range potential removed. ]
{deuteron magnetic and quadrupole moments evaluated in the impulse
  approximation and including the exchange current contribution. 
The values are the same using all six different
  nucleon form factor parameterizations and three combinations of
  independent current matrix elements. Argonne $V18$ potential is used
  in the calculation.  The values labeled with WSS are
  from~\cite{Wiringa95}. The experimental values are $0.2860\pm
  0.0015 fm^2$ ~\cite{Bishop79} and $0.857406\pm 0.000001 \mu_N$
  ~\cite{Lindg65}.}
\label{momts}
\end{center}
\end{table}

\clearpage
%%%%%%%%%%%%%%%%%%%%%%%%%%%%%%%%%%%%%%%%%%%%%%%%%%%%%%

\clearpage
%%%%%%%%%%%%%%%%%%%%%%%%%%%%%%%%%%%%%%%%%%%%%%%%%%%%%%

\begin{figure}
\begin{center}
\rotatebox{270}{
\includegraphics[width=13cm]{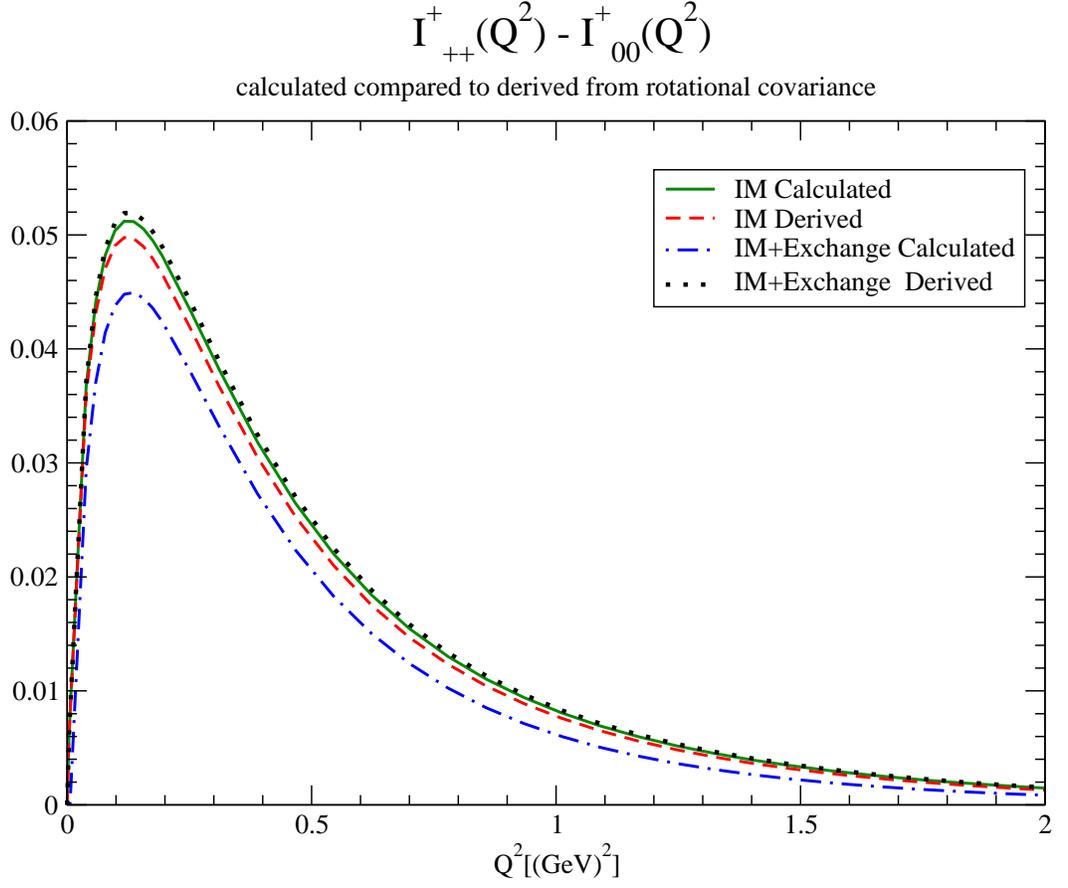}
}
\end{center}
\caption{\Large(color online) $I^{+}_{++} (0) - I^+_{00} (0)$ calculated 
directly compared to $I^{+}_{++} (0) - I^+_{00} (0)$ calculated by 
rotational convariance with and without exchange currrent I 
\label{fig1}
}
\end{figure}

\clearpage

\begin{figure}
\begin{center}
\rotatebox{270}{
\includegraphics[width=13cm]{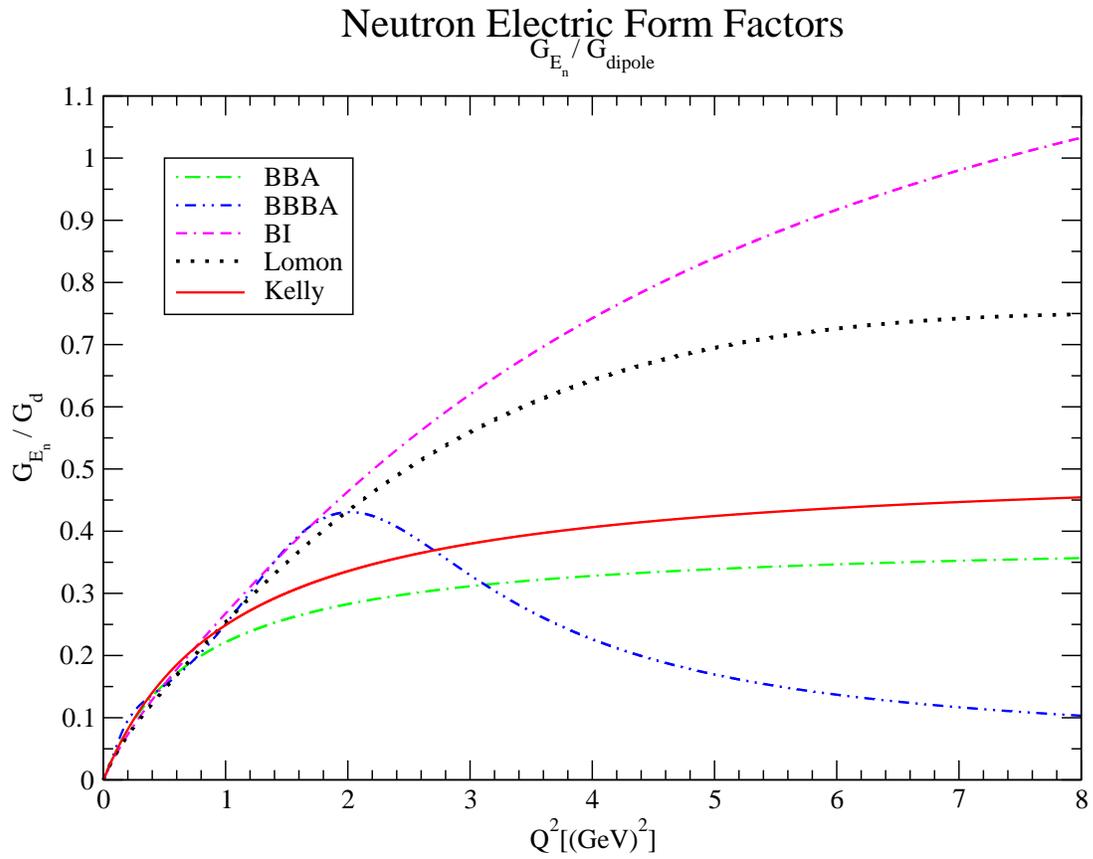}
}
\end{center}
\caption{\Large(color online) Neutron electric form factor parameterizations 
\label{fig2}
}
\end{figure}

\clearpage

\begin{figure}
\begin{center}
\rotatebox{270}{
\includegraphics[width=13cm]{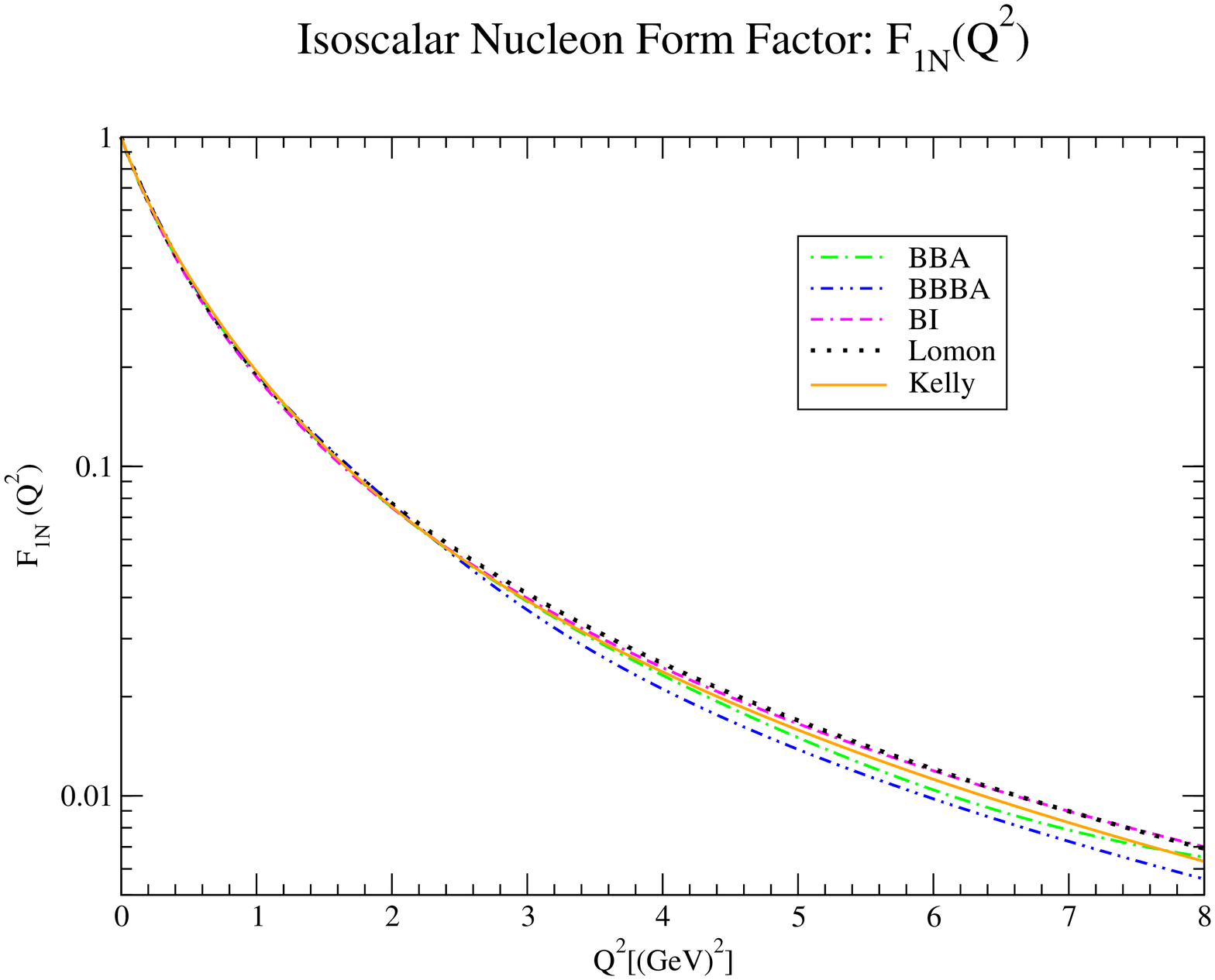}
}
\end{center}
\caption{\Large(color online) Isoscalar nucleon form factor $F_{1N}(Q^2)$ 
\label{fig3}
}
\end{figure}

\clearpage

\begin{figure}
\begin{center}
\rotatebox{270}{
\includegraphics[width=13cm]{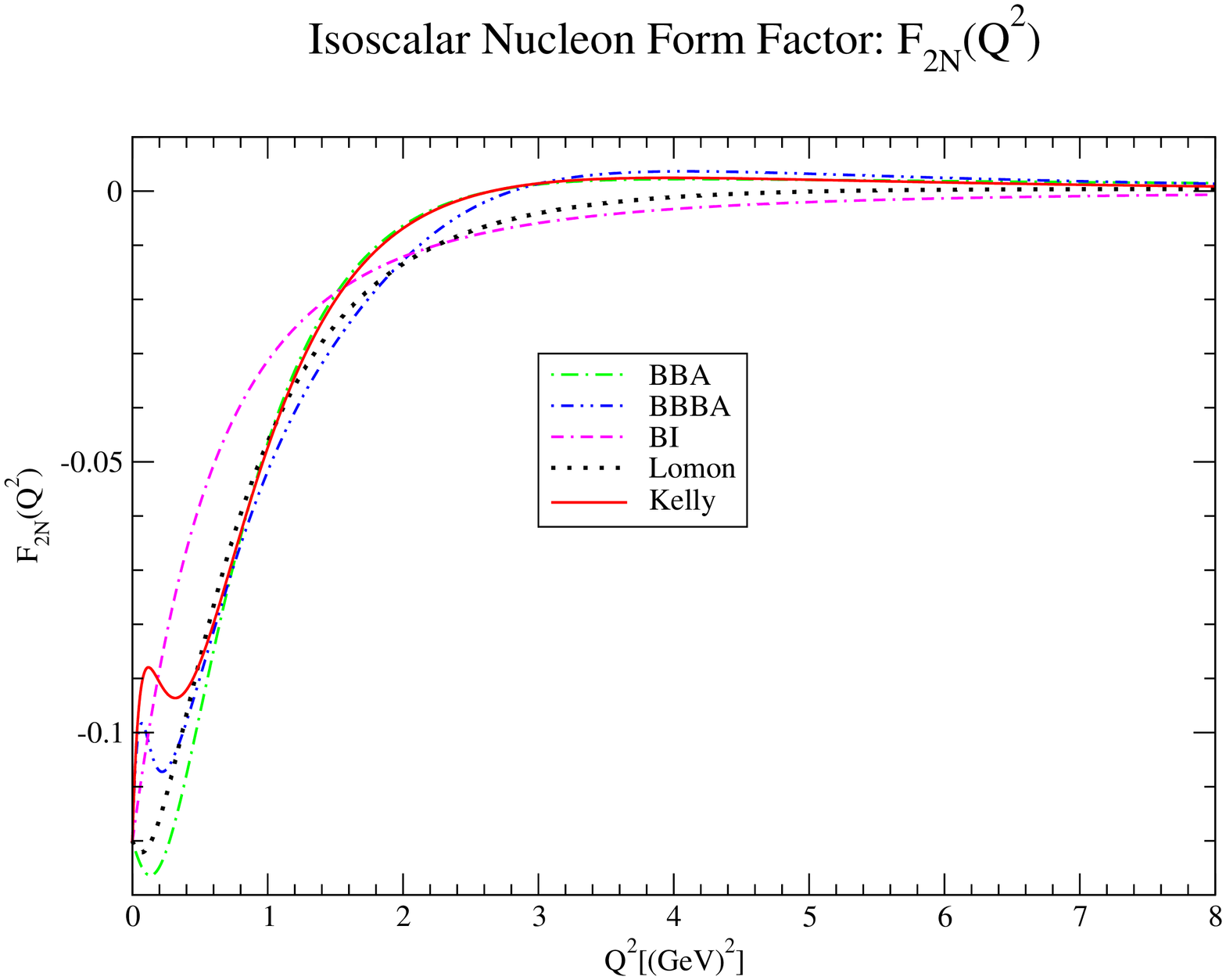}
}
\end{center}
\caption{\Large(color online) Isoscalar nucleon form factor $F_{2N}(Q^2)$ 
\label{fig4}
}
\end{figure}

\clearpage

\begin{figure}
\begin{center}
\rotatebox{270}{
\includegraphics[width=13cm]{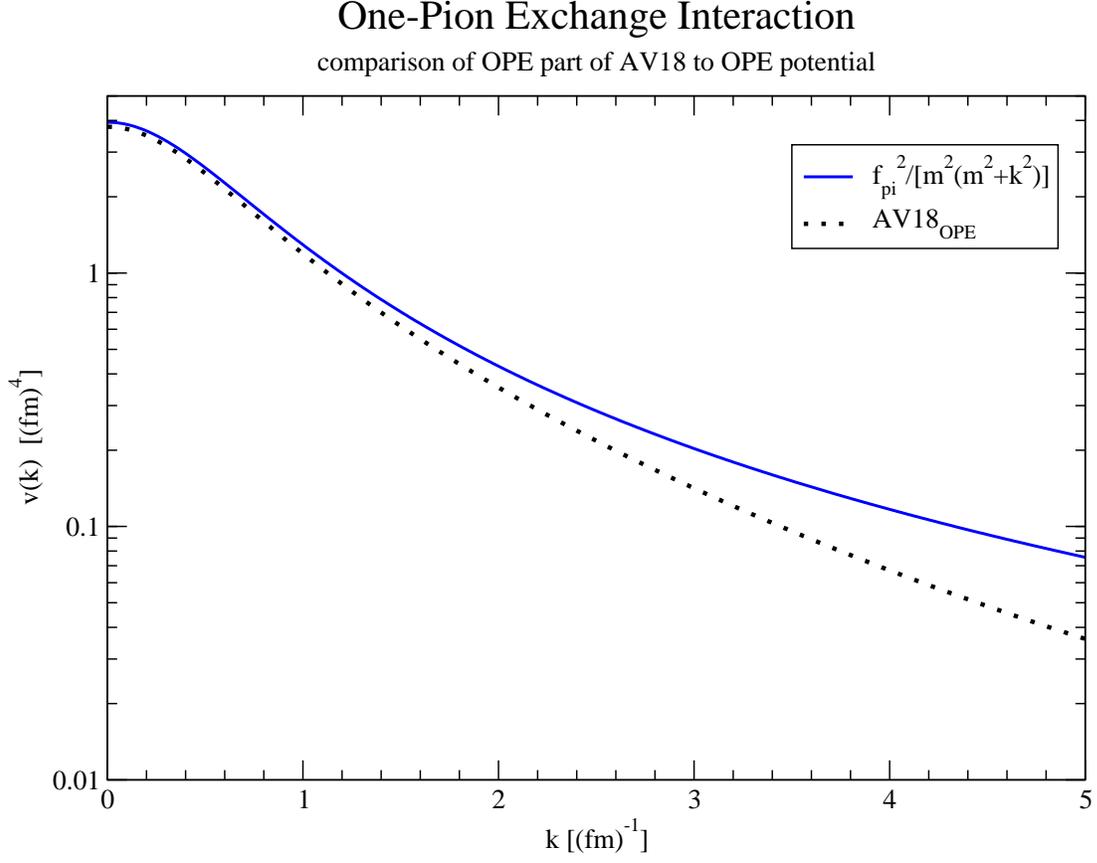}
}
\end{center}
\caption{\Large(color online) One pion-exchange interaction compared to 
one-pion exchange part of AV18 interaction 
\label{fig5}
}
\end{figure}

\clearpage

\begin{figure}
\begin{center}
\rotatebox{270}{
\includegraphics[width=13cm]{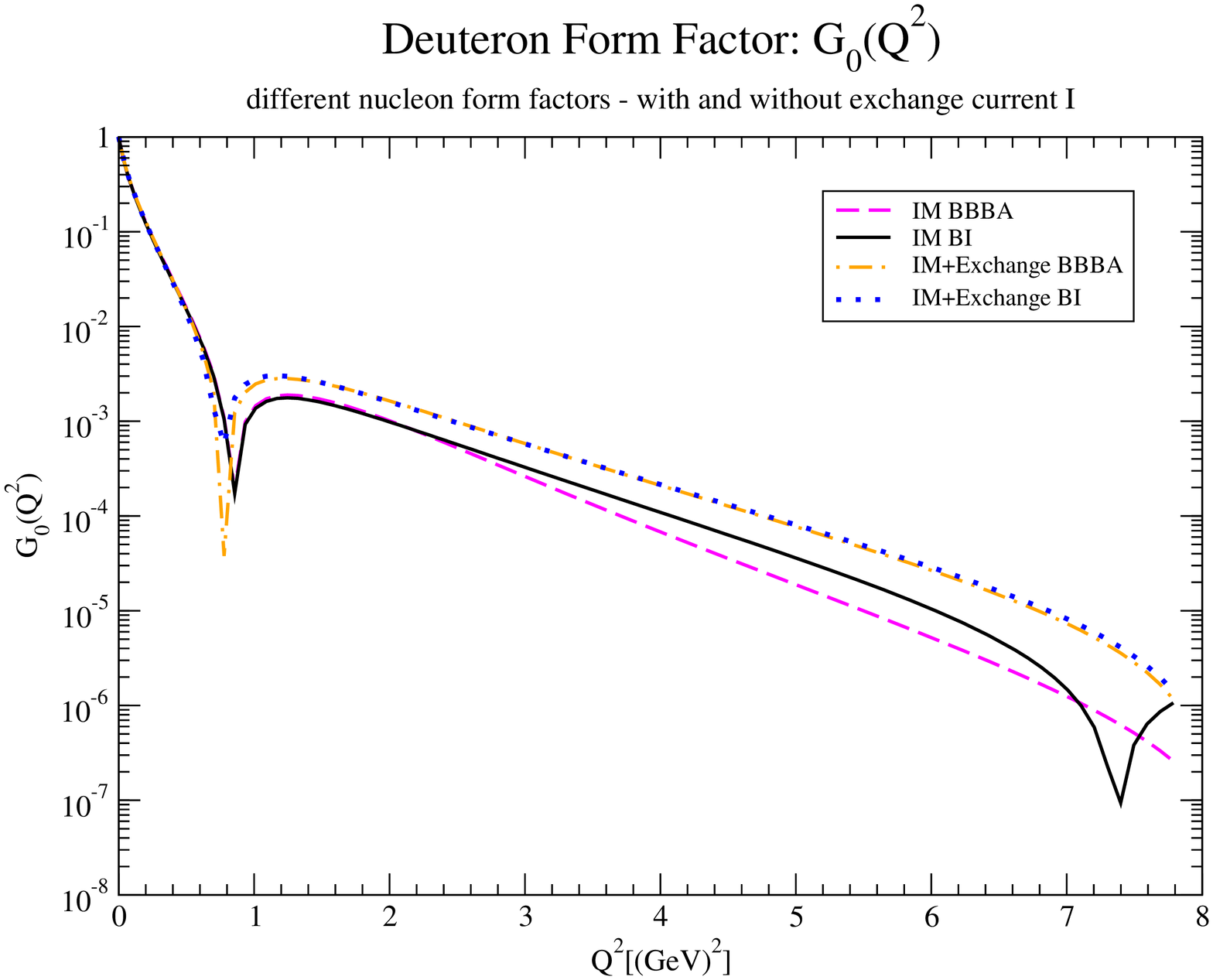}
}
\end{center}
\caption{\Large(color online) Deuteron form factor $G_0(Q^2)$; different 
nucleon form factors with and 
without exchange current I 
\label{fig6}
}
\end{figure}

\clearpage

\begin{figure}
\begin{center}
\rotatebox{270}{
\includegraphics[width=13cm]{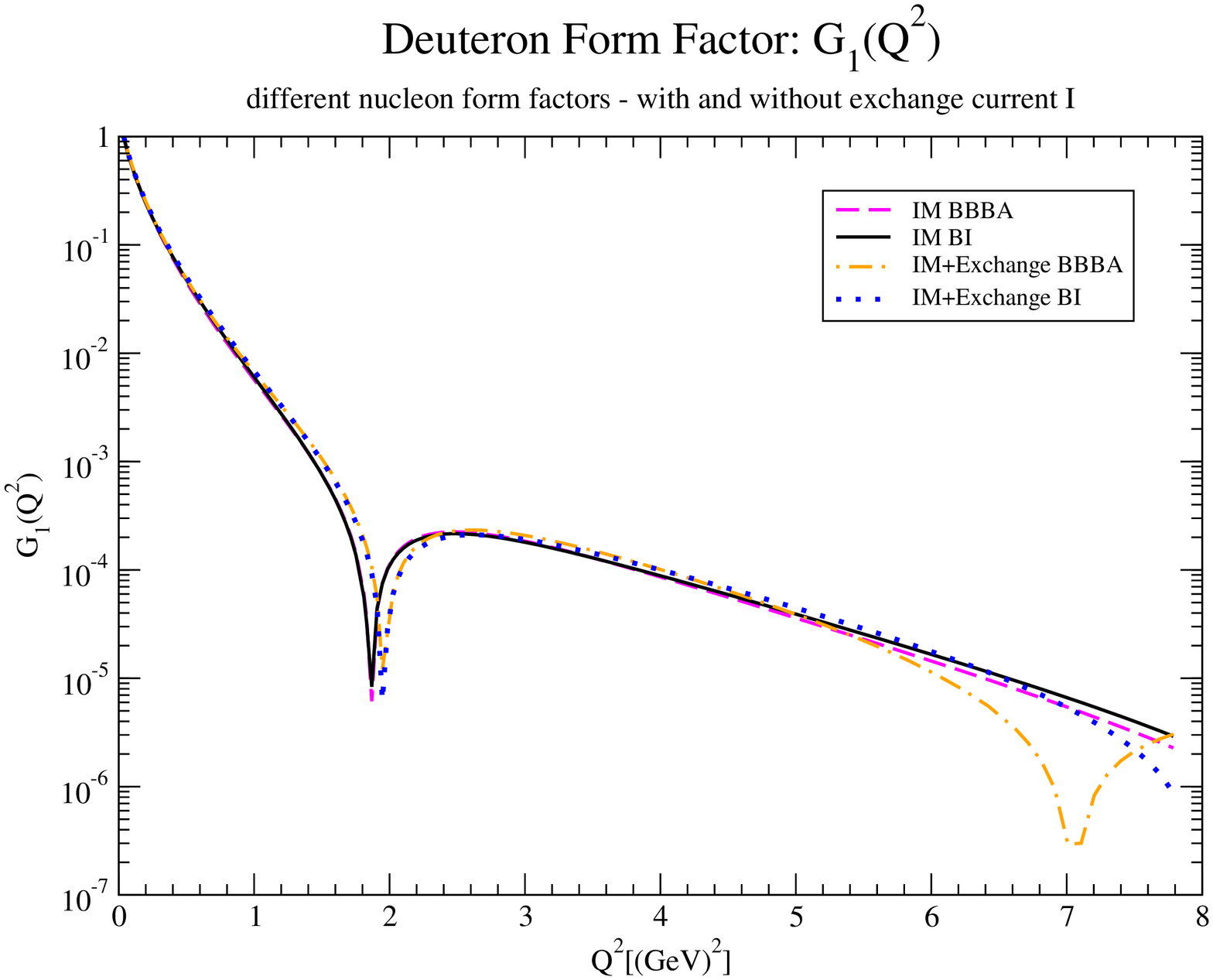}
}
\end{center}
\caption{\Large(color online) Deuteron form factor $G_1(Q^2)$; ; different 
nucleon form factors with and 
without exchange current I 
\label{fig7}
}
\end{figure}

\clearpage

\begin{figure}
\begin{center}
\rotatebox{270}{
\includegraphics[width=13cm]{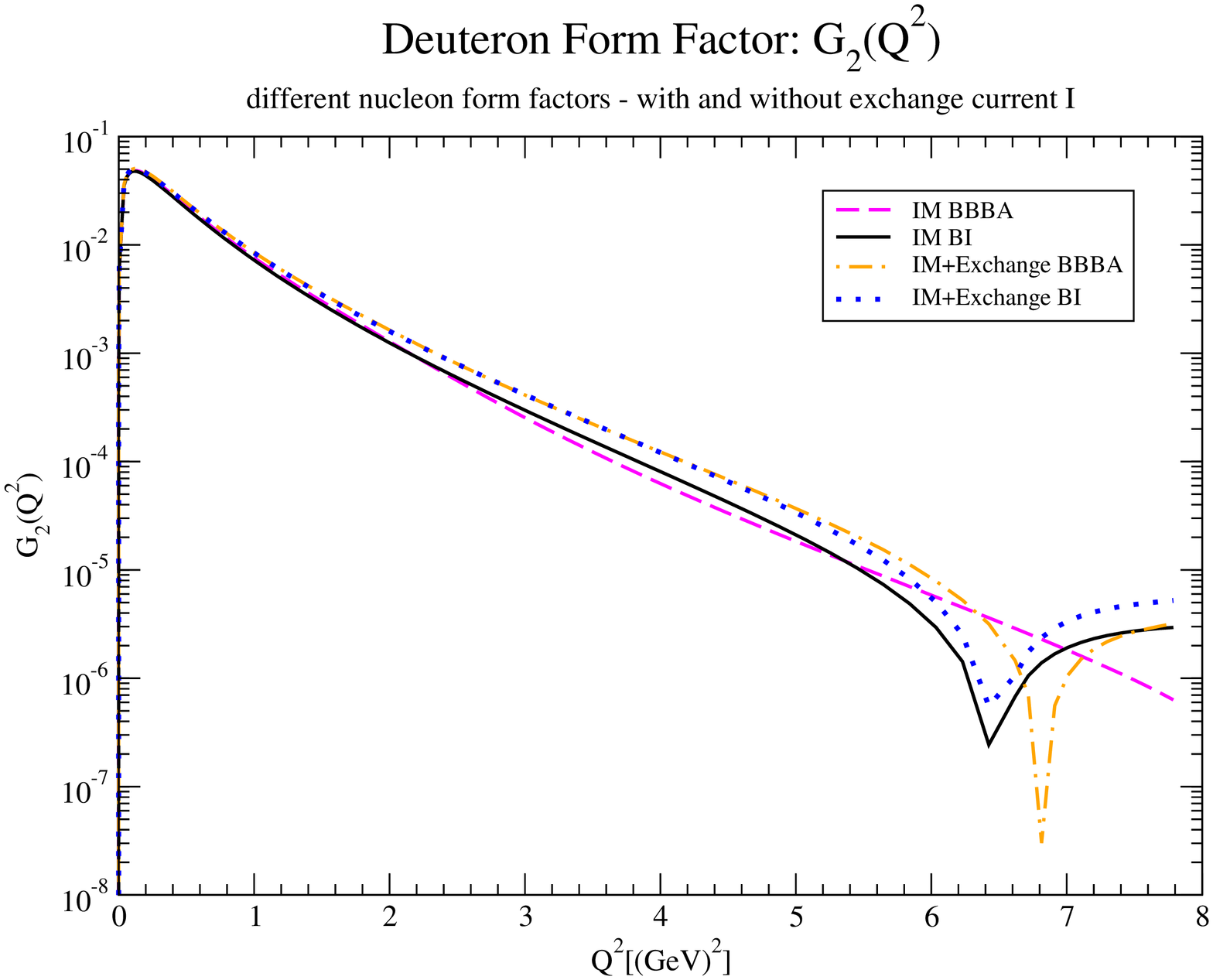}
}
\end{center}
\caption{\Large(color online) Deuteron form factor $G_2(Q^2)$; ; different 
nucleon form factors with and 
without exchange current I 
\label{fig8}
}
\end{figure}

\clearpage

\begin{figure}
\begin{center}
\rotatebox{270}{
\includegraphics[width=13cm]{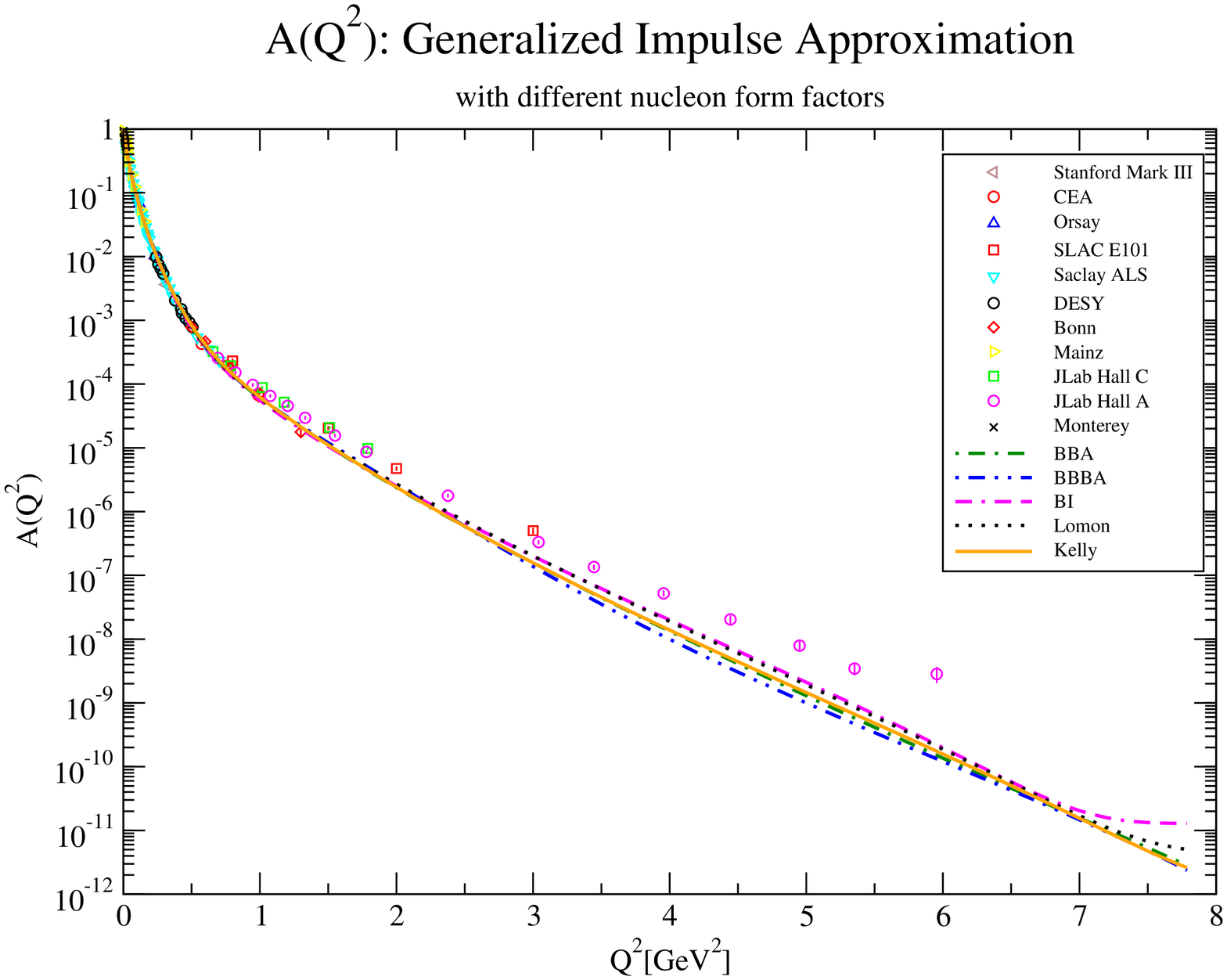}
}
\end{center}
\caption{\Large(color online) $A(Q^2)$: Generalized impulse approximation 
with different nucleon form factors
\label{fig9}
}
\end{figure}

\clearpage

\begin{figure}
\begin{center}
\rotatebox{270}{
\includegraphics[width=13cm]{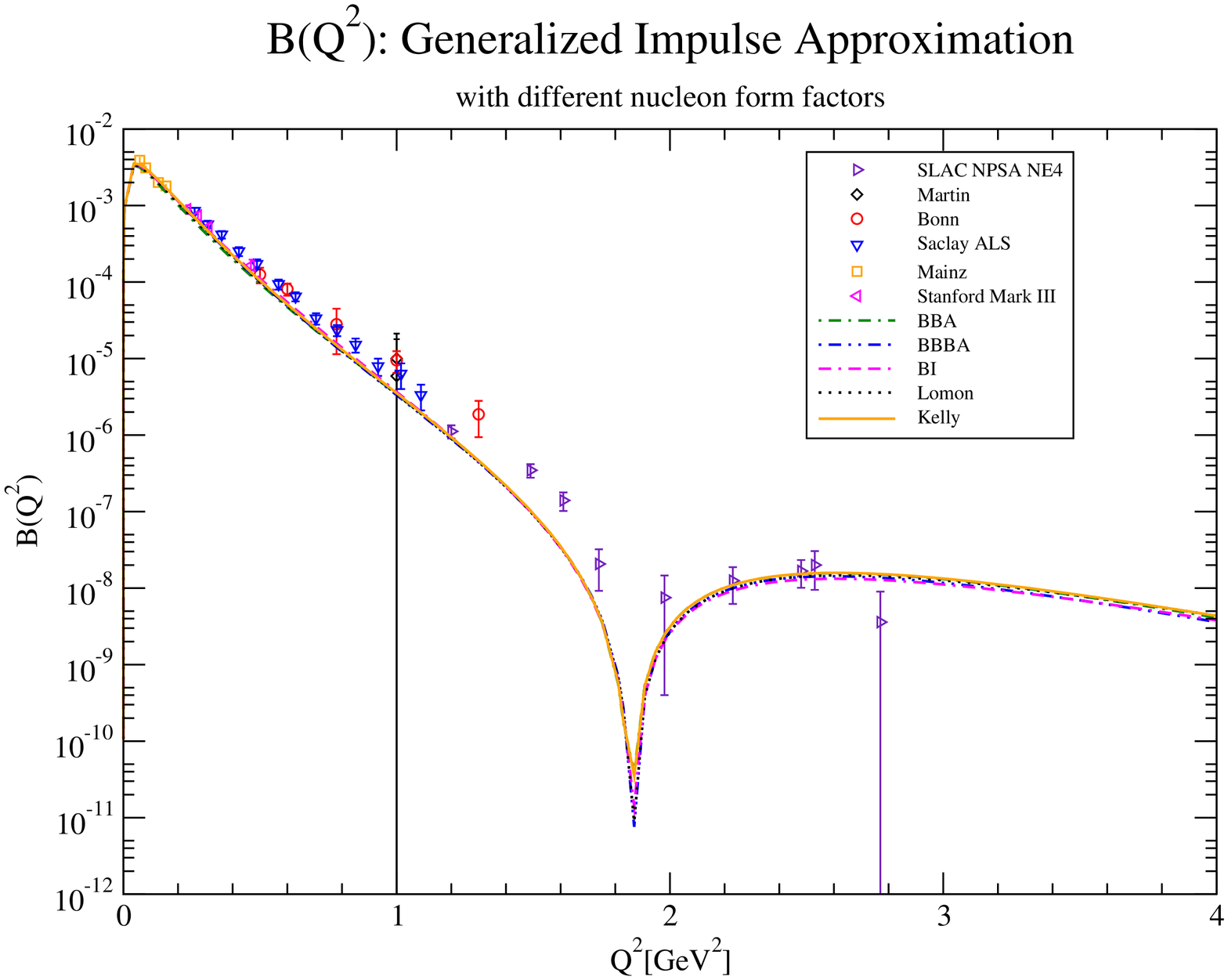}
}
\end{center}
\caption{\Large(color online) $B(Q^2)$: Generalized impulse approximation  
with different nucleon form factors
\label{fig10}
}
\end{figure}

\clearpage

\begin{figure}
\begin{center}
\rotatebox{270}{
\includegraphics[width=13cm]{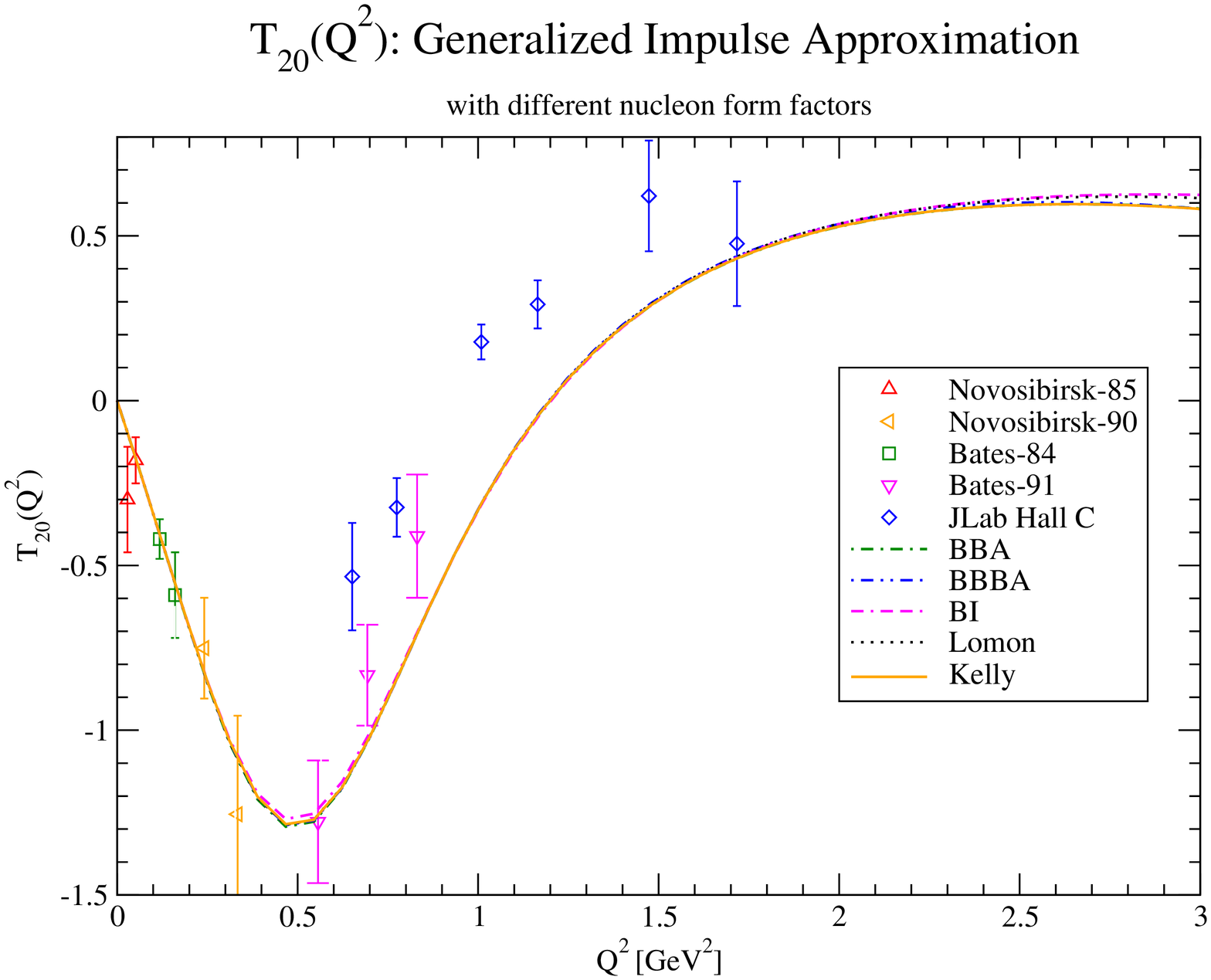}
}
\end{center}
\caption{\Large(color online)$T_{20}(Q^2, 70^o)$: Generalized impulse 
approximation with different nucleon form factors 
\label{fig11}
}
\end{figure}

\clearpage

\begin{figure}
\begin{center}
\rotatebox{270}{
\includegraphics[width=13cm]{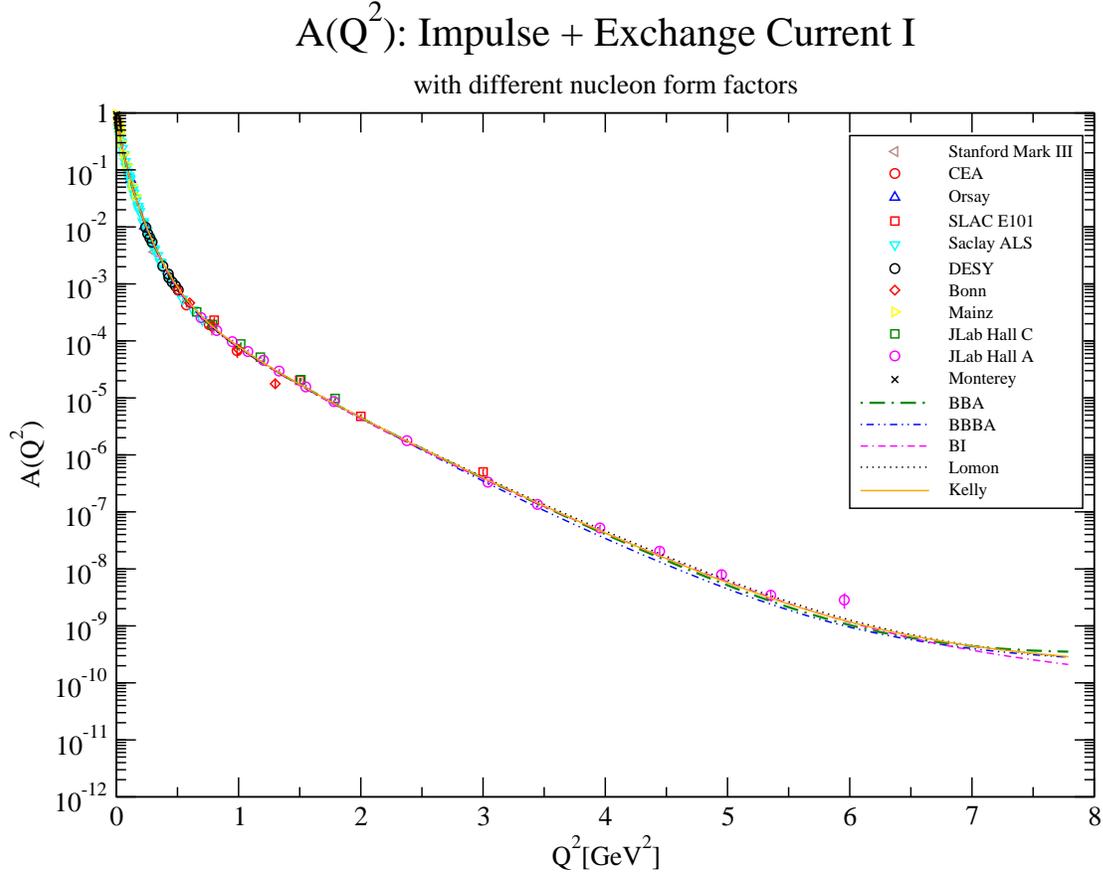}
}
\end{center}
\caption{\Large(color online) $A(Q^2)$: Generalized impulse approximation plus exchange current I
\label{fig12}
}
\end{figure}

\clearpage

\begin{figure}
\begin{center}
\rotatebox{270}{
\includegraphics[width=13cm]{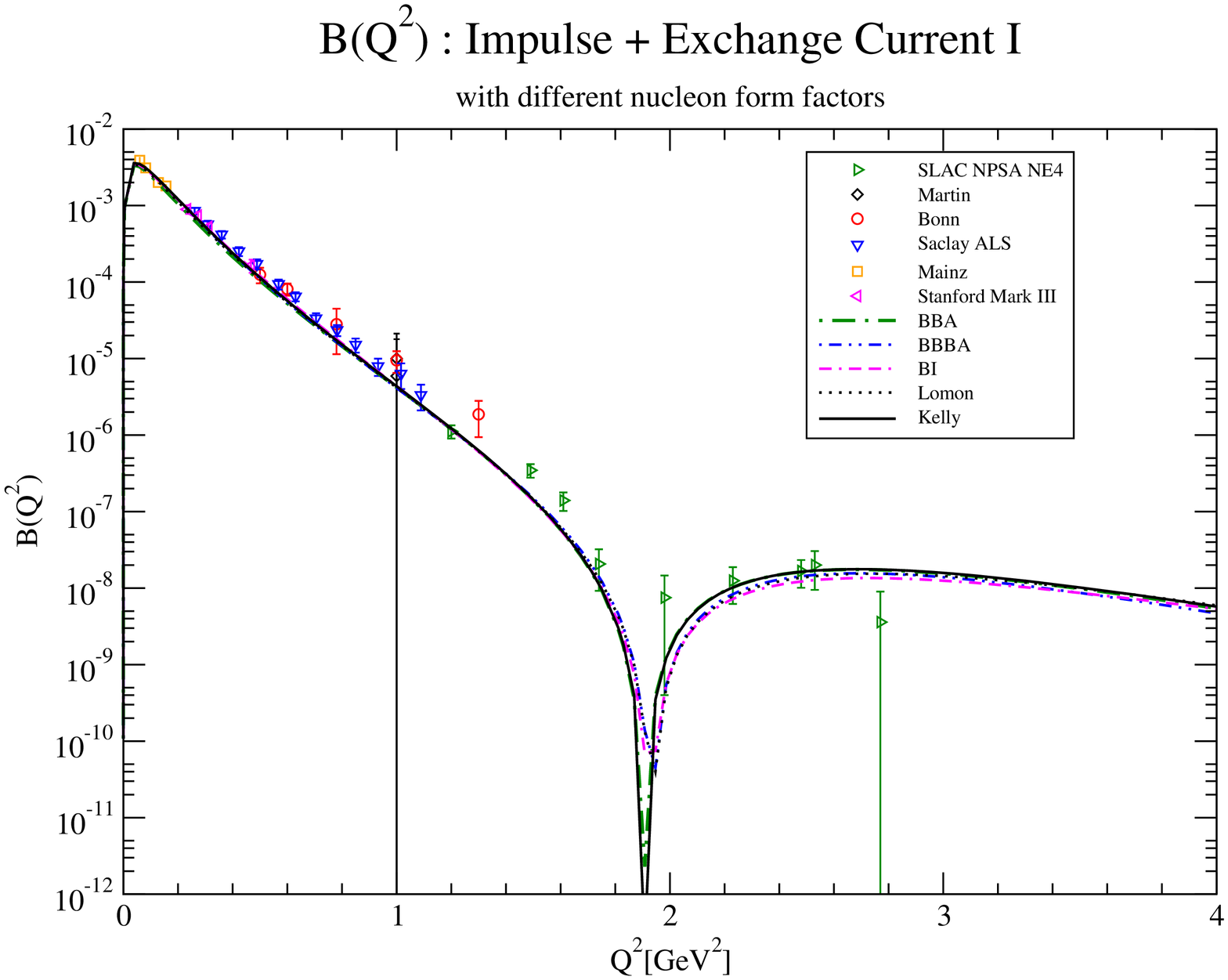}
}
\end{center}
\caption{\Large(color online) $B(Q^2)$: Generalized impulse approximation plus exchange current I with differnet nucleon form factors
\label{fig13}
}
\end{figure}

\clearpage

\begin{figure}
\begin{center}
\rotatebox{270}{
\includegraphics[width=13cm]{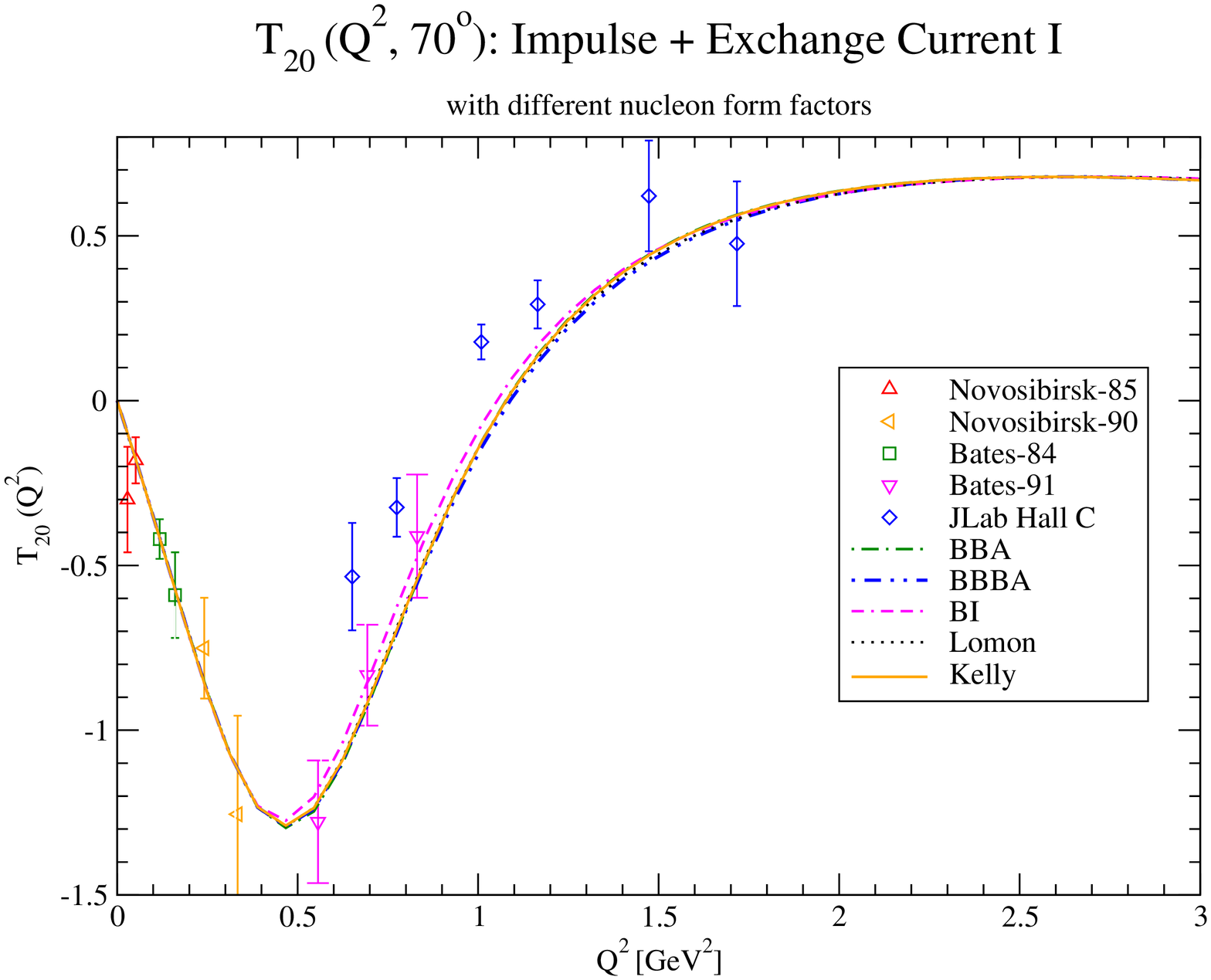}
}
\end{center}
\caption{\Large(color online) $T_{20}(Q^2,70^o)$: Generalized impulse approximation plus exchange current I with different nucleon form factors
\label{fig14}
}
\end{figure}

\clearpage

\begin{figure}
\begin{center}
\rotatebox{270}{
\includegraphics[width=13cm]{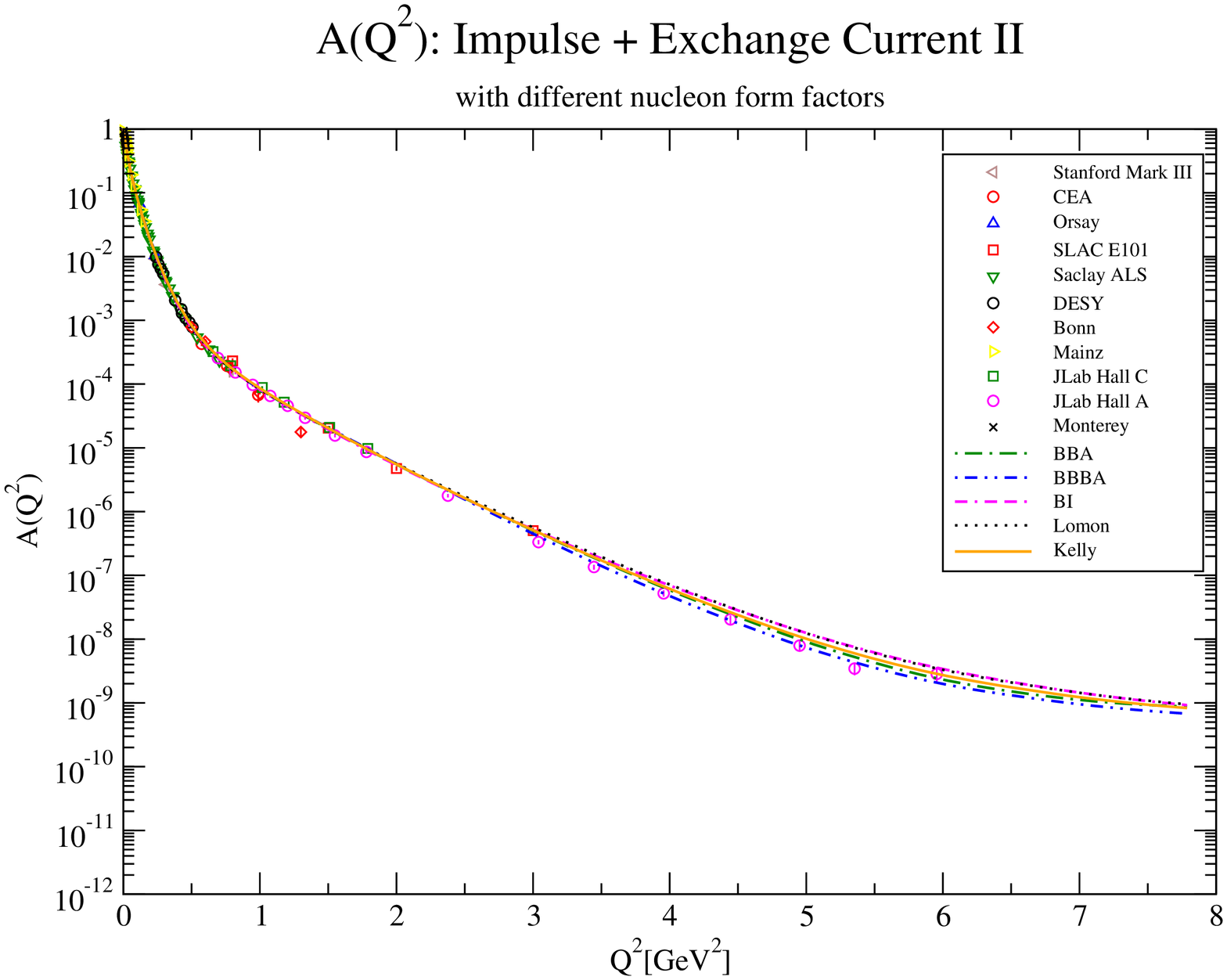}
}
\end{center}
\caption{\Large(color online) $A(Q^2)$: Generalized impulse approximation 
plus exchange current II with different nucleon form factors
\label{fig15}
}
\end{figure}

\clearpage

\begin{figure}
\begin{center}
\rotatebox{270}{
\includegraphics[width=13cm]{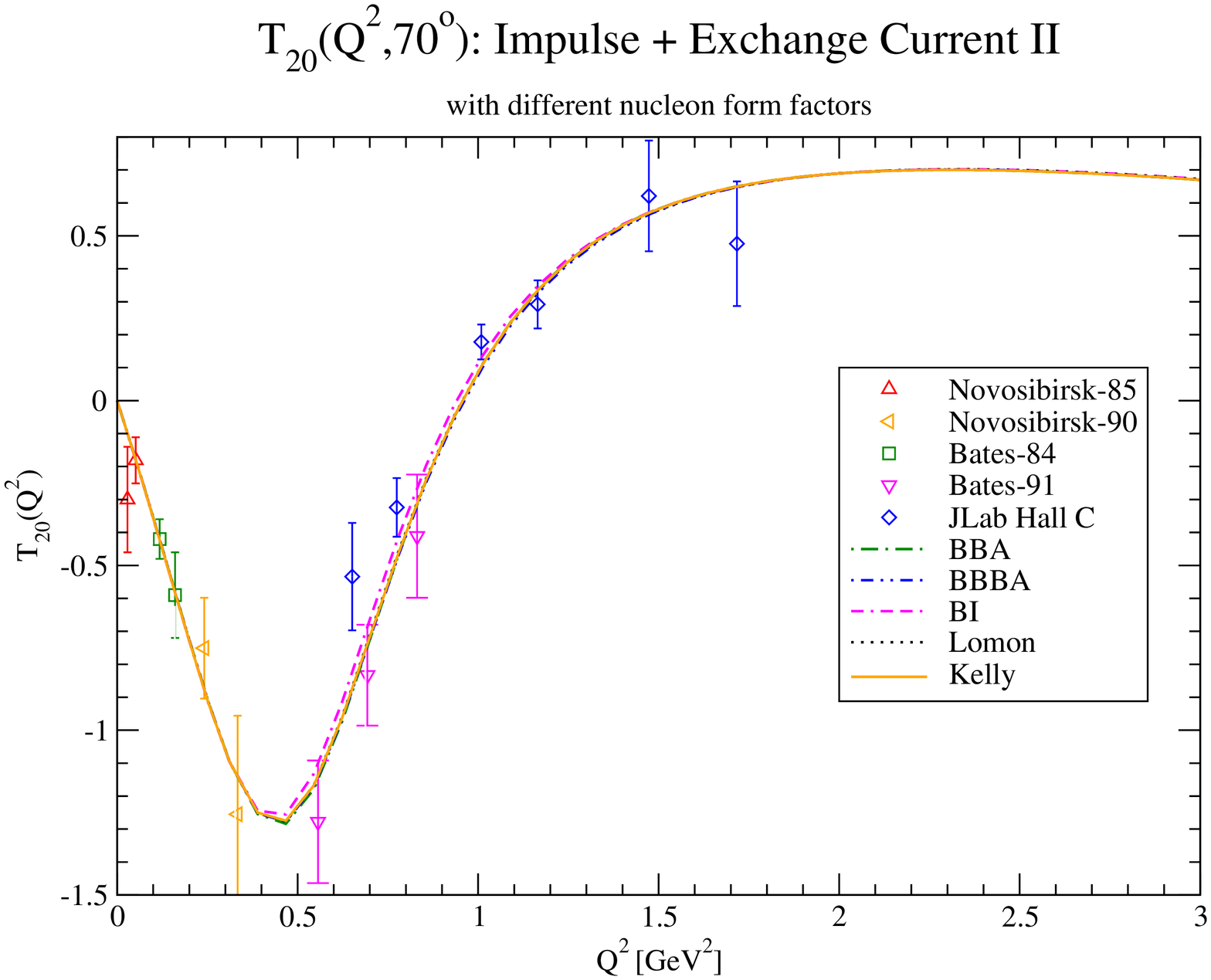}
}
\end{center}
\caption{\Large(color online) $T_{20}(Q^2,70^o)$: Generalized impulse approximation plus exchange current II with different nucleon form factors 
\label{fig16}
}
\end{figure}

\clearpage

\begin{figure}
\begin{center}
\rotatebox{270}{
\includegraphics[width=13cm]{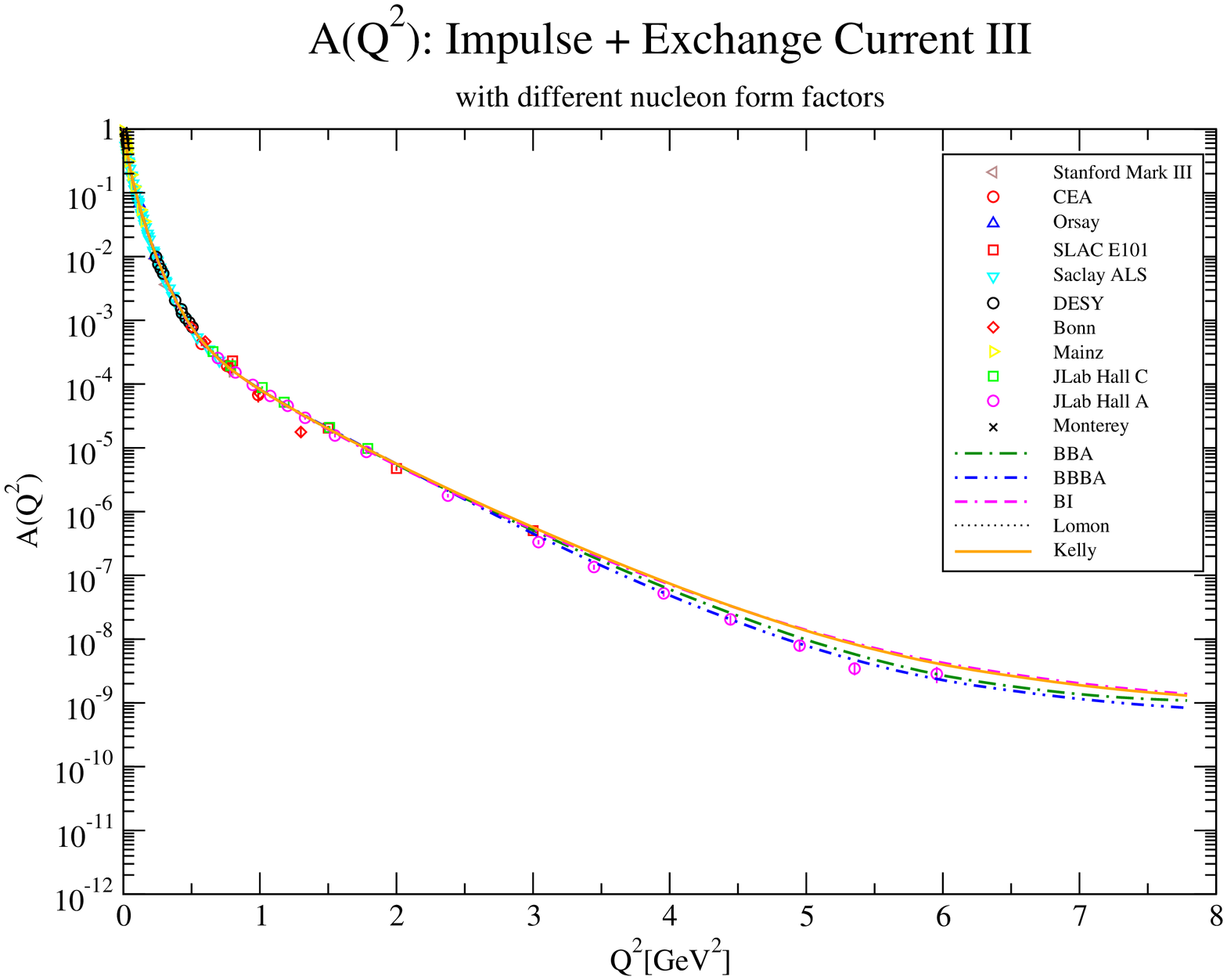}
}
\end{center}
\caption{\Large(color online) $A(Q^2)$: Generalized impulse approximation plus exchange current III with different nucleon form factors 
\label{fig17}
}
\end{figure}

\clearpage

\begin{figure}
\begin{center}
\rotatebox{270}{
\includegraphics[width=13cm]{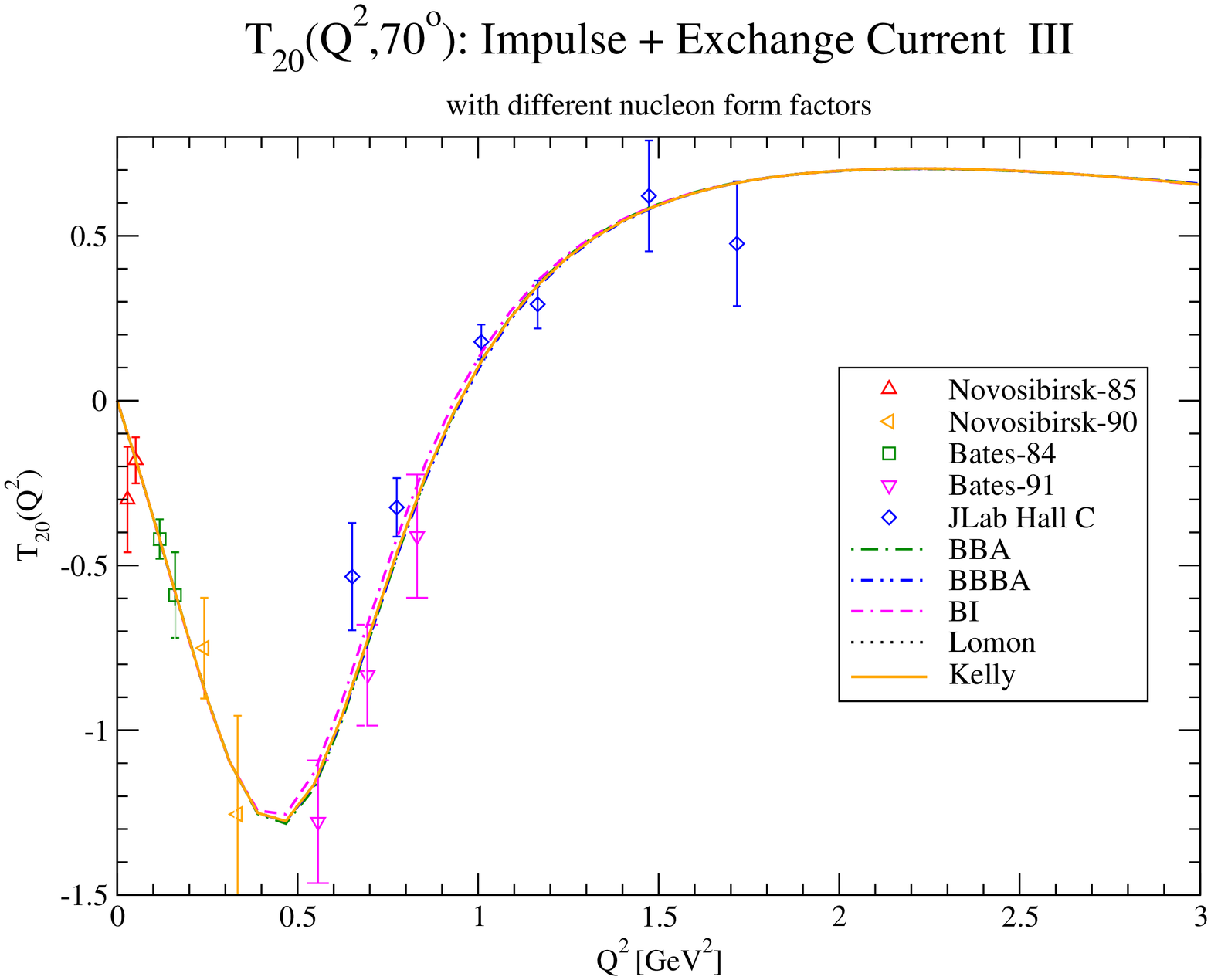}
}
\end{center}
\caption{\Large(color online) $T_{20}(Q^2,70^o)$: Generalized impulse approximation plus exchange current III   
\label{fig18}
}
\end{figure}

\clearpage

\begin{figure}
\begin{center}
\rotatebox{270}{
\includegraphics[width=13cm]{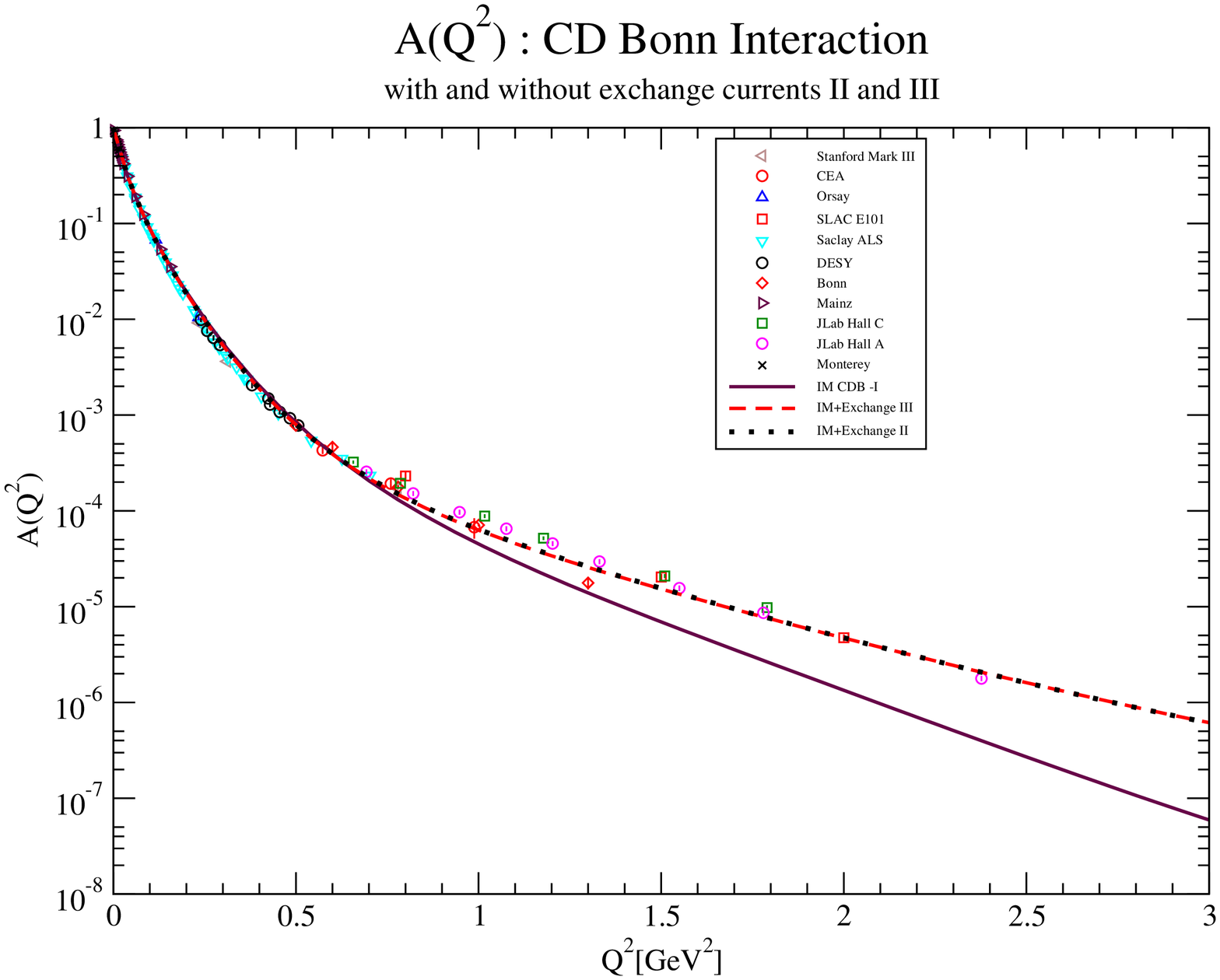}
}
\end{center}
\caption{\Large(color online) $A(Q^2)$: CD Bonn with and without exchange 
currents II and III
\label{fig19}
}
\end{figure}

\clearpage

\begin{figure}
\begin{center}
\rotatebox{270}{
\includegraphics[width=13cm]{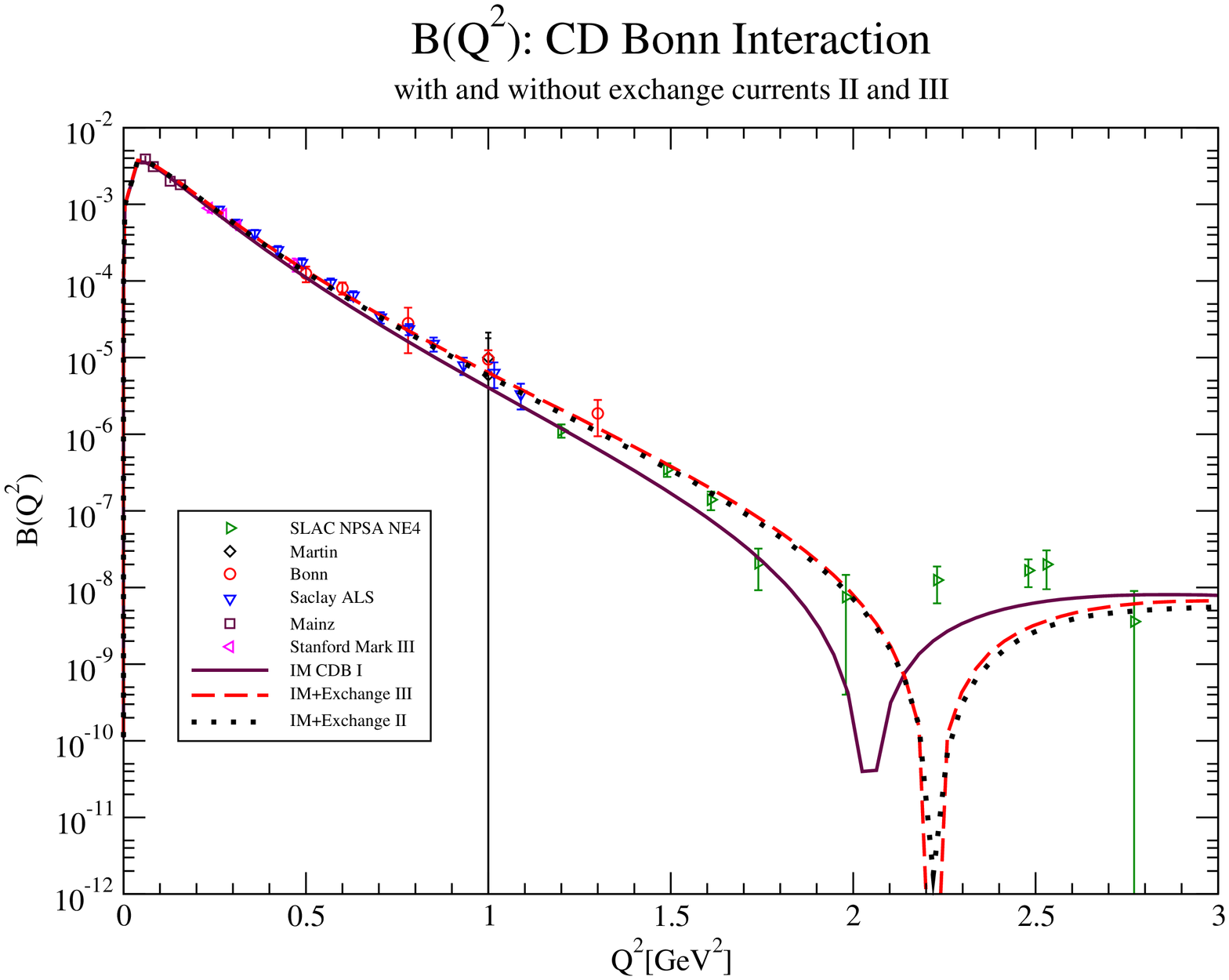}
}
\end{center}
\caption{\Large(color online) $B(Q^2)$: CD Bonn with and without 
exchange currents II and III 
\label{fig20}
}
\end{figure}

\clearpage

\begin{figure}
\begin{center}
\rotatebox{270}{
\includegraphics[width=13cm]{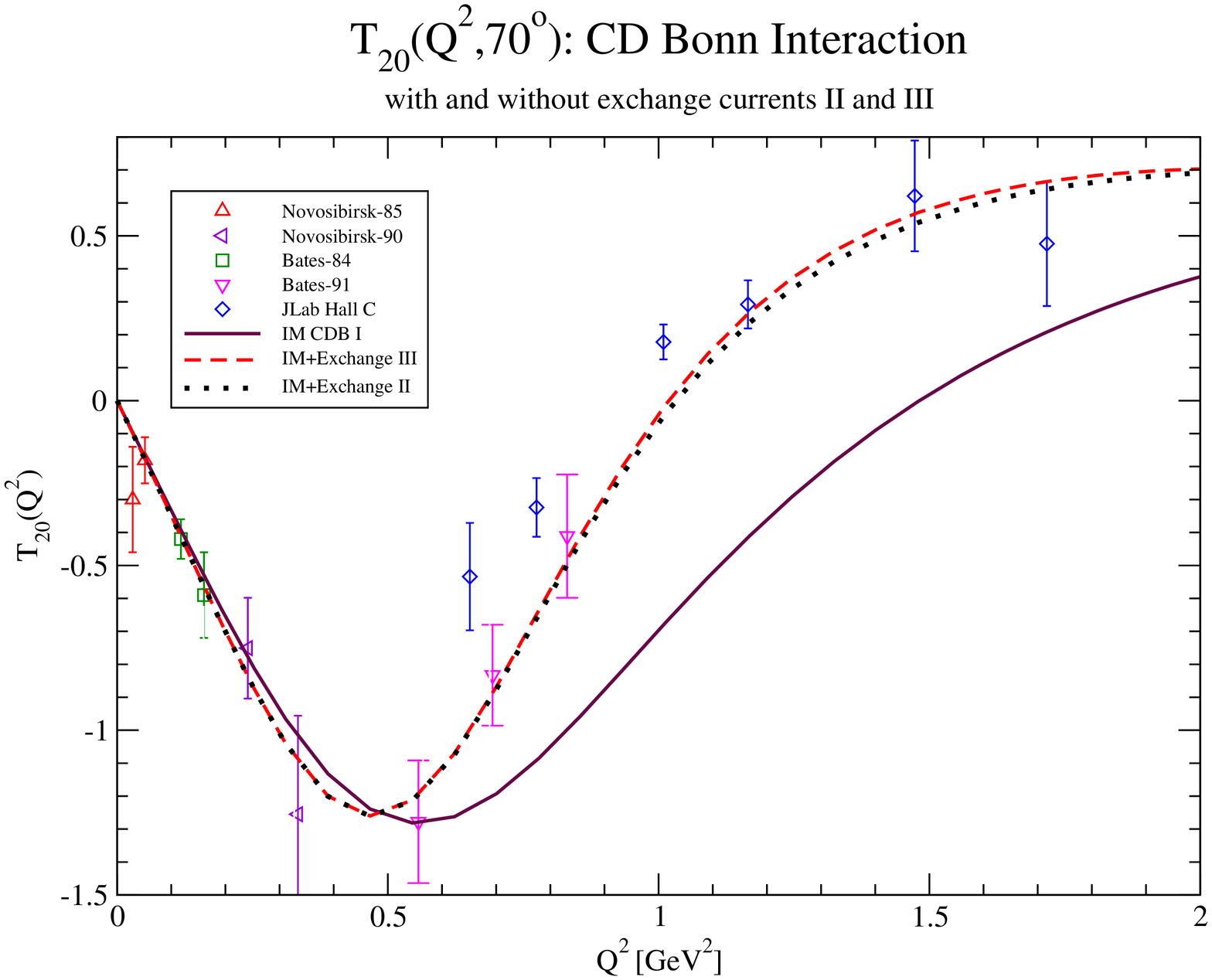}
}
\end{center}
\caption {\Large(color online) $T_{20}(Q^2,70^o)$: CD Bonn with and without 
exchange currents II and III
\label{fig21}
}
\end{figure}

\clearpage

\begin{figure}
\begin{center}
\rotatebox{270}{
\includegraphics[width=13cm]{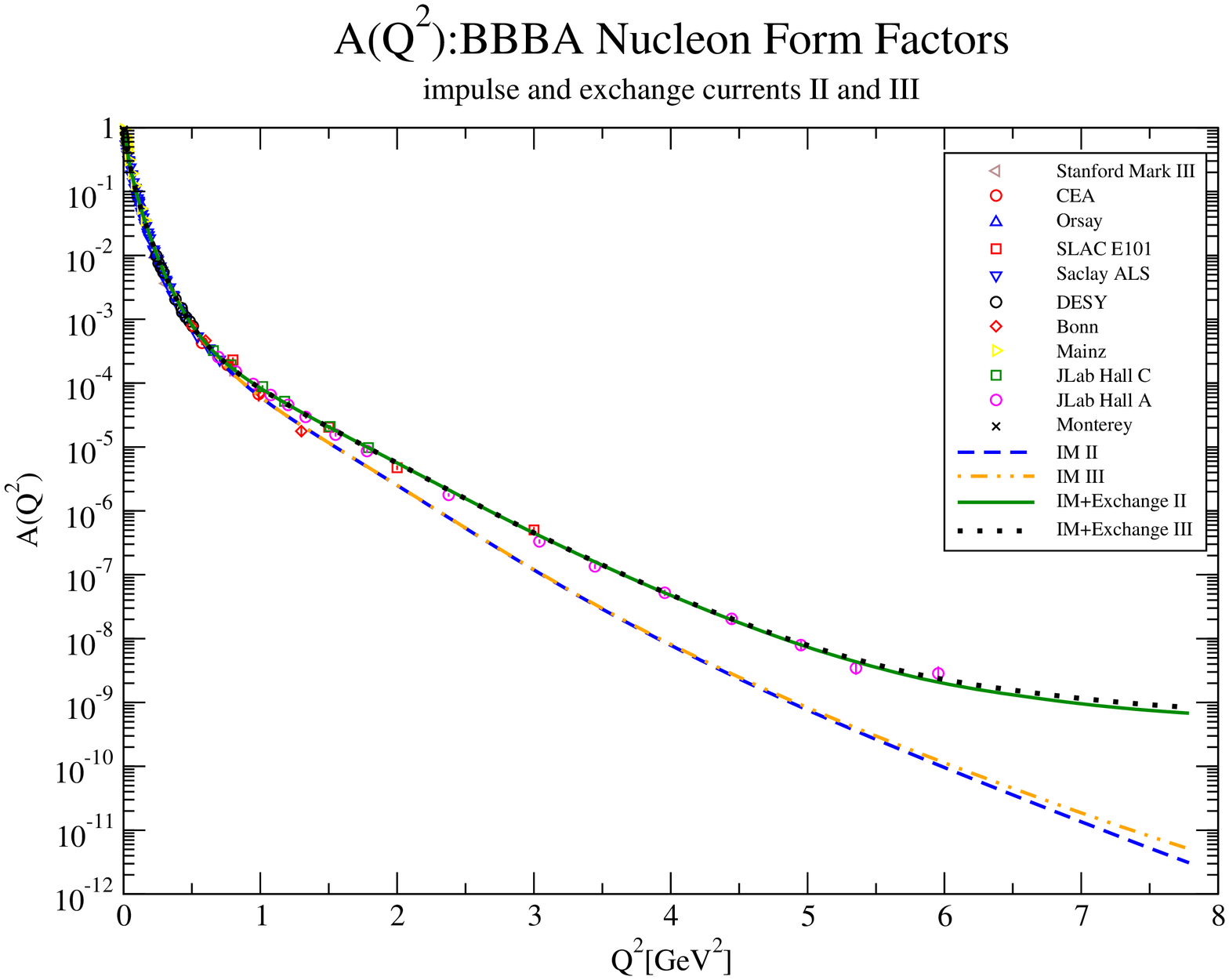}
}
\end{center}
\caption{\Large(color online) $A(Q^2)$, AV18, BBBA form factors 
with and without exchange current II and III 
\label{fig22}
}
\end{figure}

\clearpage

\begin{figure}
\begin{center}
\rotatebox{270}{
\includegraphics[width=13cm]{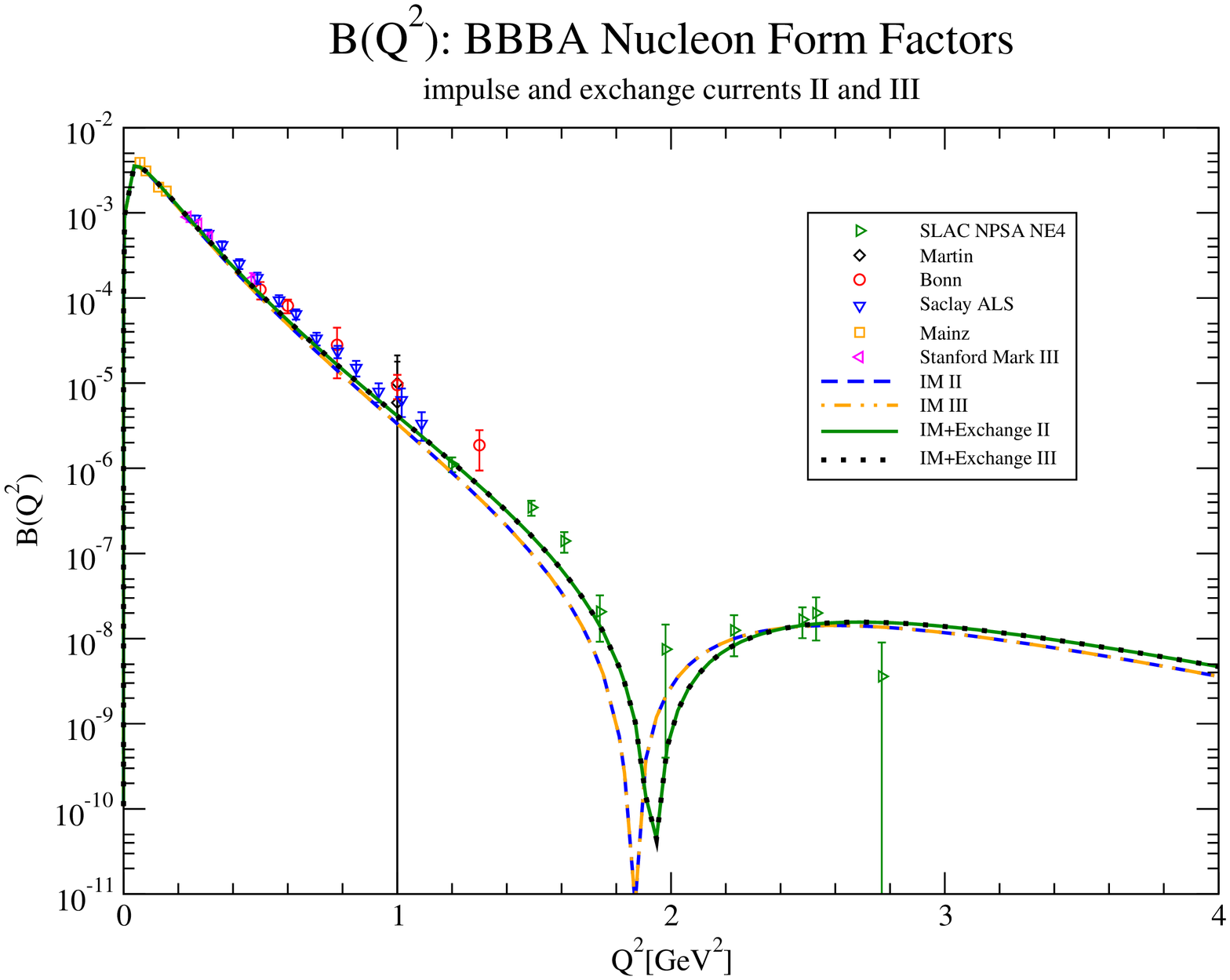}
}
\end{center}
\caption{\Large(color online) $B(Q^2)$ , AV18, BBBA form factors 
with and without exchange current II and III
\label{fig23}
}
\end{figure}

\clearpage

\begin{figure}
\begin{center}
\rotatebox{270}{
\includegraphics[width=13cm]{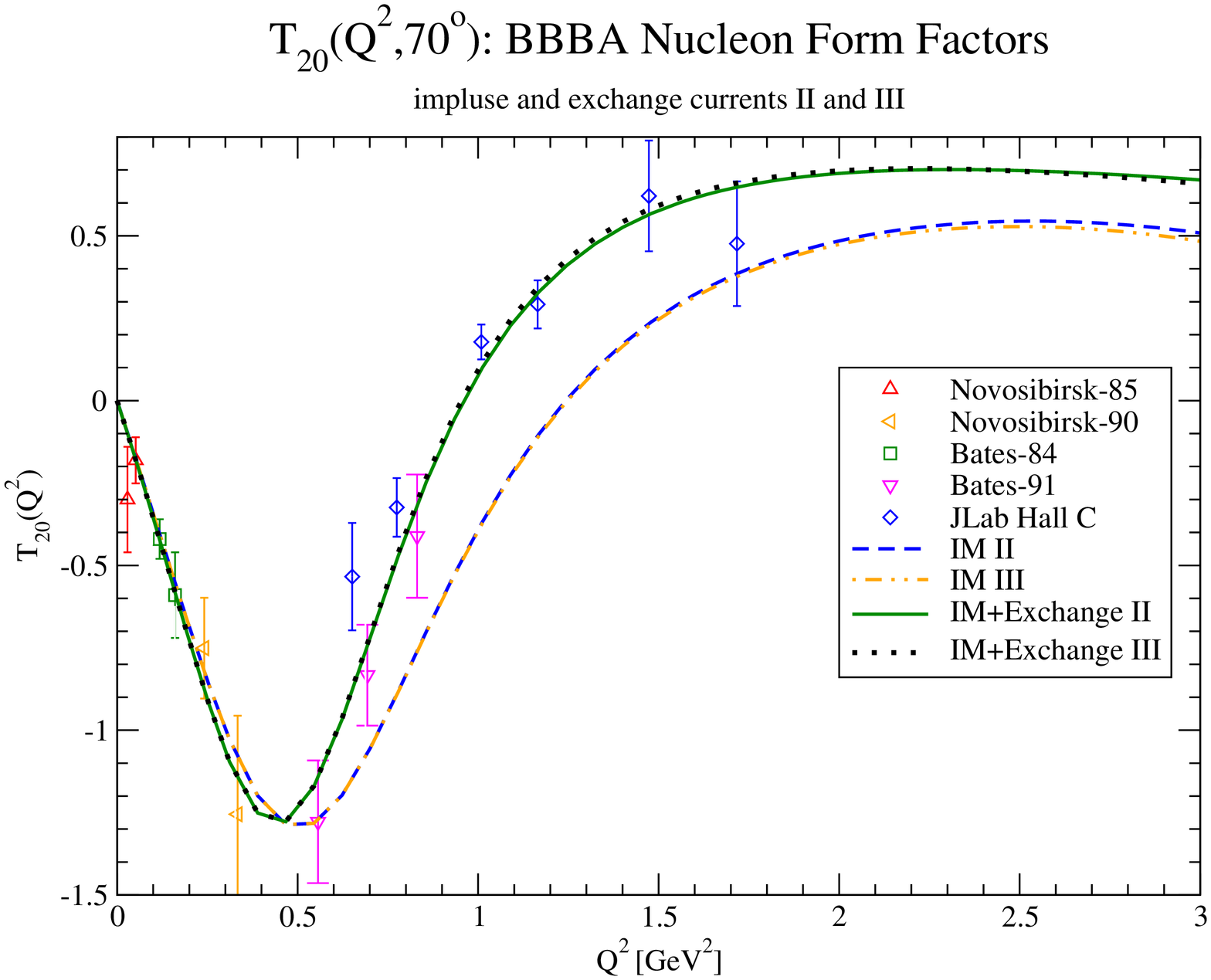}
}
\end{center}
\caption{\Large(color online) $T_{20}(Q^2,70^o)$ , AV18, BBBA 
form factors with and without exchange current II and III
\label{fig24}
}
\end{figure}

\clearpage

\begin{figure}
\begin{center}
\rotatebox{270}{
\includegraphics[width=13cm]{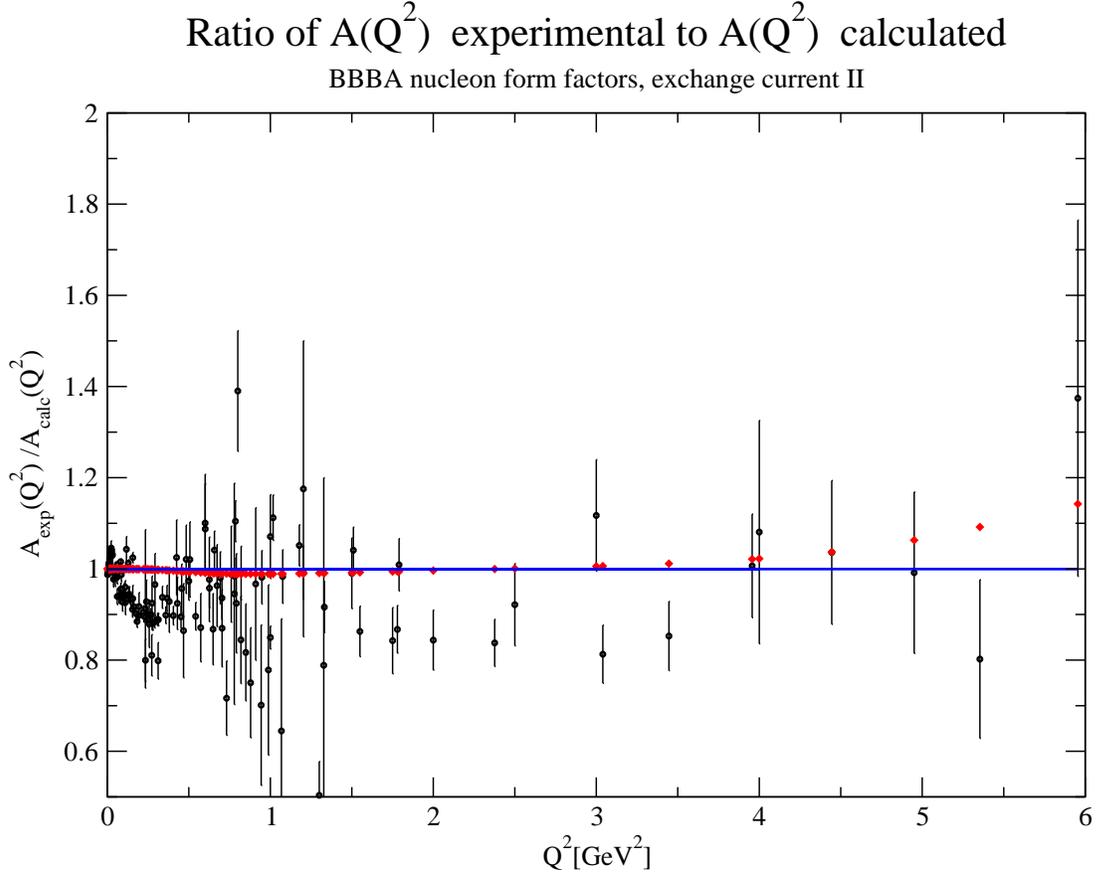}
}
\end{center}
\caption{\Large(color online) Ratio of experimental values of $A(Q^2)$ to 
calculated values of $A(Q^2)$ with exchange current II 
\label{fig25}
}
\end{figure}

\end{document}